\documentclass[11pt,a4paper]{article}
\usepackage[english]{babel}
\usepackage[utf8]{inputenc}
\usepackage{jheppub}
\usepackage{braket}
\usepackage{amsmath}
\usepackage{array, url}
\usepackage{fancyhdr}
\usepackage{amsfonts}
\usepackage{amsthm}
\usepackage{mathtools}
\usepackage{bm}
\usepackage{titlesec}
\usepackage{graphicx}
\usepackage{color}
\usepackage{tensor}
\usepackage{mathrsfs}
\usepackage{etoolbox}
\usepackage{tikz}
\usetikzlibrary{positioning}
\usepackage{tensor}
\usepackage[title]{appendix}
\numberwithin{equation}{section}

\newcommand{\be}{\begin{equation}}
\newcommand{\ee}{\end{equation}}
\newcommand{\half}{{1\over2}}

\def\ba{\begin{eqnarray}}
\def\ea{\end{eqnarray}}

\def\/{\over}
\def\ie{\emph{i.e.} }

\def\C{\mathbb{C}}

\def\R{\mathbb{R}}

\def\cC{\mathcal{C}}

\def\cF{\mathcal{F}}
\def\cG{\mathcal{G}}

\def\cO{\mathcal{O}}

\def\cR{\mathcal{R}}

\def\bD{\mathbb{D}}

\def\bN{\mathbb{N}}

\def\dg {\dagger}
\def\p{\partial}

\def\ov {\over}
\def\ov{\over}

\def\rn{\rangle}
\def\ln{\langle}

\def\t{\theta}
\def\s{\sigma}
\def\e{\epsilon}

\def\vphi{\varphi}
\def\a{\alpha}
\def\b{\beta}
\def\d{\delta}
\def\k{\kappa}
\def\g {\gamma}
\def\la {\lambda}

\def\z{\zeta}

\def\n {\nabla}
\def\L{\Lambda}

\def\ra{\rightarrow}

\def\Tr{\mathrm{Tr}}

\def\r{\mathrm}
\def\hc{\text{h.c.}}

\def\_{\hspace{2cm}}
\def\-{\\\notag}
\def\={&=&}

\newcommand{\bea}{\begin{eqnarray}}
	\newcommand{\eea}{\end{eqnarray}}

\newcommand{\bpm}{\begin{pmatrix}}
	\newcommand{\epm}{\end{pmatrix}}

\newcommand{\bit}{\begin{itemize}}
	\newcommand{\eit}{\end{itemize}}

\newcommand{\ben}{\begin{enumerate}}
	\newcommand{\een}{\end{enumerate}}

\newcommand\bsp{\begin{split}}
	\newcommand\esp{\end{split}}

\def\le{\left}
\def\ri{\right}

\def\qq{\qquad}

\def\rn{\rangle}
\def\ln{\langle}

\def\t{\theta}
\def\s{\sigma}
\def\e{\epsilon}

\def\vphi{\varphi}
\def\a{\alpha}
\def\b{\beta}
\def\d{\delta}
\def\k{\kappa}
\def\g {\gamma}
\def\la {\lambda}

\def\z{\zeta}

\def\Tr{\mathrm{Tr}}

\def\r{\mathrm}
\def\hc{\text{h.c.}}
\def\tx{ x}

\def\Label#1{\label{#1}%
	\smash{\hbox to0pt{\raise1ex\hbox{\tiny[#1]}\hss}}}
\def\noLabels{\let\Label=\label}
\def\nobbibitem{\let\bbibitem=\bibitem}

\newcommand{\beqn}{\begin{eqnarray}}
	\newcommand{\eeqn}{\end{eqnarray}}

\def\le{\left}
\def\ri{\right}

\def\qq{\qquad}

\numberwithin{equation}{section}

\newcommand{\subf}[2]{%
	{\small\begin{tabular}[t]{@{}c@{}}
			#1\\#2
	\end{tabular}}%
}

\usepackage{tabu}

\def\sde{\epsilon}

\newcommand{\overbar}[1]{\mkern 1.5mu\overline{\mkern-1.5mu#1\mkern-1.5mu}\mkern 1.5mu}
\DeclareMathOperator\tr{Tr}

\title{Quantum information geometry of driven CFTs}

\author[\flat]{Jan~de Boer,}
\author[\natural]{Victor~Godet,}
\author[\dagger,\sharp]{Jani~Kastikainen,}
\author[\sharp]{Esko~Keski-Vakkuri}

\affiliation[\flat]{Institute for Theoretical Physics and Delta Institute for Theoretical Physics, University of
Amsterdam, PO Box 94485, 1090GL, Amsterdam, The Netherlands}
\affiliation[\natural]{International Centre for Theoretical Sciences (ICTS-TIFR), Tata Institute of Fundamental Research, Shivakote, Hesaraghatta, Bangalore 560089, India}
\affiliation[\dagger]{Université Paris Cité, CNRS, Astroparticule et Cosmologie, F-75013 Paris, France}
\affiliation[\sharp]{Helsinki Institute of Physics and Department of Physics, P.O.Box 64, FIN-00014 University of Helsinki, Finland}

\emailAdd{j.deboer@uva.nl}
\emailAdd{victor.godet@icts.res.in}
\emailAdd{jani.kastikainen@helsinki.fi}
\emailAdd{esko.keski-vakkuri@helsinki.fi}

\abstract{Driven quantum systems exhibit a large variety of interesting and sometimes exotic phenomena. Of particular interest are  driven conformal field theories (CFTs) which describe quantum many-body systems at criticality. In this paper, we develop both a spacetime and a quantum information geometry perspective on driven 2d CFTs. We show that for a large class of driving protocols the theories admit an alternative but equivalent formulation in terms of a CFT defined on a spacetime with a time-dependent metric. We prove this equivalence both in the operator formulation as well as in the path integral description of the theory. A complementary quantum information geometric perspective for driven 2d CFTs employs the so-called Bogoliubov--Kubo--Mori (BKM) metric, which is the counterpart of the Fisher metric of classical information theory, and which is obtained from a perturbative expansion of relative entropy. We compute the BKM metric for the universal sector of Virasoro excitations of a thermal state, which captures a large class of driving protocols, and find it to be a useful tool to classify and characterize different types of driving. For Möbius driving by the $\text{SL}(2,\mathbb{R})$ subgroup, the BKM metric becomes the hyperbolic metric on the disk. We show how the non-trivial dynamics of Floquet driven CFTs is encoded in the BKM geometry via Möbius transformations. This allows us to identify ergodic and non-ergodic regimes in the driving. We also explain how holographic driven CFTs are dual to driven BTZ black holes with evolving horizons. The deformation of the black hole horizon towards and away from the asymptotic boundary provides a holographic understanding of heating and cooling in Floquet CFTs.}

\arxivnumber{2306.00099}

\begin{document}
\maketitle

\section{Introduction}

Non-equilibrium quantum systems take many forms, and their physical understanding remains challenging, particularly when they involve a large number of degrees of freedom. Controlled studies of non-equilibrium physics are possible by driving the system with an external force. The engineering of periodically driven (Floquet) systems in the laboratory has led to the discovery of new non-equilibrium phases of matter \cite{kitagawa_topological_2010,else_floquet_2016}, the properties of which are currently an active area of research (see \cite{mori_floquet_2022} for a review).

Intriguingly, some properties of non-equilibrium systems can be understood using equilibrium concepts. Two standard relations, the Jarzynski equality \cite{jarzynski_nonequilibrium_1997,jarzynski_equilibrium_1997} and the Crooks fluctuation theorem \cite{crooks_entropy_1999}, relate equilibrium free energy differences to non-equilibrium work expectation values, and have been extended to quantum systems \cite{tasaki_jarzynski_2000,kurchan_quantum_2001}. Moreover, progress has been made in understanding slowly driven systems that remain near thermal equilibrium during their evolution. In this case, the information geometry of equilibrium (Gibbs) states has been used to characterize the amount of dissipated work during slow non-equilibrium processes in both classical \cite{salamon_thermodynamic_1983,hoffmann_measures_1989,diosi_thermodynamic_1996,sivak_thermodynamic_2012} and quantum systems \cite{deffner_nonequilibrium_2011,zulkowski_geometry_2012,zulkowski_optimal_2015,scandi_thermodynamic_2019,miller_work_2019,abiuso_geometric_2020}. This paper aims in part to extend these methods to driven conformal field theories.

The idea of applying tools of differential geometry to statistics and thermodynamics has a long history \cite{weinhold_metric_1975,ruppeiner_thermodynamics_1979}, dating back to the work of Rao on parameter estimation \cite{rao_information_1992}. Classical information geometry of a system is obtained by endowing its space of probability distributions, also known as the statistical manifold, with a Riemannian metric. The Fisher information metric, which is the Hessian matrix of the Kullback--Leibler divergence \cite{kullback_information_1951}, is a natural metric to consider and has played a central role in classical information geometry due to its invariance under sufficient statistics, a result known as Chentsov's theorem \cite{cencov_statistical_2000,campbell_extended_1986,ay_information_2015,bauer_uniqueness_2016}.

In the context of quantum systems, each point on the statistical manifold represents a possible density matrix of the quantum system \cite{helstrom_quantum_1969,balian_dissipation_1986,braunstein_statistical_1994,hasegawa_non-commutative_1995,uhlmann_density_1993,uhlmann_geometric_1995,petz_geometries_1996}. However, unlike in classical information geometry, there is no quantum generalization of Chentsov's theorem, and there exists an infinite set of (monotone) metrics that decrease under completely positive trace preserving maps \cite{morozova_markov_1991,petz_monotone_1996}. One such metric is the Bogoliubov--Kubo--Mori (BKM) metric \cite{kubo_statistical-mechanical_1957,mori_transport_1965,petz_bogoliubov_1993}, which is a quantum generalization of the Fisher metric and is obtained by expanding the quantum relative entropy $ S(\rho\lVert \sigma) $ to second order in state perturbations \cite{ingarden_information_1981,petz_riemannian_1996,lesniewski_monotone_1999}. The BKM metric has been used in non-equilibrium quantum thermodynamics \cite{scandi_thermodynamic_2019,abiuso_geometric_2020} and its holographic dual has also been studied in the AdS\slash CFT correspondence \cite{lashkari_canonical_2016,banerjee_connecting_2017}.\footnote{In the holography literature, the BKM metric is often referred to as the quantum Fisher information metric.}

In this work, we develop quantum information geometry in the context of driven two-dimensional conformal field theory (CFT) at finite temperature, where the driving is associated to diffeomorphisms. Applying a diffemorphism to spatially deform a CFT is motivated as a model of a gapless inhomogeneous system at criticality, {\em e.g.}, a spin chain with site-dependent couplings controlled by an enveloping function breaking translational invariance. This has led to an extensive study of so-called rainbow, sine-squared, and Möbius deformations of CFTs \cite{Vitagliano_2010,Katsura_2012,Ram_rez_2014,Ramirez:2015yfa,Ishibashi:2015jba,Ishibashi:2016bey,Okunishi:2016zat,Wen:2016inm,Rodriguez-Laguna_2017,Tamura:2017vbx,Tonni:2017jom,Tada:2017wul,Alba:2018ime,langmann_diffusive_2019,Gluza:2022dqu}. 
By altering the diffeomorphism with time, one obtains non-equilibrium models, {\em e.g.,} quenches \cite{Wen:2018vux,MacCormack:2018rwq,Goto:2021sqx} and Floquet driving \cite{Wen:2018agb,Lapierre:2019rwj}. While there are many experimental platforms for studying Floquet driving, there are also novel developments in quantum simulating conformal field theories by atomic systems. For example, the Ising CFT can be simulated by a chain of Rydberg atoms \cite{bernien2017probing,Keesling_2019,Borish_2020,PhysRevB.104.235109}. Perhaps future developments will allow for quantum simulations of CFTs incorporating driving with spatial inhomogeneities. 

For this work, it is useful to make a clear distinction between various concepts. First, we focus on a family of CFT states that are connected to the thermal state by the action of diffeomorphisms $f$ on a circle, denoted Virasoro states $\sigma_f$. Among those states,
we are interested in trajectories (paths) $\sigma_{f_t}$ associated with a one-parameter family
of diffeomorphisms. Such trajectories can then be relevant in two different types of 
dynamical processes:  {\em Type I Virasoro processes}, where the trajectories result from a unitary time evolution from an initial thermal state of a driven {\em closed} CFT, and
{\em Type II Virasoro processes}, which arise
when the driving is applied by a family of diffeomorphisms $f_t$ to an {\em open} CFT
with Hamiltonian $H_{f_t}$, coupled to a heat bath at constant temperature. 
 
In Type I Virasoro processes, at every instant $t$ the CFT state $\sigma_t$ is a Virasoro trajectory state, $\sigma_t=\sigma_{f_t}$. In this setting, we show in Section \ref{sec:virasoroCFT} how to reinterpret such driving geometrically as driving induced by a time-dependent spacetime metric. In other words, we develop a more detailed Lorentzian construction of the Euclidean one presented in \cite{caputa_quantum_2019,Erdmenger:2021wzc}. The driving itself may be rapid, or it can be very slow so that the corresponding background spacetime metric is approximately flat. In Type II Virasoro processes, the time evolution of an initial thermal state may lead to non-unitary non-equilibrium dynamics, in particular the actual evolved state $\sigma_t$ is difficult to compute. As we discuss in Section \ref{sec:virasorogeometry}, driven open quantum systems coupled to a heat bath may involve a new time scale associated with relaxation into thermal equilibrium. If the driving is sufficiently slow with respect to the relaxation
time scale, the near-equilibrium slow driving may be approximated by a so-called step-equilibration process \cite{nulton_quasistatic_1985,anders_thermodynamics_2013,scandi_thermodynamic_2019,miller_work_2019,abiuso_geometric_2020}. When the slowly driven open system is a CFT, the Virasoro 
trajectories become relevant again, now  as a computational tool for quantifying dissipated work; a more detailed explanation is left to Section \ref{sec:virasorogeometry}.

In the Type I context, the operator formulation \cite{Erdmenger:2021wzc} of the spacetime driving appears to break diffeomorphism invariance of the CFT, because the stress tensor operator expectation value is fixed by the Virasoro algebra as a function of time, but it is not conserved due to explicit time-dependence (which was already anticipated in \cite{Oblak:2017ect}). To resolve this puzzle, we formulate the spacetime driving process using the path integral.  We consider the generating functional of stress tensor correlation functions, the Polyakov action \cite{POLYAKOV1981207}, which yields a conserved stress tensor. Using a novel Lorentzian decomposition of the Polyakov action, we obtain the counterterms that must be added to the operator result to restore diffeomorphism invariance and yield a conserved stress tensor one-point function with the correct trace anomaly. Similar shifts in renormalization scheme have appeared in the past in Euclidean signature and are related to holomorphic factorization \cite{knecht_shifting_1990,lazzarini_integrating_1998, Verlinde:1989hv,verlinde_conformal_1990}.

The Virasoro states generated by spacetime driving span an infinite-dimensional information geometry $\text{Diff}_+S^{1}\slash \text{U}(1)$ which we equip with the BKM information metric, generalizing previous studies on finite-dimensional systems. We compute the explicit form of the metric, find it to be universal up to the central charge and the stress tensor one-point function of the CFT. The resulting  metric belongs to the family of right-invariant homogenous Sobolev $\dot{H}^2$ metrics on $\text{Diff}_+S^{1}\slash \text{U}(1)$. Having constructed the BKM information geometry of Virasoro states, we consider two types of applications: minimization of work dissipation in slowly driven close-to-equilibrium open CFTs (Type II processes), and ergodicity and complexity in unitarily driven Floquet CFTs (Type I processes).

For open systems, the BKM information geometry of a system can be used to find slow-driving protocols that minimize the amount of work dissipated by non-equilibrium effects: driving protocols correspond to trajectories (paths) on the space of Gibbs states and geodesics of the BKM metric minimize dissipation \cite{scandi_thermodynamic_2019,abiuso_geometric_2020}. So far, this method has only been applied to finite-dimensional quantum systems (see for example \cite{mehboudi_thermodynamic_2022}). We generalize the method to slowly driven open CFTs,  where the above trajectories are the Type II Virasoro processes, 
and for which optimal protocols (geodesics) can be found explicitly using Euler--Arnold theory \cite{arnold_sur_1966,arnol2013mathematical,khesin_euler_2003} (see also \cite{Flory:2020dja,erdmenger_complexity_2020,basteiro_quantum_2022,Oblak:2020jek}). For example, in the infinite temperature limit, the BKM metric reduces to the homogeneous Sobolev $\dot{H}^1$ metric on $\text{Diff}_+S^{1}\slash \text{U}(1)$ whose geodesic equation is the Hunter--Saxton equation with known explicit solutions \cite{hunter_dynamics_1991,2002math.....10397K,Lenells1,Lenells2}.

For unitarily driven CFTs in the Type I setting, the information geometry gives a framework to study complexity and ergodicity. 
For this purpose, we consider both periodic Floquet driving and aperiodic Fibonacci and random driving, in the context of the models developed in \cite{Wen:2018agb,Fan:2019upv,Lapierre:2019rwj,Wen:2020wee,Wen:2021mlv}. The prototype examples studied in the literature involve the Möbius subgroup $\text{SL}(2,\mathbb{R})$ of $\text{Diff}_+S^{1}$ which already possesses a rather rich dynamics. It was found in \cite{Wen:2018agb,Wen:2020wee,Han:2020kwp} that Möbius processes can be in two different phases depending on the long time behavior: oscillatory in the non-heating phase and exponential in the heating phase. Floquet driving with more general Virasoro Hamiltonians has also been considered \cite{Fan:2020orx,moosavi_inhomogeneous_2019,Das:2021gts}. The relation between Virasoro process and spacetime metrics explains some of the observations in \cite{Lapierre:2019rwj,PhysRevB.103.224303} related to the ``emergent Rindler space''. Effects related to dissipation and chaos were considered in \cite{Wen:2022pyj,Choo:2022lgm,Gritsev2018}. Floquet CFTs are particularly interesting because they are relevant experimentally, see for example \cite{DissipativeFloquetNature,FloquetDissipationPRX}.
Non-equilibrium dynamics in closed two-dimensional conformal field theories (CFTs) have been extensively studied in the literature \cite{moosavi_inhomogeneous_2019}. In this work, on one hand the information geometry gives a nice and physically motivated visualization of the state evolution during the driving. On the other hand, we can extract and compute various measures of complexity and ergodicity to characterize the different phases and transitions between them.

For CFTs with a holographic gravity dual, we develop a dual interpretation of unitary Type I driving, generalizing previous studies \cite{MacCormack:2018rwq,caputa_quantum_2019,Flory:2020dja,Erdmenger:2021wzc, Goto:2021sqx, Caputa:2020mgb,Caputa:2022zsr,Das:2022pez} (see also \cite{auzzi_time-dependent_2012,auzzi_periodically_2013} for related work). In our case, the driving starts
with the CFT in a thermal state, corresponding to a BTZ black hole in the bulk. The driving leads to an evolution of the shape of the black hole horizon. The horizon oscillates in the non-heating phase. In the heating phase, it deforms towards a point of the boundary. This corresponds to an emergent hot spot in the CFT, a point that is heating while the rest of the system is cooling down \cite{Goto:2021sqx}. This gives a holographic interpretation of the conformal Floquet refrigerator \cite{Wen:2022pyj}.

The structure of the paper is as follows. In the rest of the introduction, we will give a brief summary of our main results. In Section \ref{sec:virasoroCFT}, we introduce the class of universal CFT states of interest, Virasoro states, and show how they arise from driving the CFT with a time-dependent background spacetime metric. Then in Section \ref{sec:virasorogeometry}, we formulate the quantum information geometry of Virasoro states, derive the BKM metric and show how lengths of trajectories on the BKM geometry compute work dissipation in Type II driven open CFTs. In Section \ref{sec:mobiusgeometry}, we return to Type I driving, and restrict our attention to the $\r{SL}(2,\mathbb{R})$ Möbius subspace of the information geometry and apply our formalism to study phases of periodically driven Floquet CFTs. In Section \ref{sec:holographicdual}, we give a holographic interpretation of our results. We conclude with a discussion in Section \ref{sec:discussion} where we also give some future directions. 

The connection between Virasoro states and the Lie group $\text{Diff}_+S^1$ is reviewed in detail in Appendix \ref{app:virasoro}. In Appendix \ref{app:polyakovdecomposition}, we present a complete treatment of the decomposition of the Polyakov action in different parametrizations of the background spacetime metric. A proof of positivity of relative entropy between Virasoro states is relegated to Appendix \ref{app:nonneg}.

\subsection{Summary of results}

We summarize first our results on driving a CFT, then move on to highlight results 
for Type I and II processes.

\paragraph{Virasoro processes from spacetime metrics.} In this work, we consider a CFT$_2$ on the Lorentzian cylinder $\R\times S^1$ at finite temperature. Given two diffeomorphisms of the circle $f$ and $\overbar{f}$, we can define the Virasoro  Hamiltonian \cite{Wen:2016inm,Okunishi:2016zat}
\be\label{defHam}
H_{f,\overbar{f}} = \int_{0}^{2\pi} d x^- \,\frac{ T_{--}( x^-)}{f'( x^-)}+\int_{0}^{2\pi} d x^+ \,\frac{ T_{++}( x^+)}{\overbar{f}'( x^+)} \equiv H_{f} + \overbar{H}_{\overbar{f}}~,
\ee
where $x^\pm$ are lightcone coordinates in which the metric is conformally flat. The corresponding Gibbs state defines a Virasoro state
\be
\sigma_{f,\overbar{f}} =\frac{e^{-\beta H_{f}}}{\tr{e^{-\beta H_{f}}}}\otimes \frac{e^{-\overbar{\beta} \overbar{H}_{\overbar{f}}}}{\tr{e^{-\overbar{\beta} \overbar{H}_{\overbar{f}}}}}~.
\ee
Virasoro states are in one-to-one correspondence with elements $(f,\overbar{f})$ of $\text{Diff}_+S^1\times \text{Diff}_+S^1$ and they can be obtained from the thermal state by unitary action of the Virasoro group. We can define a Virasoro trajectory by considering a one-parameter family of Virasoro states 
\be
\s_t = \s_{f_t,\overbar{f}_t}
\ee
corresponding to a  path $(f_t,\overbar{f}_t)$ on the diffeomorphism group $\text{Diff}_+S^1\times \text{Diff}_+S^1$ parametrized by $t$. Such a trajectory can be physically realized in a Type I process with the driving Hamiltonian 
\be
H_t = -  \int_0^{2\pi}d x^-\,v_t( x^-)\, T_{--}( x^-)+\int_0^{2\pi} d x^+\,\overbar{v}_t( x^+)\, T_{++}( x^+) 
\ee
written in terms of the tangent vectors $(v_t,\overbar{v}_t)$ to the trajectory of inverse diffeomorphisms $(F_t,\overbar{F}_t)$ where $F_t = f_t^{-1}$ and $\overbar{F}_t = \overbar{f}_t^{-1}$. Interestingly, this driving is equivalent to putting the CFT on a generic time-dependent spacetime metric. Building on the ideas of \cite{MacCormack:2018rwq,caputa_quantum_2019,Erdmenger:2021wzc}, we make this correspondence precise and show that the general metric 
\be
ds^2 = e^{\omega}(d\phi +\nu\, dt)(d\phi+\overbar{\nu}\,dt) = e^{\varphi}\,d x^-d x^+
\label{introprocessmetric}
\ee
drives the CFT along a  Virasoro trajectory $(f_t,\overbar{f}_t)$, determined uniquely from the two functions $\nu,\overbar{\nu}$ as given in \eqref{nutrajectory}. More precisely, we show that unitary time evolution along slices of constant $t$ produces  the state $\sigma_t$ on each timeslice. We show this explicitly by using operator methods clarifying some of the results of \cite{Erdmenger:2021wzc}. Expectation values of the stress tensor operator on each constant $t$ slice are fixed by the Virasoro algebra to be
\bea
\tr{(\sigma_t\,  T_{--})} \= f'_t( x^-)^2\,\langle T\rangle_{\beta}-\frac{c}{24\pi}\{f_t(\tx^-), x^-\}~,\\
\tr{(\sigma_t\,  T_{++})}  \=\overbar{f}'_t( x^+)^2\,\langle T\rangle_{\overbar{\beta}}-\frac{c}{24\pi}\{\overbar{f}_t(\tx^+),\tx^+\}~,
\eea
and $\tr{(\sigma_t\,  T_{+-}}) = 0$. In these expressions, the Schwarzian  derivative does not act on the $t$ label. However, the explicit time-dependence implies that this stress tensor is not conserved in the background metric and does not satisfy the correct Weyl anomaly. In the path integral description, the Polyakov action $W[g]$ provides the generating function for stress tensor correlators. Using a new ``Lorentzian'' decomposition of the Polyakov action
\begin{equation}
    W[g] = \Gamma[\nu] + \overbar{\Gamma}[\overbar{\nu}] +I_{\text{ct}}[\omega,\nu,\overbar{\nu}]~,
\end{equation}
we can distinguish different contributions to the one-point function $\langle  T_{ab} \rangle_g$. The operator expectation values $\tr{(\sigma_t\,  T_{ab})}$ are corrected by the contribution from a local counterterm $I_{\text{ct}}[\omega,\nu,\overbar{\nu}]$. The Hamiltonian picture gives the answer consistent with holomorphic factorization in $\nu$ and $\overbar{\nu}$ but insisting on holomorphic factorization breaks the two-dimensional diffeomorphism symmetry. We derive the additional contributions that need to be added to restore this symmetry and which complete $\tr{(\sigma_t\, T_{ab})}$ to make it conserved and with the correct Weyl anomaly.

\paragraph{BKM geometry of Virasoro states.} The space of Virasoro states can be equipped with the Bogoliubov--Kubo--Mori (BKM) information metric obtained from a perturbative expansion of relative entropy. The relative entropy between two Virasoro states is
\begin{equation}
S(\sigma_{f_2}\lVert \sigma_{f_1}) = \frac{c\beta}{48\pi}\int_{0}^{2\pi}d x\,\biggl[\biggl(\frac{\mathcal{F}''( x)}{\mathcal{F}'( x)}\biggr)^{2}+\frac{48\pi\langle T\rangle_{\beta}}{c}\,[\mathcal{F}'( x)^{2}-1]\biggr]
\label{ScurlyFsummary}
\end{equation}
where $\langle T\rangle_{\beta}$ is the stress tensor one-point function in the thermal state, $c$ is the central charge, and $\mathcal{F}$ is the composition 
\begin{equation}
\mathcal{F} = f_2\circ f_1^{-1}
\end{equation}
where $f_{1,2}$ are two diffeomorphisms of the circle. The BKM metric at the point $f = \text{id} $ on $\text{Diff}_+S^1$ contracted with a pair of tangent vectors $u,v$ follows to be
\begin{equation}\label{endresult}
\mathcal{G}_{f=\text{id}}(u,v) = \frac{c\beta}{24\pi}\int_{0}^{2\pi}d x\,\biggl[u''( x)\,v''( x)+\frac{48\pi\langle T\rangle_{\beta}}{c}\,u'( x)\,v'( x)\biggr]~.
\end{equation}
It turns out that unitary invariance of relative entropy makes the BKM metric right-invariant under the action of $\text{Diff}_+S^1$. The end result (\ref{endresult}) can be identified as a two-parameter family of homogenous $\dot{H}^2$ Sobolev metrics. These results extend previous studies of relative entropy between perturbative close mixed states \cite{faulkner_bulk_2015,lashkari_modular_2016,sarosi_relative_2016,sarosi_relative_2017,faulkner_nonlinear_2017,may_holographic_2018,ugajin_perturbative_2018}.

Trajectories in the BKM geometry can be used to quantify work dissipation in Type II driven open quantum systems in contact with a heat bath \cite{scandi_thermodynamic_2019}. This involves the BKM action
\be
S_\r{BKM}(t)  = \int^{t}_{0} ds\, \cG_{f_s}(\dot{f}_s,\dot{f}_s)
\ee
which is the action of a point particle associated with the trajectory in the BKM geometry. When the CFT is coupled to a bath and when the driving is slow, the non-equilibrium dynamics of the CFT can be approximated by a step-equilibration process \cite{scandi_thermodynamic_2019}. In this case, the BKM action gives the total amount of dissipated work at time $t$:
\be
W_\r{diss}(t) = S_\r{BKM}(t)~.
\ee
Optimal slowly driven processes are then obtained by minimizing the BKM action which corresponds to geodesics in the BKM geometry. The geodesic BKM distance between two points gives a notion of circuit complexity. It is a particular measure for the minimal number of gates, built from Virasoro generators, required to map a source state to a target state.

\paragraph{Möbius processes and Floquet CFT.}

We will also consider Möbius states that correspond to Virasoro states $\sigma_f$ where the diffeomorphism $f$ belongs to the $\text{SL}(2,\mathbb{R})$ subgroup of $\text{Diff}_+S^1$. The space of Möbius states is the unit disk $\r{SL}(2,\R)\slash \text{U}(1) = \mathbb{D} $ and the BKM metric becomes the hyperbolic Poincaré metric on the disk
\be
\bD = \{z\in \C\mid |z|<1\},\quad ds^2_\r{BKM} =\frac{c\b(\g+1)}{24}\frac{4 |dz|^2}{(1-|z|^2)^2},\quad |z|<1~,
\ee
with a multiplicative factor involving the thermal one-point function of the stress tensor $\g = 48\pi \ln T\rn_\b /c$. As a result the Möbius trajectories can be represented by trajectories on the Poincaré disk. We study in detail the case of periodic unitary driving, {\em i.e.} Floquet CFTs \cite{Wen:2018agb,Fan:2019upv,Wen:2020wee,Wen:2021mlv}. These Type I processes correspond to iterations of Möbius transformations on the BKM geometry and have a solvable albeit rich dynamics. Viewing the Möbius processes as repeated measure-preserving transformations of the state space also makes contact with ergodic theory. 

The BKM representation gives a clear distinction between the two possible phases \cite{Wen:2018agb} of Möbius processes known as heating and non-heating phases. The heating phase corresponds to trajectories diverging towards the boundary of the unit disk while they remain bounded in the non-heating phase, see Figure \ref{Fig:KMphases}. In the heating phase, we observe a linear growth of the complexity defined by the geodesic BKM distance, while in the non-heating phase, the complexity oscillates due to the process being confined to the region $\cR$. Within the non-heating phase, we identify ergodic and non-ergodic regimes. For generic parameters, the process is ergodic in the sense that its trajectory is confined and becomes dense in an available region $\cR$ of the BKM geometry. The ergodic behavior is characterized by the period angle $\t$ as illustrated in Figure \ref{Fig:ergodic}. Non-ergodic regimes with periodic trajectories correspond to $\t$ becoming a rational multiple of $\pi$.

The BKM action $S_\r{BKM}$ is an interesting quantity to characterize both Type I and II processes. It diverges exponentially in the heating phase and grows linearly in the non-heating phase. For the latter phase, we can define the finite BKM rate
\be
\eta_\r{BKM} = \lim_{t\to\infty } \frac{S_\r{BKM}(t)}{t}
\ee
which characterize processes in the non-heating phase. Another quantity that can be defined from the BKM geometry is the hyperbolic area of the region $\cR$. We compute and compare these information quantities which measure  the ``size'' of the region of the state space covered by the process.

\paragraph{Holographic dual.} The holographic dual of a Type I Virasoro process is an AdS$_3$ spacetime with the time-dependent boundary metric \eqref{introprocessmetric}, a setup also discussed in \cite{MacCormack:2018rwq, Erdmenger:2021wzc,Das:2022pez,Liska:2022vrd}. In our case,
the dual spacetime can be described as a BTZ black hole with a dynamical horizon corresponding to the effect of a time-dependent bulk diffeomorphism. We show that the curve parametrizing the location of the horizon in a quasi-static approximation in a constant time slice at time $t$ is given by
\be
r_t(x) = {\pi f_t'(x)\/\b} \sqrt{1-{\b^2\/4\pi^2} {f_t''(x)^2\/f_t'(x)^4}}~.
\ee
For Type I Möbius processes, the holographic dual of the heating phase corresponds to the black hole deforming towards a point of the asymptotic boundary, as depicted in Figure \ref{Fig:processSpacetime}. This produces a heating point in the CFT while the rest of the system cools down \cite{Goto:2021sqx}. Let $d_t$ be the distance between the horizon and the conformal boundary at time $t$. We have 
\be
d_\text{hor}(t)  = d_\text{hor}(0)- \la_L t~
\ee
so the distance decreases linearly with a rate equal to the Lyapunov exponent. Interestingly, the Möbius trajectory in the BKM geometry approximately mirrors the position of the black hole in the dual spacetime.

\section{ Virasoro processes in conformal field theory}\label{sec:virasoroCFT}

In this section we define Virasoro trajectories: sequences  of Virasoro states corresponding to a path on $\r{Diff}_+S^1/\r{U}(1)$ (or more precisely its universal cover). We will focus on unitary processes where the trajectory is realized by a time-dependent Hamiltonian. We will see that this is equivalent to putting the CFT on a time-dependent metric. In the next section, we will discuss a different way to realize a Virasoro trajectory using a step-equilibration process, corresponding to slowly driving the CFT in contact with a heat bath.

\subsection{Virasoro states, orbits, and trajectories}\label{subsec:virstatedefs}

We are interested in the action of diffeomorphisms of $S^1$ in the Hilbert space of a CFT on the Lorentzian cylinder $\mathbb{R}\times S^1$ equipped with a flat metric (a general curved metric is considered in the next section). In distinction to previous studies \cite{caputa_quantum_2019,erdmenger_complexity_2020,Erdmenger:2021wzc,erdmenger_complexity_2022}, we are interested in orbits of diffeomorphisms generated from Gibbs states rather than pure states. 

We begin with the group of orientation preserving diffeomorphisms
\begin{equation}
\text{Diff}_+S^{1} = \{h\colon S^{1} \rightarrow S^{1} \,\lvert\,h \text{ is smooth and orientation-preserving}\}
\end{equation}
and its universal cover
\begin{equation}
\widetilde{\text{Diff}}_+S^{1} = \{f\colon \mathbb{R}\rightarrow \mathbb{R} \,\lvert\,f(x+2\pi) = f(x) +2\pi,\, f'(x)>0,f\text{ is smooth}\}
\label{universalcover}
\end{equation}
which we use to establish an isomorphism between the Lie algebra of the tangent vectors $T_{\text{id}}\text{Diff}_+S^{1}$ and the algebra of vector fields $ \text{Vect}\,S^{1} $ (further details of this section are discussed in Appendix \ref{app:virasoro}). The central extension of this Lie algebra, represented in the Hilbert space $ \mathcal{H} $ of a CFT on a spatial $S^1$, is the Virasoro algebra\footnote{The form of the algebra that appears in radial quantization is obtained by $ L_0^{\text{rad}} = L_0 + \frac{c}{24} $.}  generated by the operators $ L_n $ with the
commutation relations
\begin{equation}
[L_n,L_m] = (n-m)\,L_{n+m}+\frac{c}{12}n^{3}\,\delta_{n,-m}
\label{virasoroalgebra}
\end{equation}
with the hermiticity condition $ L_n^{\dagger} = L_{-n} $. More precisely, a CFT has two copies of commuting Virasoro algebras that live in tensor products, one $L_n\otimes\mathbf{1}$ in the right-moving and
one in the left-moving sector $\mathbf{1}\otimes L_n$. The generator of time and space translations {\em i.e.} the Hamiltonian and the momentum are given by
\begin{equation}
H  = L_0\otimes \mathbf{1} + \mathbf{1}\otimes L_0, \quad P  = L_0\otimes \mathbf{1} - \mathbf{1}\otimes L_0.
\label{HandP}
\end{equation}
A Gibbs thermal equilibrium state is then
\begin{equation}
\sigma_{\beta,\overbar{\beta}} \equiv \sigma_\beta \otimes \sigma_{\overbar{\beta}} = \frac{e^{-\beta L_0}}{\tr e^{-\beta L_0}} \otimes \frac{e^{-\overbar{\beta}L_0}}{\tr e^{-\overbar{\beta}L_0}}
\end{equation}
where the parameters $\beta,\overbar{\beta}>0$ can be converted to a temperature $T$ and a chemical potential for angular momentum $P$. 

In Lorentzian signature, the conformal group of the cylinder $S^1\times \mathbb{R}$ equipped with the flat metric is isomorphic to $\widetilde{\text{Diff}}_+S^{1}\times \widetilde{\text{Diff}}_+S^{1}$ \cite{kong_algebraic_2009,Schottenloher:2008zz} whose Lie algebra coincides with $\text{Vect}\,S^1\times \text{Vect}\,S^1 $ \cite{oblak_bms_2017}. The action of the conformal group can be represented by unitary operators:
\begin{equation}
\widetilde{\text{Diff}}_+S^{1}\times \widetilde{\text{Diff}}_+S^{1} \ni (F,\overbar{F}) \mapsto V_{F,\overbar{F}} = V_F\otimes \overbar{V}_{\overbar{F}},
\end{equation}
where $V_F$ acts on the right-movers while the complex conjugate representation $\overbar{V}_{F}$ acts on the left-movers (see Appendix \ref{app:virasoro} for details). We can thus consider the orbit of the conformal group:
\begin{equation}
\text{Vir}_{\beta ,\overbar{\beta}} \equiv \left\{V_F\, \sigma_\beta V^\dagger_F  \otimes \overbar{V}_{\overbar{F}}\, \sigma_{\overbar{\beta}}\overbar{V}^\dagger_{\overbar{F}}  \,\Big\lvert\, F,\overbar{F} \in \widetilde{\text{Diff}}_+S^{1} \right\} \equiv \text{Vir}_\beta \times \text{Vir}_{\overbar{\beta}}\ ,
\end{equation}
we call $\text{Vir}_{\beta,\overbar{\beta}}$ a {\em Virasoro orbit}, and its elements are {\em Virasoro states}
\begin{equation}
    \sigma_{f,\overbar{f}} = \sigma_f\otimes  \sigma_{\overbar{f}},\quad \sigma_f = V_F\,\sigma_\beta\,V_F^{\dagger}.
    \label{virasorostatedefinition}
\end{equation}
Because the thermal state $\sigma_{\beta,\overbar{\beta}}$ is invariant under rotations $x\mapsto x+c$, Virasoro states
are in one-to-one correspondence with $\widetilde{\text{Diff}}_+S^{1}/\,\mathbb{R}\times \widetilde{\text{Diff}}_+S^{1}/\,\mathbb{R}$, where $\mathbb{R}$ is the universal cover of
the group $\text{Rot}\,S^1=\text{U}(1)$. Due to the unitary action, {\em e.g.} von Neumann entropy stays
constant in each orbit. Later, we will equip the Virasoro orbits with a quantum information metric derived from relative entropy. These are the quantum statistical manifolds where we can
study quantum information geometry with detailed computational control.

There is an alternative way to present the Virasoro states \eqref{virasorostatedefinition}. Using the operator
\begin{equation}
T(x) = \frac{1}{2\pi}\sum_{n=-\infty}^{\infty}L_n\,e^{in x},
\label{Tphi}
\end{equation}
we can define the Virasoro Hamiltonian
\begin{equation}
    H_f \equiv \int_0^{2\pi}dx\,\frac{T(x)}{f'(x)}~.
    \label{virasorohamiltoniandefinition}
\end{equation}
In Appendix \ref{app:virasoro}, we show in detail that they can be obtained by unitary action from $L_0$ as
\begin{equation}
    H_f = V_F\,L_0\,V_F^\dagger
\end{equation}
where somewhat counterintuitively $F=f^{-1}$ is the inverse diffeomorphism and we are neglecting an additive constant that cancels in normalized states. The fact that the action with $F=f^{-1}$ generates \eqref{virasorohamiltoniandefinition} with $f'( x)$ in the denominator explains the notation $\sigma_f$. As a result, a Virasoro state can be defined as a Gibbs state of the Virasoro Hamiltonian, with the equality
\begin{equation}
\sigma_{f,\overbar{f}} = \frac{e^{-\beta H_f}}{\tr{e^{-\beta H_f}}}\otimes \frac{e^{-\overbar{\beta}H_{\overbar{f}}}}{\tr{e^{-\overbar{\beta}H_{\overbar{f}}}}} = V_{F,\overbar{F}} \,\sigma_{\beta,\overbar{\beta}} V^\dagger_{F,\overbar{F}}.
\label{virasorostate}
\end{equation}
Now the chiral light-ray components of the CFT stress tensor are given in terms of \eqref{Tphi} as
\begin{equation}
     T_{--}( x^-) = T( x^-)\otimes \mathbf{1},\quad  T_{++}( x^+) = \mathbf{1}\otimes T(- x^+),
    \label{TtildeT}
\end{equation}
where $x^{\pm}$ denote light-ray coordinates in which the flat metric on $S^1\times \mathbb{R}$ takes the form $ds^2 = dx^-dx^+$. We define the {\em Virasoro Hamiltonian} as
\begin{equation}
H_{f,\overbar{f}} = H_f\otimes \mathbf{1}  + \mathbf{1}\otimes H_{\overbar{f}} \equiv \int_{0}^{2\pi} d x^- \,\frac{ T_{--}( x^-)}{f'( x^-)}+\int_{0}^{2\pi} d x^+ \,\frac{ T_{++}( x^+)}{\overbar{f}'( x^+)}.
\label{HfT}
\end{equation}
and the Virasoro angular momentum as
\begin{equation}
    P_{f,\overbar{f}} = H_f\otimes \mathbf{1} - \mathbf{1}\otimes H_{\overbar{f}}.
\end{equation}
A Virasoro state \eqref{virasorostate} can thus be alternatively viewed as an equilibrium Gibbs state for a {\em deformed CFT} whose Hamiltonian is equal to $H_{f,\overbar{f}}$ and angular momentum to $P_{f,\overbar{f}}$. When $f = \overbar{f} = \text{id}$ are identity diffeomorphisms, \eqref{HandP} is recovered.

We will also be interested in paths along Virasoro orbits. Consider two one-parameter families of Virasoro Hamiltonians $H_{f_t}$ and $H_{\overbar{f}_t}$ where we take
\begin{equation}
    f_0 = \overbar{f}_0 = \text{id},\quad f_1 = f,\quad \overbar{f}_1 = \overbar{f}.
\end{equation}
The Hamiltonians define a family of Virasoro states which we call a {\em Virasoro trajectory}
\begin{equation}
\sigma_{f_t,\overbar{f}_t} = \frac{e^{-\beta H_{f_{t}}}}{\tr{e^{-\beta H_{f_t}}}}\otimes \frac{e^{-\overbar{\beta}H_{\overbar{f}_{t}}}}{\tr{e^{-\overbar{\beta}H_{\overbar{f}_t}}}}, \quad t \in [0,1],
\label{virasorotrajectory}
\end{equation}
such that the initial and final states coincide with $ \sigma_{\beta,\overbar{\beta}} $ and $\sigma_{f,\overbar{f}}$ respectively.

The trajectory $\sigma_{f_t} = V_{F_t}\,\sigma_\beta\,V_{F_t}^\dagger$ with $F_t = f_t^{-1}$ ($t$ being a parameter) is generated by a time-dependent Hermitian operator $ G_t $ via
\begin{equation}
\dot{V}_{F_t} = -i\,G_t\,V_{F_t}
\label{processgenerator}
\end{equation}
where the unitary representation $V_{F_t}$ creating the Virasoro state is explicitly
\begin{equation}
V_{F_t} = \mathcal{T}\exp{\biggl(-i\int_{0}^{t} ds\,G_{s}\biggr)}
\end{equation}
with time decreasing from the left to the right. A Virasoro trajectory is thus determined completely by specifying the initial state $\sigma_\beta$ and the generator $G_t$ which takes the form\footnote{In \eqref{generatorGt}, there is also an irrelevant additive constant whose effect is to produce an overall phase in $V_{F_t}$ which cancels in states.}
\begin{equation}
    G_t = \int_{0}^{2\pi}d x\,v_t( x)\,T( x),\quad v_t( x) = (\dot{F}_t\circ F^{-1}_t)( x),
    \label{generatorGt}
\end{equation}
where $v_t \in T_{F_t}\text{Diff}_+S^1$ is the tangent vector of the curve $F_t\in \text{Diff}_+S^1$ (see Appendix \ref{app:virasoro} for more details). As a result, the unitary operator creating a Virasoro state $\sigma_f$ from $\sigma_\beta$ satisfies $V_F = V_{F_1} $, so we can view $V_{F_t}$ as a continuous circuit of gates that synthesizes the state $\sigma_f$.

Virasoro trajectories \eqref{virasorotrajectory} can arise from unitary time-evolution with a time-dependent driving Hamiltonian $H_t = G_t$ in a closed CFT which we call a Type I Virasoro process. In this case, the state $\sigma_t$ obtained by unitary evolution from the initial state $ \sigma_{\beta,\overbar{\beta}} $ of the CFT coincides with the trajectory at all times $\sigma_t = \sigma_{f_t,\overbar{f}_t} $. Virasoro trajectories can also appear in a CFT which is an open system interacting with an environment and whose state evolution $\sigma_t$ is determined by a master equation. In Type II Virasoro processes, the actual state evolution $\sigma_t$ is hard to compute, but the trajectory \eqref{virasorotrajectory} determines the family of equilibrium states $\pi_t$ of the CFT as $\pi_t = \sigma_{f_t,\overbar{f}_t} $. Details of Type II processes are presented in Section
\ref{subsec:dissipation}.

\subsection{Virasoro process in CFT$ _2 $ driven by a spacetime metric}\label{sec:hamiltonianderivation}

In this section, we will show how a Type I Virasoro process can be understood as a CFT in a spacetime with a time-dependent background metric. We interpret the 
time-dependent metric driving the system, and show that this gives rise to a time-dependent Hamiltonian with the unitary time-evolution leading to a Virasoro trajectory. This expands on the ideas of \cite{Erdmenger:2021wzc}.

\paragraph{Driving Hamiltonian.} We put the CFT$_2$ on the cylinder $\mathbb{R}\times S^1 $ parameterized by $ \phi^A = (t,\phi) $ where $ \phi \sim \phi + 2\pi $ and $t\in \mathbb{R}$ is the time. The metric on the cylinder on which the CFT lives is taken to be of the general form
\begin{equation}
ds^2 = g_{AB}\,d\phi^A d\phi^B = e^{\omega}(d\phi + \nu dt)(d\phi + \overbar{\nu} dt)
\label{CFTg}
    \end{equation}
with three functions $ \omega,\nu$ and $\overbar{\nu}$. This parametrization of the metric is closely related to the Beltrami parametrization in light-ray coordinates $ \phi^{\pm} = \phi \pm t $, see Appendix \ref{app:polyakovdecomposition} for details.

We introduce new coordinates $x^a = (\tau,x)$, such that $ x\sim x+2\pi$ and $\tau\in \mathbb{R}$, on the cylinder and we define $x^\pm =x \pm\tau$. They are related to the original coordinates $\phi^\pm$ by a diffeomorphism 
\begin{equation}
 x^{a} = F^{a}(\phi^-,\phi^+)
\label{tildediffeo}
\end{equation}
chosen so that the metric \eqref{CFTg} is conformally flat\footnote{We adopt the notation where components of a tensor field in two coordinate systems are distinguished only by their indices (for example $g_{AB}$ and $g_{ab}$).}
\begin{equation}
ds^{2} = g_{ab}\, dx^a dx^b =  e^{\varphi}\,d x^{-}d x^{+}~.
\label{metricTilde}
\end{equation}
The relation between the diffeomorphism \eqref{tildediffeo} and the parameters $ \nu,\overbar{\nu} $ is given by
\begin{equation}
\nu = \frac{\partial_t F^{-}}{\partial_\phi F^{-}}, \quad \overbar{\nu} = \frac{\partial_t F^{+}}{\partial_\phi F^{+}},
\label{nus}
\end{equation}
and the relation between the Weyl factors is
\begin{equation}
\varphi = \omega -\log{(\partial_{\phi}F^{-})}-\log{(\partial_{\phi}F^{+})}.
\label{varphimetric}
\end{equation}
The requirement of the diffeomorphism \eqref{tildediffeo} being orientation-preserving is $\nu < \overbar{\nu}$ which we will assume from now on. The flat Minkowski metric is then obtained by setting $ \nu = -1 $ and $ \overbar{\nu} =1 $ corresponding to $  x^{\pm} =F^{\pm}(\phi^-,\phi^+) = \phi^\pm $.

In the conformally flat coordinates $ x^{a}$, the components $  T_{ab}(x^-,x^+) $ of the stress tensor can be written as
\begin{equation}
 T_{--}( x^-) = T( x^-)\otimes \mathbf{1},\quad  T_{++}( x^+) = \mathbf{1}\otimes T(- x^+),\quad  T_{-+}( x^-,x^+)= 0,
\label{Tppmmvirasoro}
\end{equation}
as an operator acting on the two chiral sectors\footnote{The minus sign in the exponent in $  T_{++} $ also appears in \cite{moosavi_inhomogeneous_2019}. It is consistent with Wick rotation $ \tau = -i\tau_E $ under which $  x^{-} \rightarrow w = x + i\tau_E $ and $  x^{+} \rightarrow \overbar{w} = x - i\tau_E $ so that $  T_{--} $ becomes the holomorphic stress tensor (that depends only on $ w $) and $  T_{++} $ the antiholomorphic one (that depends only on $ \overbar{w} $).} and where $T(\phi)$ is defined in \eqref{Tphi}. The stress tensor components in the original coordinates $\phi^A$ are then obtained as
\begin{equation}
T_{AB}(\phi^-,\phi^+) = \frac{\partial F^a}{\partial \phi^A}\frac{\partial F^{b}}{\partial \phi^B}\, T_{ab}( x^-,x^+)
\label{diffeogT}~,
\end{equation}
which relates the stress tensor operators in the $\phi^\pm$ coordinates and $x^\pm$ coordinates. Notice that this equation does not have a central term, because $T_{ab}$ is an operator valued tensor field that transforms as a normal tensor under diffeomorphisms.

The driving Hamiltonian is the Noether charge associated with the timelike vector field $\partial_t$ and takes the form\footnote{The sign of $ H_t $ is fixed by the requirement that \eqref{finaldriving} reduces to $ L_0+\overbar{L}_0 $ for the flat metric.} \cite{Erdmenger:2021wzc}
\begin{equation}
H_{t}  = -\int_{0}^{2\pi} d\phi \sqrt{-g}\, T_{t}^{\;\;\,t}~.
\label{generalHamiltonian}
\end{equation}
By using \eqref{CFTg} and \eqref{diffeogT} it becomes
\begin{equation}
H_t = \int_{0}^{2\pi} d\phi\,\left[-(\partial_\phi F^-)(\partial_t F^{-})\, T_{--}( x^{-})+(\partial_\phi F^{+})(\partial_t F^{+})\, T_{++}( x^{+}) \right],
\label{finaldriving}
\end{equation}
where the Weyl factor $ \omega $ has cancelled, because \eqref{generalHamiltonian} is Weyl invariant. This matches with the expression in \cite{Erdmenger:2021wzc} where $ H_t $ arises as the generator of time-translations along a non-trivial foliation of the cylinder in a fixed flat background metric. In our case, $ H_t $ is the generator along a fixed foliation, but with a non-trivial background metric. These two interpretations of $ H_t $ are of course equivalent.

\paragraph{Virasoro trajectories.} We will now show that unitary evolution in the metric $g$ generates a time-dependent Virasoro state (a Virasoro trajectory) on constant $t$ slices of the Lorentzian cylinder $\R \times S^1$. To this end, consider a curve
\begin{equation}
t\mapsto (F_t,\overbar{F}_{t})\in \widetilde{\text{Diff}}_+S^1\times \widetilde{\text{Diff}}_+S^1
\label{doubledtrajectory}
\end{equation}
on the conformal group parametrized by $t$. This curve determines a unique two-dimensional diffeomorphism \eqref{tildediffeo} via the formula
\begin{equation}
 x^- = F^-(\phi^-,\phi^+) = F_{t}(\phi),\quad  x^+ =F^+(\phi^-,\phi^+) = \overbar{F}_{t}(\phi)~.
\label{diffeoprocess}
\end{equation}
As a result, the curve \eqref{doubledtrajectory} determines (up to the Weyl factor) a two-dimensional metric of the form \eqref{CFTg} with
\begin{equation}
    \nu = \frac{\dot{F}_t(\phi)}{F'_t(\phi)},\qquad \overbar{\nu} =\frac{\dot{\overbar{F}}_t(\phi)}{\overbar{F}'_t(\phi)}.
    \label{nutrajectory}
\end{equation}
The argument can also be reversed: given a two-dimensional metric $g$ of the form \eqref{CFTg}, it determines a unique curve \eqref{doubledtrajectory} on the conformal group via \eqref{nutrajectory}. Hence we have a one-to-one mapping between conformal classes of two-dimensional metrics on $S^1\times \mathbb{R}$ and curves on $\widetilde{\text{Diff}}_+S^1\times \widetilde{\text{Diff}}_+S^1$.

An alternative parametrization of the two-dimensional diffeomorphism in terms of the Virasoro trajectory is to write $ x^- = G_t(\phi^-)$ and $ x^+ = \overbar{G}_t(\phi^+)$ as in \cite{Erdmenger:2021wzc} which is obtained from \eqref{diffeoprocess} using $G_t = F_t \circ r_{t}$ and $\overbar{G}_t = \overbar{F}_t\circ r_{-t}$ where $r_t(\phi) = \phi+t$ is a translation along the circle.   They both lead to the same state evolution, because $F$ and $F\circ r_t$ produce the same Virasoro state $\sigma_F$ under unitary action on the thermal state. Here we choose the parametrization \eqref{diffeoprocess} in which the tangent vectors \eqref{uubarnunubar} take a more natural form.

Consider now the time evolution of the CFT in the background metric \eqref{CFTg}. Starting from an initial thermal state $ \sigma_{\beta,\overbar{\beta}} $ at $t=0$, the state of the CFT at a later time is
\begin{equation}
\sigma_t =  U_t\, \sigma_{\beta,\overbar{\beta}}\,U_t^{\dagger},
\label{unitaryevolution}
\end{equation}
where the unitary evolution is generated by the Hamiltonian operator \eqref{finaldriving} as 
\begin{equation}
U_t =  \mathcal{T}\exp{\biggl(i\int_{0}^{t} ds\,H_{s}\biggr)}.
\label{CFTunitary}
\end{equation}
Using \eqref{uubarnunubar}, the driving Hamiltonian \eqref{finaldriving} can be written as
\begin{equation}
H_t = -  \int_0^{2\pi}d x^-\,v_t( x^-)\, T_{--}( x^-)+\int_0^{2\pi} d x^+\,\overbar{v}_t( x^+)\, T_{++}( x^+) 
\label{finalHam}
\end{equation}
where we performed changes of integration variables from $\phi$ to $x^\pm$ in the two integrals respectively. The tangent vectors of the curve $(F_t,\overbar{F}_t)$ are given by
\begin{equation}
v_t(x^-) = (\dot{F}_t \circ F_t^{-1})(x^-) ,\quad \overbar{v}_t(x^+) = (\dot{\overbar{F}}_t \circ \overbar{F}_t^{-1})(x^+),
\label{uubarnunubar}
\end{equation}
where the inverse $F_t^{-1}$ is defined with $t$ being a parameter. We also denote by $\dot{F}_t$ the diffeomorphism $\phi\mapsto \p_t F_t(\phi)$.

We identify \eqref{finalHam} as the generator of the Virasoro trajectory \eqref{doubledtrajectory} on the space of density matrices so that \eqref{CFTunitary} is a tensor product of two Virasoro unitaries up to a phase
\begin{equation}
U_t \propto V_{F_t}\otimes \overbar{V}_{\overbar{F}_{t}}
\label{CFTUt}
\end{equation}
where the Virasoro unitary is
\begin{equation}
V_{F_t} = e^{i\alpha(F_t)}\,\mathcal{T}\exp{\biggl(-i\int_{0}^{t} ds\,\int_0^{2\pi}d x^-\,u_{s}( x^-)\, T( x^-)\biggr)}
\end{equation}
and we used \eqref{Tppmmvirasoro}. The expression for the projective phase $ \alpha(F_t) $ is given in \eqref{Virasorophase}. In \eqref{CFTUt}, the unitary $\overbar{V}_{F}$ is the complex conjugate representation of the Virasoro group and it has the same expression as $ V_F $, but with $ i\rightarrow -i $ everywhere (including the Fourier expansion of the stress tensor). See Appendix \ref{app:virasoro} for more details. 

As a result, the state \eqref{unitaryevolution} of the CFT at time $t$ is a Virasoro states
\begin{equation}
\sigma_t = \sigma_{f_t,\overbar{f}_t} = \frac{e^{-\beta H_{f_t}}}{\tr{e^{-\beta H_{f_t}}}}\otimes\frac{e^{-\overbar{\beta} H_{\overbar{f}_t}}}{\tr{e^{-\overbar{\beta} H_{\overbar{f}_t}}}},
\label{virasorotrajectoryCFT}
\end{equation}
where the phase in \eqref{CFTUt} has cancelled. Hence the unitary evolution in the background metric \eqref{CFTg} evolves the system along a trajectory in the space of Virasoro states. We have a one-to-one mapping between conformal classes of metrics and Virasoro trajectories.

The information about the Virasoro trajectory is contained in the diffeomorphism \eqref{metricTilde} that maps the general metric \eqref{CFTg} to its conformally flat form. The Weyl factor, which determines curvature of the metric, does not play a role in the determination of the Virasoro trajectory. Hence the same Virasoro trajectory \eqref{virasorotrajectoryCFT} is generated regardless if the background metric is curved or not. This answers a puzzle regarding curved backgrounds in \cite{Erdmenger:2021wzc}.

The evolution in the background metric is not necessarily restricted to start from a thermal state, but it can start from any initial state. However, time-evolution in the metric will always generate the action of a Virasoro unitary operator on the initial state dressing it with Virasoro hair. In this work, we choose the initial state to be a thermal state for two reasons. First, it is a full-rank state so that the relative entropy, which is the basis of the BKM information geometry, is well-defined. Second, it gives a stress tensor expectation value which is just a constant.

\paragraph{Stress tensor expectation value.} The state can be diagnosed from the one-point function of the stress tensor. Let $ f_t = F^{-1}_t $ denote the inverse of $ F_t $ at fixed $ t $. Then the Virasoro algebra implies that (see Appendix \ref{app:virasoro})
\begin{equation}
V_{f_t} T_{--}( x^-) V_{f_t}^\dagger = f_t'( x^-)^2\,  T_{--}(f_t( x^-))- \frac{c}{24\pi}\{f_t( x^-), x^-\}
\end{equation}
where the Schwarzian is computed at fixed $ t $ and similarly for the left-moving sector. Using cyclicity of the trace, we get
\begin{align}
\tr{(\sigma_t\,  T_{--})} &= f'_t( x^-)^2\,\langle T\rangle_{\beta}-\frac{c}{24\pi}\{f_t(\tx^-), x^-\}\nonumber\\
\tr{(\sigma_t\,  T_{++})}  &=\overbar{f}'_t( x^+)^2\,\langle T\rangle_{\overbar{\beta}}-\frac{c}{24\pi}\{\overbar{f}_t(\tx^+),\tx^+\}\nonumber\\
\tr{(\sigma_t\,  T_{-+})} &= 0,
\label{Ttildeval}
\end{align}
where we have defined
\begin{equation}
    \langle T\rangle_{\beta} \equiv \tr{(\sigma_{\beta}\,T(x))} = \frac{1}{2\pi}\,\tr{(\sigma_{\beta}L_0)},
    \label{Tbeta}
\end{equation}
and used \eqref{TtildeT}. The cross component $ \tr{(\sigma_t\,  T_{-+})} $ vanishes since $  T_{-+} $ is identically zero as an operator. The expectation values of the stress tensor $ \tr{(\sigma_t\,T_{AB})} $ in the original coordinates $\phi^A$ are obtained from \eqref{Ttildeval} by using \eqref{diffeogT}.

Because the Schwarzian derivatives in \eqref{Ttildeval} are computed at fixed $ t $, the stress tensor one-point function is not conserved in the background metric \eqref{CFTg}: $\n^a \,\tr{(\sigma_t\,  T_{ab})} \neq 0$. In the coordinates $x^\pm$ using \eqref{Ttildeval}, the covariant derivative of the expectation value is given by
\begin{equation}
\frac{\partial}{\partial x^+}\tr{(\sigma_t\,  T_{--}(x^-))} \neq 0, \quad \frac{\partial}{\partial x^-} \tr{(\sigma_t\,  T_{++}(x^+))} \neq 0,
\end{equation}
which are non-zero since $t = t(x^-,x^+)$. The non-conservation of the expectation value reflects the breakdown of diffeomorphism invariance which can be restored after adding suitable counterterms, as we will see in the next section.

\subsection{Virasoro process from the CFT path integral}\label{sec:Polyakov}

In this section, we explain how to obtain the Virasoro trajectory of the previous section from the path integral of the CFT. In particular, we relate expectation values of the stress tensor in Virasoro states to one-point functions in a background metric.

\paragraph{Stress tensor one-point function.} We will start by relating the expectation values \eqref{Ttildeval} of the previous section to one-point functions $\langle T_{ab}\rangle_g$ obtained from the path integral. The one-point function can be computed from the Lorentzian Polyakov action\footnote{The propagator here is the Feynman propagator obtained as analytic continuation of the Euclidean time-ordered propagator.} \cite{POLYAKOV1981207}
\begin{equation}
	W[g] = -\frac{c}{192\pi}\int d^{2}x\sqrt{-g}\,R\,\frac{1}{\Box}R
	\label{polyakov}
\end{equation}
as a functional derivative
\begin{equation}
	\langle T_{ab}\rangle_{g} = \frac{2}{\sqrt{-g}}\frac{\delta W}{\delta g^{ab}}.
	\label{Wvariation}
\end{equation}
Due to diffeomorphism invariance and Weyl non-invariance of the Polyakov action, it follows that
\begin{equation}
	\nabla^{a}\langle T_{ab} \rangle_{g} = 0, \qq g^{ab}\,\langle T_{ab}\rangle_{g} = \frac{c}{24\pi}R~.
 \label{conservation}
\end{equation}
The stress tensor in the $(x^+,x^-)$ coordinate system is then
\begin{align}\label{Ttildepp}
\langle  T_{--}\rangle_{g}   &=-\frac{1}{(\partial_\phi F^{-})^{2}}\left( \frac{\delta}{\delta \nu} + \frac{1}{\overbar{\nu}-\nu}\frac{\delta}{\delta \omega}\right) W~,\\
\label{Ttildemm}
\langle  T_{++}\rangle_{g}   &=\frac{1}{(\partial_\phi F^{+})^{2}}\left( \frac{\delta}{\delta \overbar{\nu}} - \frac{1}{\overbar{\nu}-\nu}\frac{\delta}{\delta \omega}\right) W~,\\
\langle  T_{-+}\rangle_{g}  &=-\frac{1}{(\partial_\phi F^{-})(\partial_\phi F^{+})}\frac{1}{\overbar{\nu}-\nu} \frac{\delta}{\delta \omega} W~.
\label{pathintegralvevs}
\end{align}
In the parametrization of the metric \eqref{CFTg} in terms of $ \omega,\nu,\overbar{\nu} $, the Polyakov action decomposes as (see Appendix \ref{app:polyakovdecomposition})
\begin{equation}
W[g] = \Gamma[\nu] + \overbar{\Gamma}[\overbar{\nu}] + K[\nu,\overbar{\nu}] + I_{\text{L}}[\omega,\nu,\overbar{\nu}]
\label{nunubardecomposition}
\end{equation}
where $ I_{\text{L}}[\omega,\nu,\overbar{\nu}] $ is the Liouville action in the background metric $ \hat{g}_{ab} \equiv e^{-\omega}g_{ab} $ with the Weyl factor removed and
\begin{gather}
\Gamma[\nu]  = -\frac{c}{48\pi}\int d\phi dt\, \nu\,\partial_\phi^{2}\log{(\partial_\phi F^{-})} , \quad \overbar{\Gamma}[\overbar{\nu}] = \frac{c}{48\pi}\int d\phi dt\, \overbar{\nu}\,\partial_\phi^{2}\log{(\partial_\phi F^{+})},\\ K[\nu,\overbar{\nu}] = -\frac{c}{48\pi}\int d\phi dt\,\frac{1}{\overbar{\nu}-\nu}\,[(\partial_\phi \nu)+(\partial_\phi \overbar{\nu})]^{2}.
\end{gather}
Here $ \Gamma[\nu] $ and $ \overbar{\Gamma}[\overbar{\nu}] $ are non-local functionals of $ \nu,\overbar{\nu} $, because $ F^{\pm} $ depend non-locally on $ \nu,\overbar{\nu} $ as a solution of the first order differential equation \eqref{nus}. On the other hand, the remaining two terms $ K[\nu,\overbar{\nu}] $ and $ I_{\text{L}}[\omega,\nu,\overbar{\nu}] $ are local in components $ \omega,\nu,\overbar{\nu} $ of the metric. The formula \eqref{nunubardecomposition} is analogous to the decomposition of $ W[g] $ in the Beltrami parametrization $ \mu,\overbar{\mu} $ of the metric in light-ray coordinates $ \phi^{\pm} $ that can be found in \cite{Verlinde:1989hv,verlinde_conformal_1990,nguyen_holographic_2021}. Both of the decompositions are derived in detail in Appendix \ref{app:polyakovdecomposition}.

We can now see that the Liouville term $ I_{\text{L}}[\omega,\nu,\overbar{\nu}] $ in \eqref{nunubardecomposition} encodes the Weyl anomaly and leads to a non-zero $ \langle T_{-+}\rangle_{g} $. On the other hand due to $ K[\nu,\overbar{\nu}] $, the Polyakov action does not holomorphically factorize in $ \nu,\overbar{\nu} $, the ``holomorphic anomaly'', which leads to a mixing between the left and right-moving sectors in $ \langle  T_{\pm\pm}\rangle_{g} $. Hence the one-point functions \eqref{pathintegralvevs} do not agree with the result \eqref{Ttildeval} of the previous section for operator expectation values. Regardless, it turns out there exists a renormalization scheme in which the one-point function agrees with \eqref{Ttildeval}.

The right renormalization scheme is obtained by removing the Weyl and holomorphic anomalies by subtracting the local counterterm
\begin{equation}
I_{\text{ct}}[\omega,\nu,\overbar{\nu}] = K[\nu,\overbar{\nu}] + I_{\text{L}}[\omega,\nu,\overbar{\nu}]
\label{countertermnunubar}
\end{equation}
from action of the CFT. However, this comes at a cost: the subtraction breaks diffeomorphism invariance of the effective action which is a diffeomorphism anomaly. Hence we are trading diffeomorphism invariance for holomorphic factorization in $ \nu,\overbar{\nu} $. This is completely analogous to restoring holomorphic factorization in the Beltrami parametrizaion $\mu,\overbar{\mu}$ which also leads to a diffeomorphism anomaly \cite{knecht_shifting_1990}.

Assuming the counterterm \eqref{countertermnunubar} is subtracted from the action, the one-point functions become
\begin{equation}
\langle  T_{--}( x^{-})\rangle_{g}   = -\frac{1}{F_t'(\phi)^{2}}\frac{\delta \Gamma}{\delta \nu}, \quad \langle  T_{++}( x^{+})\rangle_{g}=\frac{1}{\overbar{F}_t'(\phi)^{2}}\frac{\delta \overbar{\Gamma}}{\delta \overbar{\nu}}, \quad \langle  T_{-+}\rangle_{g}=0,
\label{derivativesafter}
\end{equation}
where we used \eqref{diffeoprocess}. One can show that\footnote{The proof of an analogous formula in the Beltrami parametrization can be found in \cite{nguyen_holographic_2021}. The result \eqref{Gammaderivatives} in $\nu,\overbar{\nu}$ parametrization can be proven in the same way using formulae provided in Appendix \ref{app:polyakovdecomposition}.}
\begin{equation}
\frac{\delta \Gamma[\nu]}{\delta \nu(t,\phi)} = -\frac{c}{24\pi}\{F_t(\phi),\phi\}, \quad \frac{\delta \overbar{\Gamma}[\overbar{\nu}]}{\delta \overbar{\nu}(t,\phi)} = \frac{c}{24\pi}\{\overbar{F}_t(\phi),\phi\},
\label{Gammaderivatives}
\end{equation}
where derivatives in the Schwarzians are taken at fixed $ t $. Substituting to \eqref{derivativesafter} gives
\begin{equation}
\langle  T_{--}( x^{-})\rangle_{g} = -\frac{c}{24\pi}\{f_t( x^{-}), x^{-}\}, \quad \langle  T_{++}( x^{+})\rangle_{g} = -\frac{c}{24\pi}\{\overbar{f}_t( x^{+}), x^{+}\}
\label{pathintegralresult}
\end{equation}
where $ f_t = F_{t}^{-1} $ and we used $ \{F_t(\phi),\phi\} = -F_t'(\phi)^{2}\,\{f_t(F_t(\phi)),F_t(\phi)\}$ combined with $  x^{-} = F_t(\phi) $ (and similarly for the left-movers). This matches with the Schwarzian part of the result \eqref{Ttildeval} of the previous section. Hence only in the diffeomorphism invariance breaking renormalization scheme with the local counterterm \eqref{countertermnunubar} subtracted, the path integral derivation matches with the operator derivation.

\paragraph{Including the initial state.} In the above analysis, we only reproduced the Schwarzian part of the expectation value $ \tr{(\sigma_t\, T_{ab})} $ \eqref{Ttildeval}. We can also derive the initial thermal state contribution by keeping part of the Weyl factor $\omega$ in the background metric as physical. To this end, we write the background metric \eqref{CFTg} as
\begin{equation}
g_{ab}\,dx^{a}dx^{b} = e^{\hat{\omega}}e^{\chi+\overbar{\chi}}\,(d\phi + \nu dt)(d\phi + \overbar{\nu} dt)
\label{CFTginitial}
\end{equation}
where $\hat{\omega} = \omega - \chi-\overbar{\chi}$ and
\begin{equation}
\chi = \log{\partial_\phi D(\phi)}, \quad \overbar{\chi} = \log{\partial_\phi\overbar{D}(\phi)},
\label{chichibar}
\end{equation}
are given in terms of two functions $D,\overbar{D}$. In Appendix \ref{app:polyakovdecomposition}, we show that the Polyakov action \eqref{polyakov} can be alternatively decomposed as
\begin{equation}
W[g] = \Gamma[\nu,\chi] + \overbar{\Gamma}[\overbar{\nu},\overbar{\chi}] + K[\nu,\overbar{\nu},\chi,\overbar{\chi}] + I_{\text{L}}[\hat{\omega};\nu,\overbar{\nu},\chi,\overbar{\chi}]
\label{altdecomposition}
\end{equation}
where $ I_{\text{L}}[\hat{\omega};\nu,\overbar{\nu},\chi,\overbar{\chi}] $ is the Liouville action of $ \hat{\omega} $ in the background metric $e^{-\hat{\omega}}g_{ab}$ and
\begin{align}
\Gamma[\nu,\chi] &= \Gamma[\nu] +\frac{c}{24\pi}\int d\phi dt\,\nu\,\{D(\phi),\phi\}\nonumber\\
\overbar{\Gamma}[\overbar{\nu},\overbar{\chi}] &= \overbar{\Gamma}[\overbar{\nu}]-\frac{c}{24\pi}\int d\phi dt\,\overbar{\nu}\,\{\overbar{D}(\phi),\phi\}\nonumber\\ K[\nu,\overbar{\nu},\chi,\overbar{\chi}] &= -\frac{c}{48\pi}\int d\phi dt\,\frac{1}{\overbar{\nu}-\nu}\,[(\mathcal{D}_{\phi} \nu)+(\overbar{\mathcal{D}}_{\phi} \overbar{\nu})]^{2}.
\label{genGammasimpletext}
\end{align}
Here $\mathcal{D}_\phi\nu = \partial_\phi\nu + (\partial_\phi\chi)\,\nu $ and $\overbar{\mathcal{D}}_{\phi} \overbar{\nu} = \partial_\phi\overbar{\nu} + (\partial_\phi \overbar{\chi})\,\overbar{\nu}$ are ``covariant derivatives''. The decomposition \eqref{altdecomposition} is a generalization of the decomposition \eqref{nunubardecomposition} to include a background Weyl factor $\chi+\overbar{\chi}$ and a similar decomposition in the Beltrami parametrization was first derived in \cite{lazzarini_integrating_1998} (see also Appendix \ref{app:polyakovdecomposition}).

Now instead of the counterterm \eqref{countertermnunubar}, we subtract the local counterterm
\begin{equation}
I_{\text{ct}}[\hat{\omega},\nu,\overbar{\nu},\chi,\overbar{\chi}] \equiv K[\nu,\overbar{\nu},\chi,\overbar{\chi}] + I_{\text{L}}[\hat{\omega};\nu,\overbar{\nu},\chi,\overbar{\chi}]
\label{countertermnunubargen}
\end{equation}
from the Polyakov action which leaves us with
\begin{equation}
    W[g] = \Gamma[\nu,\chi] + \overbar{\Gamma}[\overbar{\nu},\overbar{\chi}]
\end{equation}
which using \eqref{pathintegralvevs} gives the one-point functions
\begin{align}
    \langle  T_{--}( x^{-})\rangle_{g} &= -\frac{c}{24\pi}\,\{D(\phi),\phi\}\,f_t'( x^{-})^2 -\frac{c}{24\pi}\{f_t( x^{-}), x^{-}\},\nonumber\\
    \langle  T_{++}( x^{+})\rangle_{g} &= -\frac{c}{24\pi}\,\{\overbar{D}(\phi),\phi\}\,\overbar{f}_t'( x^{+})^2 -\frac{c}{24\pi}\{\overbar{f}_t( x^{+}), x^{+}\}.
    \label{1pointinitialstate}
\end{align}
Now if we choose
\begin{equation}
    D(\phi) = \tanh{\biggl(\sqrt{\frac{12\pi\langle T\rangle_{\beta}}{c}}\,\phi\biggr)},\quad \overbar{D}(\phi) = \tanh{\biggl(\sqrt{\frac{12\pi\langle T\rangle_{\smash[b]{\overbar{\beta}}}}{c}}\,\phi\biggr)},
\end{equation}
whose Schwarzian derivatives are
\begin{equation}
    \frac{c}{24\pi}\,\{D(\phi),\phi\} = -\langle T\rangle_\beta,\quad \frac{c}{24\pi}\,\{\overbar{D}(\phi),\phi\} = -\langle T\rangle_{\overbar{\beta}},
\end{equation}
the one-point functions \eqref{1pointinitialstate} become
\begin{equation}
    \langle  T_{--}( x^{-})\rangle_{g} = \tr{(\sigma_t\,  T_{--})},\quad \langle  T_{++}( x^{+})\rangle_{g} = \tr{(\sigma_t\,  T_{++})}
\end{equation}
where the expectation values on the right-hand side are given by \eqref{Ttildeval}. Hence the CFT path integral and the Polyakov action encode the initial thermal state in the Weyl factors \eqref{chichibar} of the background metric.

These Weyl factors can be understood as follows. The Liouville field \eqref{varphimetric} can be written as
\begin{equation}
    \varphi = \omega + \log{f'_t( x^{-})} + \log{\overbar{f}'_t( x^{+})}.
\end{equation}
where we used \eqref{diffeoprocess} and that $f_t = F_t^{-1}$. Now if we split $\omega = \hat{\omega} + \chi +\overbar{\chi}$, this can be written as 
\begin{equation}
    \varphi =\hat{\omega}+\log{(D\circ f_t)'( x^{-})} + \log{(\overbar{D}\circ \overbar{f}_t)'( x^{+})}.
    \label{liouvillevarphi}
\end{equation}
Hence including the Weyl factor is equivalent to replacing $f_t$ and $\overbar{f}_t$ with $D\circ f_t$ and $\overbar{D}\circ \overbar{f}_t$ respectively. Since $f_0 = \overbar{f}_0 = \text{id}$, the metric \eqref{CFTg} in the vicinity of the initial time $t = 0$ takes the form
\begin{equation}
    ds^2 = e^{\hat{\omega}}\,D'( x^-)\,\overbar{D}'( x^+)\,d x^-d x^+ = e^{\hat{\omega}}\,dX^-dX^+
\end{equation}
where we have defined a new set of coordinates $X^- = D( x^-)$ and $X^+ = \overbar{D}( x^+)$. Hence the functions $D,\overbar{D}$ parametrize the freedom in the choice of the conformal frame at the beginning of the driving process. 

The initial conformal frame is determined by the initial state. The thermal expectation value satisfies
\begin{equation}
    \langle T\rangle_{\beta} = 
    \begin{dcases}
    \frac{c\pi}{12\beta^2},\quad &\beta \rightarrow 0\\
        -\frac{c}{48\pi},\quad &\beta \rightarrow \infty,  
    \end{dcases}
    \label{thermal1pointvalues}
\end{equation}
where $\beta\rightarrow \infty$ is the vacuum expectation value and the high-temperature $\beta \rightarrow 0$ value follows from modular invariance (we call it the Cardy limit). Hence, for example, if the initial state is a vacuum state, we have
\begin{equation}
    X^- = D(x^-) = \tan{\biggl(\frac{x^-}{2}\biggr)},\quad X^+ = \overbar{D}(x^+) = \tan{\biggl(\frac{x^+}{2}\biggr)},
    \label{vacuumD}
\end{equation}
which is the conformal diffeomorphism that maps the diamond $ x^\pm \in (-\pi,\pi)$ inside the Lorentzian cylinder to Minkowski space parametrized by light-ray coordinates $X^\pm \in \mathbb{R}$ \cite{besken_local_2020}. By Wick rotation $ x^- = w \in \mathbb{C}$ and $ x^+ = \overbar{w}\in \mathbb{C}$, \eqref{vacuumD} can be identified as the map from the Euclidean cylinder to the complex plane.

\paragraph{Restoring diffeomorphism invariance.} Based on the above analysis, it is now clear what we have to add to the expectation values \eqref{Ttildeval} to make them conserved in the background metric \eqref{CFTg}: we have to add the local counterterm \eqref{countertermnunubargen}. Explicitly the relation in the $ x^\pm$ coordinate system is
\begin{align}
\langle  T_{--}\rangle_{g}  &= \tr{(\sigma_t\,  T_{--})}+\frac{1}{(\partial_\phi F^-)^{2}}\left( \frac{\delta}{\delta \nu} + \frac{1}{\overbar{\nu}-\nu}\frac{\delta}{\delta \hat{\omega}}\right) I_{\text{ct}}[\hat{\omega},\nu,\overbar{\nu},\chi,\overbar{\chi}]\nonumber\\
\langle  T_{++}\rangle_{g}  &= \tr{(\sigma_t\,  T_{++})}-\frac{1}{(\partial_\phi F^+)^{2}}\left( \frac{\delta}{\delta \overbar{\nu}} - \frac{1}{\overbar{\nu}-\nu}\frac{\delta}{\delta \hat{\omega}}\right) I_{\text{ct}}[\hat{\omega},\nu,\overbar{\nu},\chi,\overbar{\chi}]\nonumber\\
\langle  T_{-+}\rangle_{g} &= -\frac{1}{(\partial_\phi F^-)(\partial_\phi F^+)}\frac{1}{\overbar{\nu}-\nu} \frac{\delta}{\delta \hat{\omega}} I_{\text{ct}}[\hat{\omega},\nu,\overbar{\nu},\chi,\overbar{\chi}]
\label{1pointEV}
\end{align}
which is conserved and whose trace gives the Weyl anomaly \eqref{conservation}. It is remarkable, but probably expected, that there exists a renormalization scheme in which the results of the operator formulation of the CFT are recovered from the path integral.

The expression \eqref{1pointEV} is a rewriting of the Liouville stress tensor
\begin{equation}
\langle T_{ab}\rangle_g ={c\/24\pi}\left[{1\/2}\,\p_a\varphi\,\p_b\varphi +\n_a\n_b\varphi + g_{ab} \le(\square\varphi - {1\/4} g^{ab}\, \p_a\varphi\, \p_b\varphi\ri)\right]~,
\end{equation}
where the Liouville field is given by \eqref{liouvillevarphi} and it is not a simple sum of left- and right-moving terms. Hence we have identified where the expectation value $\tr{(\sigma_t\,T_{ab})}$ is hidden in the Liouville stress tensor.

\section{Information geometry of Virasoro states}\label{sec:virasorogeometry}

In this section we study quantum information geometry of Virasoro states. In regard to previous studies, the main differences are that
we are considering information geometry on orbits of Gibbs states rather than pure states, and the quantum information metric is derived from relative entropy rather
than the Fubini--Study metric. The information metric we consider is known as the BKM metric and it has a natural interpretation in the quantum thermodynamics of open systems. Hence as our main application, we consider an open two-dimensional CFT coupled to a heat bath and drive the CFT along a Virasoro trajectory  as in the previous section. In this context, the BKM metric on Virasoro states can be used to measure the amount of work dissipation in a step-equilibration process.

\subsection{Bogoliubov--Kubo--Mori metric on Virasoro states}\label{subsec:BKM}

The {\em Bogoliubov--Kubo--Mori metric} (BKM metric) is defined as \cite{petz_bogoliubov_1993,petz_riemannian_1996}
\begin{equation}
\mathcal{G}_{\sigma}(\delta\sigma,\delta \rho) = -\frac{\partial^{2}}{\partial \lambda_1\partial \lambda_2}S(\sigma +\lambda_1 \delta \sigma\lVert \sigma +\lambda_2\delta \rho)\bigg\lvert_{\lambda_1=\lambda_2 = 0}
\label{BKMmetric}
\end{equation}
where $S(\rho\lVert \sigma)$ is the relative entropy
\begin{equation}
S(\rho\lVert \sigma) = \tr{[\rho\,(\log{\rho} - \log{\sigma})]}.
\label{relativeentropy}
\end{equation}
While {\em e.g.} symmetry is not obvious from the above definition, one can show that it gives a well-defined quantum information metric on the space of {\em full-rank} states 
\begin{equation}
\mathcal{S}(\mathcal{H})= \{ \sigma \in \mathcal{B}_{tc}(\mathcal{H})\,\lvert \,\sigma^{\dagger} = \sigma,\, \sigma > 0,\, \tr{\sigma} = 1\}\ ,
\end{equation}
that is, Hermitian strictly positive trace-class operators on $\mathcal{H}$ with unit trace.
The space of all quantum states $\overline{\mathcal{S}(\mathcal{H})}$ is the closed convex hull of the set of extreme points of $\mathcal{S}(\mathcal{H})$ (see {\em e.g.} \cite{ciaglia_monotone_2022}). The tangent space consists of all Hermitian operators with zero trace \cite{petz_monotone_1996}
\begin{equation}
T_{\sigma}\mathcal{S}(\mathcal{H})= \{\delta \sigma\in \mathcal{B}_{tc}(\mathcal{H})\,\lvert \, \delta \sigma^{\dagger} = \delta \sigma,\, \tr{\delta \sigma} = 0\}.
\end{equation}
Due to unitary invariance of relative entropy, the BKM metric is invariant under simultaneous unitary transformations 
\begin{equation}
\mathcal{G}_{ V\sigma V^\dagger}(V\,\delta\sigma\,V^\dagger,V\,\delta\rho\,V^\dagger) = \mathcal{G}_{\sigma}(\delta\sigma,\delta \rho)
\label{unitaryinvariance}
\end{equation}
that act on states and tangent vectors as
\begin{equation}
\sigma \rightarrow V\sigma V^\dagger,\quad \delta\sigma \rightarrow V\,\delta\sigma\,V^\dagger.
\end{equation}
where $V$ is a unitary operator. Due to the contractivity of the relative entropy \cite{muller-hermes_monotonicity_2017} the BKM metric has the nice property of being monotonic under all CPTP maps (quantum channels). It thus belongs to the family of monotonic quantum information metrics \cite{petz_monotone_1996}.

We now define the BKM metric on $\text{Vir}_{\beta}$. Denoting
\begin{equation}
\sigma_{f+\lambda_1 u} = \sigma_f +\lambda_1 \delta \sigma_u+ \mathcal{O}(\lambda_1^2),\quad  
 \sigma_{f+\lambda_2 v} = \sigma_f + \lambda_2 \delta \sigma_v + \mathcal{O}(\lambda_2^2),
\end{equation}
and using the definition \eqref{BKMmetric}, we have
\begin{equation}\label{BKMmetric2}
\mathcal{G}_{\sigma_f}(\delta \sigma_u,\delta \sigma_v) = -\frac{\partial^{2}}{\partial \lambda_1\partial \lambda_2}S(\sigma_{f + \lambda_1 u}\lVert \sigma_{f + \lambda_2 v})\bigg\lvert_{\lambda_1=\lambda_2 = 0} \ .
\end{equation}
Note that this defines a right-invariant metric on $\text{Vir}_{\beta}$. The extension for the full product manifold $\text{Vir}_{\beta}\times \text{Vir}_{\overbar{\beta}}$ follows the usual rules. Before
moving to derive explicit formulas, it is useful to consider an induced metric on 
$\widetilde{\text{Diff}}_+S^1$.

We define the map
\begin{equation}
\Phi: \widetilde{\text{Diff}}_+S^{1} \rightarrow \text{Vir}_\beta,\quad \  \Phi(f) =\sigma_f = \frac{e^{-\beta H_f}}{\tr{e^{-\beta H_f}}} \ ,
\label{maptostates}
\end{equation}
which induces a pushforward map from the tangent space $ T_{f}\text{Diff}_+S^{1} $ to $ T_{\sigma_f}\text{Vir}_\beta $ as\footnote{We have $ \tr{(\sigma_f\, \delta H_u)} = 0 $ due to the normalization $ \tr{\delta \sigma_u} = 0 $ required by $\tr{\sigma_{f+\lambda u}} = \tr{\sigma_f} = 1$.}
\begin{equation}
\Phi_* (u) =\delta \sigma_u \equiv \frac{d}{d\lambda}\frac{e^{-\beta H_{f + \lambda u}}}{\tr{e^{-\beta H_{f + \lambda u}}}}\bigg\lvert_{\lambda=0} = -\beta\int_{0}^{1}ds\,\sigma^{1-s}_f\,\delta H_u\, \sigma^{s}_f
\label{standardidentity}
\end{equation}
where the Virasoro algebra element (a \textit{Virasoro excitation}) is
\begin{equation}
\delta H_u \equiv \frac{d}{d\lambda}H_{f + \lambda u}\bigg\lvert_{\lambda=0} = T_{\text{Ad}_F[\ell_0,u\circ F]} = -\int_0^{2\pi}d x\, \frac{u'( x)}{f'( x)^2}\,T( x).
\end{equation}
Here $F = f^{-1}$ and $\ell_0( x) = 1$ is the generator of rotations of the circle. The pull-back $\Phi^*\mathcal{G}$ by \eqref{maptostates} is an induced metric that we call the BKM metric on $\widetilde{\text{Diff}}_+S^{1}$, and denote for short
\begin{equation}\label{inducedmetric}
\mathcal{G}_{f}(u,v)\equiv(\Phi^*\mathcal{G})_f(u,v)= \mathcal{G}_{\sigma_f}(\delta \sigma_u,\delta \sigma_v).
\end{equation}
The Virasoro Hamiltonian and the Virasoro excitation satisfy
\begin{equation}
H_f = V_h\,H_{f\circ h}V_{h}^{\dagger}, \quad  \delta H_u = V_h\,\delta H_{u\circ h}V_{h}^{\dagger},
\end{equation}
so that unitary invariance \eqref{unitaryinvariance} of the BKM metric implies that
\begin{equation}
\mathcal{G}_{f\circ h}(u\circ h,v \circ h) = \mathcal{G}_{f}(u,v).
\end{equation}
As a result, the BKM metric is a right-invariant metric on $\widetilde{\text{Diff}}_+S^1$. Using the definition \eqref{BKMmetric2} and \eqref{inducedmetric}, we get explicitly
\begin{equation}
\mathcal{G}_{f}(u,v) = -\frac{\partial^{2}}{\partial \lambda_1\partial \lambda_2}S(\sigma_{f + \lambda_1 u}\lVert \sigma_{f + \lambda_2 v})\bigg\lvert_{\lambda_1=\lambda_2 = 0}.
\label{BKMdiffS1}
\end{equation}
We can obtain an explicit expression for this metric by computing relative entropy between two generic Virasoro states
\begin{equation}
S(\sigma_{f_2}\lVert \sigma_{f_1}) = \tr{(\sigma_{f_2}\,\beta H_{f_1} )}-\tr{(\sigma_{f_1}\,\beta H_{f_1} )}- (S(\sigma_{f_2})-S(\sigma_{f_1}))~.
\label{relentropydecomp}
\end{equation}
Von Neumann entropy is unitary invariant $ S(\sigma_{f_2}) = S(\sigma_{f_1}) = S(\sigma_{\beta}) $ so that relative entropy is simply
\begin{equation}
S(\sigma_{f_2}\lVert \sigma_{f_1}) = \tr{(\sigma_{f_2} \,\beta H_{f_1} )}-\tr{(\sigma_{f_1}\,\beta H_{f_1} )}.
\end{equation}
We can write this as
\begin{equation}
S(\sigma_{f_2}\lVert \sigma_{f_1}) = \beta \int_{0}^{2\pi}d x\,\bigl[\tr{(\sigma_\beta\, V_{\mathcal{F}}\,T( x)\,V_{\mathcal{F}}^{\dagger})}-\langle T\rangle_\beta\bigr]
\end{equation}
where we used the composition rule \eqref{composition} of Virasoro unitaries and we defined the diffeomorphism
\begin{equation}
\mathcal{F} = f_2\circ f_1^{-1}.
\label{combineddiffeo}
\end{equation}
The thermal one-point function $\langle T\rangle_\beta$ is defined in \eqref{Tbeta}. Using the transformation law \eqref{stresstransform}, we get
\begin{equation}
S(\sigma_{f_2}\lVert \sigma_{f_1}) = \beta\int_{0}^{2\pi}d x\,\biggl([\mathcal{F}'( x)^{2}-1]\,\langle T\rangle_{\beta} - \frac{c}{24\pi}\{\mathcal{F}( x), x\}\biggr).
\label{1strelcircle}
\end{equation}
Since the relative entropy depends only on the composition \eqref{combineddiffeo}, it is right-invariant which is equivalent to unitary invariance of relative entropy. From the above formula it is not immediately clear that $ S(\sigma_{f_2}\lVert \sigma_{f_1}) \geq 0 $, but we prove this in Appendix \ref{app:nonneg} using the inequality
\begin{equation}
\langle T\rangle_{\beta} \geq -\frac{c}{48\pi}
\end{equation}
and the so called \textit{average lemma} \cite{schwartz_projectively_1992,balog_coadjoint_1998} which is an inequality for the Schwarzian derivative of an element of $ \widetilde{\text{Diff}}_+S^{1} $. In fact, $ S(\sigma_{f_2}\lVert \sigma_{f_1}) $ can be identified exactly with the Hamiltonian $H(\mathcal{F})$ of the Alekseev--Shatasvili action \cite{cotler_theory_2019} so that non-negativity of relative entropy amounts to non-negativity of energy $ H(\mathcal{F}) \geq 0 $.

Using the expression for the Schwarzian and integrating by parts, \eqref{1strelcircle} can be written as
\begin{equation}
S(\sigma_{f_2}\lVert \sigma_{f_1}) = \frac{c\beta}{48\pi}\int_{0}^{2\pi}d x\,\biggl[\biggl(\frac{\mathcal{F}''( x)}{\mathcal{F}'( x)}\biggr)^{2}+\frac{48\pi\langle T\rangle_{\beta}}{c}\,[\mathcal{F}'( x)^{2}-1]\biggr].
\label{ScurlyF}
\end{equation}
Now we can expand \eqref{ScurlyF} using
\begin{equation}
    f_2 = f + \lambda_1u,\quad f_1 = f + \lambda_2 v,
\end{equation}
and identify the coefficient of $\lambda_1\lambda_2$ as the BKM metric \eqref{BKMdiffS1}. When $ f = \text{id} $, we get the 
simple formula
\begin{equation}\label{BKMdiff}
\mathcal{G}_{\text{id}}(u,v) = \frac{c\beta}{24\pi}\int_{0}^{2\pi}d x\,\biggl[u''( x)\,v''( x)+\frac{48\pi\langle T\rangle_{\beta}}{c}\,u'( x)\,v'( x)\biggr]
\end{equation}
and the expression for the metric at an arbitrary point is simply
\begin{equation}\label{BKMdiffgen}
\mathcal{G}_{f}(u,v) = \mathcal{G}_{\text{id}}\bigl(u\circ f^{-1},v\circ f^{-1}\bigr)
\end{equation}
due to right-invariance. Note that \eqref{BKMdiff} is a two-parameter family of
homogenous\footnote{``Homogenous'' means that there is no $\int^{2\pi}_0d x\, u( x)\,v( x)$ term.} Sobolev $\dot{H}^2$ metrics, which helps to identify relevant mathematical literature.

\subsection{Dissipated work along a Virasoro trajectory}\label{subsec:dissipation}

Consider a driven quantum system with a Hamiltonian $H_t$ whose time-dependence is due to time-dependent classical source terms (also called classical forces) that are externally controlled. If the system was a closed system, the state $ \sigma_t $ at time $ t $ would be obtained from the initial state by unitary time-evolution with $U_t$ generated by $H_t$. However, we will now assume that the system is an \textit{open quantum system} so that the relationship between $ \sigma_0 $ and $ \sigma_t $ is more complicated. In general, $\sigma_t$ is determined by a first-order master equation of the type
\begin{equation}
\dot{\sigma}_t = \mathcal{L}_t(\sigma_t)
\label{evolutionequation}
\end{equation}
where $ \mathcal{L}_t $ is a superoperator (the Liouvillian) acting on density matrices. In a closed system, the evolution generator is determined by the Hamiltonian $ \mathcal{L}_t(\,\cdot\,) = i[H_t,\cdot\,] $ and the evolution is unitary, but in an open system, $ \mathcal{L}_t $ depends on the details of the interaction between the system and its surroundings.\footnote{For quantum systems weakly coupled to a heat bath, the operator $ \mathcal{L}_t $ is the Lindblad operator.} The evolution obtained by solving \eqref{evolutionequation} is called a process and is formally given by $\sigma_t = \mathcal{T}\exp\{\,\int_0^t ds\,\mathcal{L}_s\} \sigma_0$ with unitary processes being a special case.

\paragraph{Step-equilibration and slow driving.} In general, the master equation \eqref{evolutionequation} of an open system has a complicated form. However, if the system is in contact with a heat bath of inverse temperature $\beta$, one expects the state to thermalize to an equilibrium Gibbs state $\pi_t$ of the instantaneous Hamiltonian $H_t$. This is characterized by the relaxing property of $\mathcal{L}_t$ \cite{cavina_slow_2017}
\begin{equation}\label{relaxingproperty}
\lim_{s\rightarrow \infty}e^{s\mathcal{L}_t}\sigma = \pi_t \equiv \frac{e^{-\beta H_t}}{\tr{e^{-\beta H_t}}},\quad \forall \sigma, t.
\end{equation}
The relaxing property implies that when the driving rate of the system is slow compared to the thermalization rate, the system evolves adiabatically along a curve in the space of equilibrium states. To make this precise, we define the driving time scale $ t_{\text{dr}} $ associated to the driving via
\begin{equation}
\dot{H}_t = \mathcal{O}(1\slash t_{\text{dr}})
\label{slowdrivinglimit}
\end{equation}
and we assume that time derivatives of all quantities are of order $ \mathcal{O}(1\slash t_{\text{dr}}) $. To reach a given target Hamiltonian, the process will take a total time $ t_{\hspace{0.5pt}\text{f}} $ which is often assumed to be of order of the driving time scale $ t_{\hspace{0.5pt}\text{f}}\slash t_{\text{dr}}  = \mathcal{O}(1) $. The system is then slowly driven when $ t_{\text{dr}}$ is much larger than any other time scale in the system (including the thermalization time scale). We call the limit $ t_{\text{dr}} \rightarrow \infty $ the \textit{slow-driving limit}.

Consider now a system in the slow-driving limit. Under right conditions, slow driving can be viewed as the continuum limit of a discrete process called a \textit{step-equilibration process} \cite{nulton_quasistatic_1985,anders_thermodynamics_2013} (see also \cite{scandi_thermodynamic_2019,miller_work_2019,abiuso_geometric_2020}). Let the time scale
for relaxation to thermal equilibrium \eqref{relaxingproperty} be $t_{\text{eq}}$. In a step-equilibration process, one begins by discretizing its total duration $ t_{\hspace{0.5pt}\text{f}} $ into $ N $ steps of duration $ \Delta t$ slightly larger than the equilibration time scale $t_{\text{eq}}$, with $t_{\hspace{0.5pt}\text{f}}=N \Delta t$. Then the time-dependence of the Hamiltonian is discretized to a sequence of $ N $ Hamiltonians $ \{H_i\}_{i=0}^{N-1} $ such that the state of the system $ \sigma_i $ evolves in discrete steps. Each driving step begins just after the system has thermalized to equilibrium. Thus, immediately before each step $ i $, the system Hamiltonian is $ H_i $ and the state of the system is the equilibrium state $ \sigma_i = \pi_i \equiv e^{-\beta H_i}/\tr \big[e^{-\beta H_i}\big] $. The stepwise driving is implemented in the following sequence:
\begin{enumerate}
	\item A step begins with an instantaneous quench $ H_i \rightarrow H_{i+1} $ while keeping the state fixed.
	\item After the quench, the system equilibriates $ \sigma_i = \pi_i \rightarrow \pi_{i+1} $ to the thermal state of the new Hamiltonian $ H_{i+1} $, which is possible since the duration of the step $\Delta t\geq t_{\text{eq}}$.
\end{enumerate}
During thermalization, the system is governed by the master equation \eqref{evolutionequation}. Its detailed form is unimportant, as long as the generator $ \mathcal{L}_t $ satisfies \eqref{relaxingproperty} and the relaxation happens during $\Delta t$. In the process, the state of the system evolves from an equilibrium state $ \pi_i $ to the next $ \pi_{i+1} $ in discrete steps and the evolution is due to thermalization from off-equilibrium states prepared by quenches. Assuming the quench is instantaneous, each steps lasts a thermalization time $ \Delta t = t_{\text{eq}} $. This stepwise evolution appears continuous in the slow-driving limit where the driving time scale $ t_{\text{dr}} $ is much larger than the equilibration time scale,
\begin{equation}\label{slowdriving}
\lambda \equiv \frac{t_{\text{eq}}}{t_{\text{dr}}} = \mathcal{O}(1\slash N) \ll 1.
\end{equation} 
The quasistatic limit where the system would be at equilibrium at all times corresponds to $\lambda \to 0$. In contrast, in the slow-driving limit $\lambda$ is finite but very small.
In this continuum limit, the state and Hamiltonian at each step before thermalization and after the quench are relabeled $ (\pi_i,H_{i+1}) \rightarrow (\sigma_t,H_t) $. If thermalization occurs in time $ t_{\text{eq}}$, the state after thermalization coincides with an equilibrium state of $ H_t $ up to corrections: $ \sigma_{t+t_{\text{eq}}} = \pi_{t} + \mathcal{O}(\lambda^{2}) $. This is exactly implemented by approximating the detailed continuum master equation \eqref{evolutionequation} in the slow-driving limit $t_{\text{dr}}\rightarrow \infty$ by a simple form\footnote{Note that in \cite{scandi_thermodynamic_2019}, the driving time is identified with the total time $t_{\text{dr}} =  t_{\hspace{0.5pt}\text{f}}$ and the equation \eqref{markovianmaster} is written using $ s \equiv t\slash t_{\hspace{0.5pt}\text{f}} $ such that $ \dot{\sigma}_{t} = (1\slash t_{\hspace{0.5pt}\text{f}})\,\dot{\sigma}_s $ where $ \dot{\sigma}_s $ is dimensionless.} that captures the relaxation property \cite{scandi_thermodynamic_2019,abiuso_geometric_2020}
\begin{equation}
\dot{\sigma}_t = \frac{1}{t_{\text{eq}}}\,(\pi_t - \sigma_t).
\label{markovianmaster}
\end{equation}
Due to $ \ddot{\sigma}_t = \mathcal{O}(1\slash t_{\text{dr}}^{2}) $, the solution of \eqref{markovianmaster} satisfies the required property
\begin{equation}
\sigma_{t+t_{\text{eq}}} = \pi_{t} +  \mathcal{O}(\lambda^{2})
\label{markovproperty}
\end{equation}
which says that the state of slowly driven system at time $t$ is equal to the equilibrium state at time $t-t_{\text{eq}}$. Hence \eqref{markovianmaster} is an effective continuum description of the whole discrete step-equilibration process. 

It has been argued that the step-equilibration process is automatically realized in nature as an approximate description of slowly driven non-equilibrium systems \cite{salamon_thermodynamic_1983}. In what follows, we will assume that \eqref{markovianmaster} provides a good approximate description of a slowly driven open quantum systems coupled to a heat bath and the validity of this approximation is discussed further in Section \ref{sec:discussion}.

\paragraph{Dissipated work in step-equilibration.} Given a Hamiltonian $ H_t $ and a state $ \sigma_t $ determined by the equation \eqref{evolutionequation}, the internal energy and the von Neumann entropy of the open quantum system are given by
\begin{equation}\label{EandS}
E(t) = \tr{(\sigma_tH_t)}, \quad S(\sigma_t) = -\tr{(\sigma_t\log{\sigma_t})}.
\end{equation}
During driving, the classical forces performs work \textit{on the system} and the total average work is 
\begin{equation}
W = \int_{0}^{t_{\hspace{0.5pt}\text{f}}} dt\,\tr{(\sigma_t\dot{H}_t)}.
\label{totalwork}
\end{equation}
At the same time, the system is exchanging heat with its surroundings and the total average heat exchange is given by
\begin{equation}\label{heatQ}
Q = \int_{0}^{t_{\hspace{0.5pt}\text{f}}} dt\,\tr{(\dot{\sigma}_tH_t)}.
\end{equation}
These definitions are such that the change in the internal energy of the system obeys the first law of thermodynamics $ E(t_{\hspace{0.5pt}\text{f}}) - E(0) = W + Q $ \cite{alicki_quantum_1979,vinjanampathy_quantum_2016}.\footnote{For more physical interpretation of these definitions, see the review \cite{vinjanampathy_quantum_2016}.} Note that for a closed system, $Q=0$ due to $\dot{\sigma}_t=i[H_t, \sigma_t]$.

In a similar manner, we define the free energy of the system as
\begin{equation}
F(t) = E(t) - \beta^{-1}S(\sigma_t)
\end{equation}
and the total amount of useful work performed on the system as
\begin{equation}
W_{\text{eq}} = F(t_{\hspace{0.5pt}\text{f}})-F(0).
\end{equation}
This is the conservative part of the work and does not depend on how the driving is performed, but only on the initial and final state.

On top of $ W_{\text{eq}} $, the classical force driving the system has to perform extra work, because the system dissipates part of the work so that $ W > W_{\text{eq}} $. The non-conservative part defines the total amount of \textit{dissipated} work
\begin{equation}
W_{\text{diss}} =  W-W_{\text{eq}}
\end{equation}
which measures how much of the work performed on the system is dissipated. We get
\begin{equation}
W_{\text{diss}} = \int_{0}^{t_{\hspace{0.5pt}\text{f}}}dt\,[ \beta^{-1}\partial_tS(\sigma_t)-\tr{(\dot{\sigma}_tH_t)}] = \int_{0}^{t_{\hspace{0.5pt}\text{f}}}dt\,\tr{[\,\dot{\sigma}_t\, (\beta^{-1}K_t-H_t)]}
\label{dissipation}
\end{equation}
where $ K_t $ is the modular Hamiltonian defined via
\begin{equation}
\sigma_t = \frac{e^{-K_t}}{\tr{e^{-K_t}}}
\end{equation}
and we used the first law of entanglement entropy $ \partial_tS(\sigma_t) = \tr{(\dot{\sigma}_tK_t)} $ \cite{lashkari_canonical_2016}.

Consider now the step-equilibration process \eqref{markovianmaster} for which the state $ \sigma_t $ satisfies
\begin{equation}
\dot{\sigma}_t = \dot{\pi}_t+ \mathcal{O}(\lambda\slash t_{\text{dr}}), \quad \beta^{-1}K_{t} = H_t- t_{\text{eq}}\dot{H}_t +   \mathcal{O}(\lambda^{2}),
\label{stepidentities}
\end{equation}
that follow from \eqref{markovproperty} and $\lambda = t_{\text{eq}}\slash t_{\text{dr}}$. Substituting to \eqref{dissipation}, the dissipation at leading linear order in $ \lambda $ is given by\footnote{Since each time-derivative is of order $ \mathcal{O}(1\slash t_{\text{dr}}) $, the leading term in $ W_\text{diss} $ is of order $ \mathcal{O}(\lambda\,(t_{\hspace{0.5pt}\text{f}}\slash t_{\text{dr}})) = \mathcal{O}(\lambda) $.}
\begin{equation}
W_\text{diss} = -t_{\text{eq}}\int_{0}^{t_{\hspace{0.5pt}\text{f}}}dt\,\tr{(\dot{\pi}_t\dot{H}_t)} + \mathcal{O}(\lambda^{2}).
\label{Wdissmarkovian}
\end{equation}
As noticed in \cite{scandi_thermodynamic_2019}, this can be written in terms of the BKM metric \eqref{BKMmetric}. To see this, first note that we can write
\begin{equation}
\mathcal{G}_{\pi_t}(\dot{\pi}_t,\dot{\pi}_t) = -\frac{\partial^2}{\partial\lambda_1\partial\lambda_2}S(\pi_t+\lambda_1 \dot{\pi}_t \lVert \pi_t+\lambda_2 \dot{\pi}_t)\bigg\lvert_{\lambda_1=\lambda_2 = 0} = -\frac{\partial^2}{\partial\lambda_1\partial\lambda_2}S(\pi_{t+\lambda_1} \lVert \pi_{t+\lambda_2})\bigg\lvert_{\lambda_1=\lambda_2 = 0}.
\end{equation}
From \eqref{relentropydecomp}, the relative entropy is explicitly
\begin{equation}
S(\pi_{t+\lambda_1} \lVert \pi_{t+\lambda_2}) = \beta\,\tr{(\pi_{t+\lambda_1}H_{t+\lambda_2})} + \ldots
\end{equation}
where dots denote terms that do not contribute a cross term $\lambda_1\lambda_2$ to the expansion in $\lambda_1$ and $\lambda_2$. Hence we get
\begin{equation}
\mathcal{G}_{\pi_t}(\dot{\pi}_t,\dot{\pi}_t) = -\beta\,\tr{(\dot{\pi}_t\dot{H}_t)}.
\end{equation}
As a result, the total dissipated work \eqref{Wdissmarkovian} for the process \eqref{markovianmaster} at leading order in $ t_{\text{eq}} $ can be written in terms of the BKM metric as
\begin{equation}
W_{\text{diss}} = \frac{t_{\text{eq}}}{\beta}\int_{0}^{t_{\hspace{0.5pt}\text{f}}} dt\,\mathcal{G}_{\pi_t}(\dot{\pi}_t,\dot{\pi}_t) + \mathcal{O}(\lambda^{2})
\label{BKMdissipation}
\end{equation}
which was first proven in \cite{scandi_thermodynamic_2019}.\footnote{In \cite{scandi_thermodynamic_2019} the formula \eqref{BKMdissipation} is written using the dimensionless time $ s \equiv t\slash t_{\hspace{0.5pt}\text{f}} $ so that the leading term is proportional to $ \lambda $ and the integral is performed from $ s = 0 $ to $ s=1 $.} Hence, the total dissipation coincides with the action of a point-particle in the space of equilibrium density matrices with the Lagrangian given by the BKM metric at leading order in $\lambda \ll 1$. This does not mean that the system remains in equilibrium for all $ t $, but only that the dissipation can be \textit{computed} in the space of equilibrium states. One can see that the dissipation vanishes $ W_{\text{diss}}\rightarrow 0 $ when $ \lambda\rightarrow 0$ as expected for an infinitely slow reversible process.

\paragraph{Dissipation along a Virasoro trajectory.} We will now consider a Type II process, a driven CFT in contact with a heat bath,  with the CFT Hamiltonian $ H_t $ determined by a trajectory $ f_t\in \widetilde{\text{Diff}}_+S^1 $ as
\begin{equation}
H_t \equiv H_{f_t} = \int_{0}^{2\pi} dx\,\frac{T(x)}{f'_t(x)},
\label{systemhamiltonianvirasoro}
\end{equation}
where $ T(x) $ is given by \eqref{Tphi}. As a result, when the system is coupled to a heat bath, the family of equilibrium states \eqref{relaxingproperty} of the system coincides with the Virasoro trajectory
\begin{equation}
\pi_t = \sigma_{f_t} = \frac{e^{-\beta H_{f_t}}}{\tr{e^{-\beta H_{f_t}}}}.
\label{CFTequilibrium}
\end{equation}
In addition, we will assume that the state evolution of the open system is described by the step-equilibration master equation \eqref{markovianmaster}. This is the more precise statement of Virasoro trajectories again becoming relevant in Type II processes, as alluded to in the Introduction.

Now the dissipation along the Type II process is given by \eqref{BKMdissipation}. Since we have
\begin{equation}
\dot{\pi}_t = \delta \sigma_{\dot{f}_t},
\end{equation}
where $\delta\sigma_u$ is given by \eqref{standardidentity}, the work dissipated \eqref{BKMdissipation} by the system is given by
\begin{equation}
W_{\text{diss}} = \frac{t_{\text{eq}}}{\beta}\int_{0}^{t_{\hspace{0.5pt}\text{f}}} dt\,\mathcal{G}_{f_t}(\dot{f}_t,\dot{f}_t)  + \mathcal{O}(\lambda^2)
\label{WdissCFT}
\end{equation}
where the induced BKM metric \eqref{inducedmetric} on the space of Virasoro states appears. Explicitly, we get
\begin{equation}
W_{\text{diss}} = \frac{c t_{\text{eq}}}{24\pi}\int_{0}^{t_{\hspace{0.5pt}\text{f}}} dt\int_{0}^{2\pi}d\phi\,\biggl[u_t''(\phi)\,u_t''(\phi)+\frac{48\pi\langle T\rangle_{\beta}}{c}\,u_t'(\phi)\,u_t'(\phi)\biggr] + \mathcal{O}(\lambda^2),
\end{equation}
where $u_t = \dot{f}_t \circ f_t^{-1} $ is the tangent vector of the curve $f_t\in \widetilde{\text{Diff}}_+S^1$. This is our main result which writes the work dissipation along a Virasoro trajectory using the BKM metric on $ \widetilde{\text{Diff}}_+S^1 $.

An example of a quantum system whose Hamiltonian is given by \eqref{systemhamiltonianvirasoro} is a two-dimensional CFT coupled to a background spacetime metric. In this case, the Hamiltonian is a sum of the left- and right-moving Virasoro Hamiltonians and the dissipation is also a sum of two terms. Notice, however, that the Hamiltonian \eqref{systemhamiltonianvirasoro} does not coincide with the Hamiltonian \eqref{finalHam} studied in Section \ref{sec:hamiltonianderivation} which was designed to generate Virasoro states by unitary evolution: the two Hamiltonians \eqref{finalHam} and \eqref{systemhamiltonianvirasoro} are realized on constant $ t $ slices of two different metrics.

\subsection{Cost, complexity, and optimal processes}
\label{sec:ccop}

There have been previous studies of complexity of CFT states, based on using the Fubini--Study metric to define a cost function. In his original work \cite{nielsen_geometric_2005}, Nielsen considered
several possible choices of a cost function, and outlined general properties that it should satisfy. In our setting, a natural choice of a cost function would be associated with
the infinitesimal length of a segment of a trajectory, computed using the BKM metric. In this case the cost associated with a trajectory would be given by
\begin{equation}\label{thermolength}
L_\r{BKM} =\int_{0}^{t_{\hspace{0.5pt}\text{f}}} dt\,\sqrt{ \cG_{f_t}(\dot{f}_t,\dot{f}_t)}~.
\end{equation}
This is also known as the thermodynamic length \cite{scandi_thermodynamic_2019} and it reduces to  the classical thermodynamic length \cite{crooks_measuring_2007} defined in terms of the classical Fisher information metric  when $ [\sigma_t,\dot{\sigma}_t] = 0 $.

The BKM complexity between two states can be defined using the distance \eqref{thermolength} in the BKM metric. The BKM distance between two diffeomorphisms $f$ and $g$ is 
\begin{equation}
    d_\r{BKM} (f,g) = \int_{0}^{1} dt\,\sqrt{ \cG_{f_t}(\dot{f}_t,\dot{f}_t)}
\end{equation}
where we minimize over all possible paths $(f_t)_{0\leq t\leq 1}$ from $f_0=f$ to $f_1=g$. The BKM complexity between the Virasoro states $\sigma_f$ and $\sigma_g$ is defined as
\be
C_\r{BKM}(\sigma_f,\sigma_g) = d_\r{BKM}(f,g) ~.
\ee
This can be interpreted as the size of the optimal circuit made of Virasoro unitaries that maps the source state $\s_f$ to the target state $\s_g$. As a result, optimal circuits correspond to geodesics in the BKM geometry.

We can also define another quantity: the BKM action as the point particle action of a trajectory
\begin{equation}\label{BKMcost0}
S_\r{BKM} =\half \int_{0}^{t_{\hspace{0.5pt}\text{f}}} dt\,\cG_{f_t}(\dot{f}_t,\dot{f}_t)  ~.
\end{equation}
This quantity is natural as it has a physical
interpretation in terms of dissipated work \eqref{WdissCFT}, when the  trajectory is associated with
slow driving of an open system.  As minimizing both the actions \eqref{thermolength} and \eqref{BKMcost0} give rise to the same geodesic equation, the optimal circuits are also those that dissipate least when realized as step-equilibration processes.

An analogous notion of complexity have been considered in this context in \cite{Erdmenger:2021wzc,erdmenger_complexity_2022}, there based on the Fubini--Study distance. The advantage of information geometry is that it gives an unambiguous metric and does not rely on some arbitrary cost function. In distinction to previous work, here we are interested in non-equilibrium physics of driven systems, and connections to quantum thermodynamics, so we are investigating mixed states for which the BKM metric gives a natural information geometry and a physical motivation for complexity.

\subsection{Connections to Euler--Arnold theory and optimal transport}

Information geometry of Virasoro states and the BKM information metric are closely related to various results in the literature. Here we outline their connection to Euler--Arnold theory and classical information geometry.

\paragraph{Geodesic equation of the BKM metric.} The geodesic equation of the BKM metric is a special case of a general Euler--Arnold equation \cite{arnold_sur_1966,arnol2013mathematical} (see \cite{modin_geometric_2019} for a review). We start by defining an inner product on $ T_{f}\text{Diff}_+S^{1} $ by the formula
\begin{equation}
	\langle u,v \rangle_{f} \equiv \int_{0}^{2\pi} d x\,f'( x)\, u( x)\,v( x)
	\label{inner}
\end{equation}
where $ u,v \in T_{f}\text{Diff}_+S^{1} $ and it is right-invariant $ \langle u\circ g, v\circ g \rangle_{f\circ g} = \langle u,v \rangle_{f} $ by definition. In terms of the inner product, the BKM metric \eqref{BKMdiff} is given by
\begin{equation}
	\mathcal{G}_{f}(u,v) = \langle u,A_fv \rangle_{f} = \int_{0}^{2\pi} d x\,f'( x)\,u( x)\,(A_{f}v)( x)
	\label{GfAf}
\end{equation}
where the operator $ A_f $ is the \textit{inertia operator} of the BKM metric which at the identity becomes
\begin{equation}
	(A_{\text{id}}v)( x) =  \frac{c\beta}{24\pi} \partial^2_ x\,\biggl(\partial^2_ x+\frac{48\pi\langle T\rangle_{\beta}}{c}\biggr)\,v( x).
	\label{BKMinertia}
\end{equation} 
If the operator $A_f$ has suitable analytic properties \cite{kolev_local_2017}, the Euler--Arnold theory writes geodesics of the metric \eqref{GfAf} as flow lines of a Hamiltonian as explained originally in \cite{arnold_sur_1966}. If $ A_{\text{id}} $ is invertible and commutes with $ \partial_x $ (which is clearly true for the BKM metric), the geodesic equation can be written as a first-order flow equation known as the Euler--Arnold equation \cite{escher_right-invariant_2014,kolev_local_2017}
\begin{equation}
	A_{\text{id}}(\partial_tu_t) +u_t\,\partial_x(A_{\text{id}}u_t)+2\,(A_{\text{id}}u_t)\, (\partial_x u_t) = 0,
\end{equation}
where $ u_t = \dot{f}_t\circ f^{-1}_t $. Using \eqref{BKMinertia}, the geodesic equation of the BKM metric can be obtained explicitly.

\paragraph{Hunter-Saxton limit.} The BKM metric simplifies in the Cardy limit $ \beta \rightarrow 0 $ where the stress tensor expectation value $\langle T\rangle_\beta$ in a thermal state takes a universal form due to modular invariance \eqref{thermal1pointvalues} of the CFT. In this limit, the BKM metric \eqref{BKMdiff} is at leading order
\begin{equation}\label{cardylimitBKM}
	\mathcal{G}_{\text{id}}(u,v) = \frac{\pi c}{6\beta}\int_{0}^{2\pi}d x\,u'( x)\,v'( x),
\end{equation}
the inertia operator $ A_{\text{id}} = \frac{\pi c}{6\beta}\partial_{ x}^{2} $ and the geodesic equation
\begin{equation}
	\dot{u}_t''+2u_t'u_t''+u_tu_t''' = 0
\end{equation}
with $ u_t = \dot{f}_t\circ f^{-1}_t $ is the Hunter--Saxton equation \cite{Lenells1,Lenells2} which was originally investigated in the context of modeling orientation waves in nematic
liquid crystals \cite{hunter_dynamics_1991}. The Hunter--Saxton equation is, together with the Korteweg--de Vries and Camassa--Holm equations, among three special flows on the Virasoro group that are completely integrable \cite{2002math.....10397K}.  
 
\paragraph{Quantum information geometry in a CFT as a classical information geometry.} In the Cardy limit, the BKM quantum information geometry in CFT maps to the classical information geometry in a classical statistical manifold, investigated in \cite{Lenells1,Lenells2,khesin_geometry_2011}. Let us review some discussion therein. Consider the statistical manifold of probability densities in $S^1$,
\be
\text{Dens}\,S^1 = \biggl\{\frac{df}{2\pi} \equiv f'( x)\,\frac{d x}{2\pi}~\bigg|~ f'( x)>0,\ \int_{S^1} \frac{df}{2\pi} = 1\biggr\} \ .
\ee
With the projection $\pi\colon \text{Diff}_+S^1 \rightarrow \text{Dens}\,S^1,\ f( x) \mapsto f' ( x)$ one can show that the space of densities $ \text{Dens}\,S^1=\text{Diff}_+S^1/\,\text{U}(1)$. Importantly, this establishes a correspondence between our original quantum statistical manifold in a CFT, the Virasoro orbit of Virasoro
states, and the classical statistical manifold, with
\be
\bigl\{\sigma_f = V_F\, \sigma_\beta V^\dagger_F \bigr\} = \text{Vir}_\beta =\text{Diff}_+S^1/\,\text{U}(1) = \text{Dens}\,S^1 \ .
\ee
Note that a Virasoro state $\sigma_f$ corresponds to a classical probability density function $f'( x)$, properly unit normalized with respect to the measure $d x /(2\pi)$. Further, we can use this correspondence and results of \cite{Lenells1,Lenells2,khesin_geometry_2011} to characterize the 
geometry of $\r{Vir}_\beta$ with the BKM metric \eqref{cardylimitBKM} (in the Cardy limit), as follows.
The map $\Phi\colon f \mapsto h=\sqrt{f'}$ defines an isometry from $\text{Dens}\,S^1=\text{Diff}_+S^1/\,\text{U}(1)$ with a $\dot{H}_1$-metric to an open subset of the infinite-dimensional sphere
\be
 S^\infty_r = \biggl\{h\in L^2\biggl(S^1,\frac{d x}{2\pi}\biggr)~\bigg|~\int_{0}^{2\pi}\frac{d x}{2\pi}\, h( x)^2  = r^2\biggr\} \subset L^2\biggl(S^1,\frac{d x}{2\pi}\biggr)
\ee
of radius $r = 1$ with the standard $L^2$ metric. The isometry is shown by computing that the $L^2$ metric on the sphere induces the metric
\be\label{lenellsmetric}
  \langle\langle u\circ f , v\circ f \rangle\rangle = \frac{1}{4} \int_{0}^{2\pi}\frac{d x}{2\pi} \, u'( x)\,v'( x)  
\ee
in $T_f \text{Diff}_+S^1$ -- the $\dot{H}_1$-metric in $\text{Diff}_+S^1$. Rescaling the metric \eqref{lenellsmetric}, such that the prefactor $1/4\mapsto b>0$, leads to an isometry to the sphere $S^\infty_r$ of radius $r=2\sqrt{b}$. 

One can now compute Riemannian distances between two Virasoro states $\sigma_f ,\sigma_g$ in $\r{Vir}_\beta$ equipped with the BKM metric \eqref{cardylimitBKM} using the result for the distance between probability measures $df ,dg$ in the $\dot{H}^1$-metric. By the isometry, geodesic distance in the $\dot{H}^1$-metric corresponds to the geodesic distance in $S^\infty_r$ along an arc of a great circle of radius $r$. Computing the length of the arc gives the result
\be\label{sphericalHellinger}
\r{dist}(df,dg) = r\, \r{arccos}\le({1\/r^2}\int_{0}^{2\pi}d x\sqrt{f'( x)\,g'( x )} \ri)~,
\ee
With the prefactor $b=\pi^2 c/(3\beta)$ in \eqref{cardylimitBKM}, $r=\sqrt{4\pi^2 c/(3\beta)}$, we get the BKM distance
\be\label{BKMdistance}
 \r{dist}_{\text{BKM}} (\sigma_f ,\sigma_g )= \sqrt{\frac{4\pi^2 c}{3\beta}} \, \r{arccos}\le(\frac{3\beta}{4\pi^2 c}\int_{0}^{2\pi} d x\sqrt{f'( x)\,g'( x )} \ri) \ .
\ee
One can also compute the diameter of the Virasoro orbit $\r{Vir}_\beta$, defined as 
\be
\r{diam}_{\text{BKM}} (\r{Vir}_\beta ) \equiv \text{sup}\, \{\,  \r{dist}_{\text{BKM}} (\sigma_f ,\sigma_g )~|~\sigma_f,\sigma_g \in \r{Vir}_\beta\} 
\ee
with the result
\be
\r{diam}_{\text{BKM}}(\r{Vir}_\beta ) = \sqrt{4\pi^3 c\/3\b}~.
\ee
In particular, this means that the quantum information distance, computed by the BKM metric in a Virasoro  orbit, reduces to a classical information distance: the distance
\eqref{sphericalHellinger} is the {\em spherical Hellinger distance} in the statistical manifold $\text{Dens}\,S^1$. The square of the Hellinger distance of probability density
functions $p,q$ on a manifold $M$ with measure $d\mu$ with the normalization so that $M$ has measure 
$\mu (M)=1$ is defined as
\be
 \r{dist}^2_{\text{Hel}}(p,q) \equiv \int_{M}d\mu\, \left( \sqrt{p}-\sqrt{q}\right)^2 
\ee
which can be expanded as 
\be
\r{dist}^2_{\text{Hel}}(p,q) = 2\,(1-\text{BC}(p,q))
\ee
where $\text{BC}(p,q)$ is the Bhattacharyya coefficient, measuring the overlap of the 
probability distributions $p,q$, which can be used to define an angle $\alpha$ as
\be
 \text{BC}(p,q) \equiv \int_{M}d\mu\, \sqrt{pq} \equiv \cos \alpha \ .
\ee
The spherical Hellinger distance \eqref{sphericalHellinger} is this angle $\alpha$, adjusting for the 
scale factors and the radius $r$. Note also that in quantum information theory the
counterpart of the Bhattacharyya coefficient is the fidelity, and the counterpart of spherical Hellinger distance is the fidelity metric. Thus the Cardy limit BKM distance \eqref{BKMdistance} may be thought as the fidelity metric in $\r{Vir}_\beta$.

Optimal transport theory in statistics studies converting a probability distribution
to an another probability distribution with minimal cost, taken to be the distance of the distributions 
in some information metric. For slowly driven CFTs in contact with a heat bath, we have shown that optimal processes with minimal dissipation correspond to geodesics in the BKM metric.  In the Cardy limit, the optimal processes in CFT can now be interpreted as optimal transport in $\r{Dens}\,S^1$
with respect to the spherical Hellinger distance, and can be written explicitly using the known form of the geodesics \cite{Lenells2,Lenells1}. For example, the path $f_t$ corresponding to the geodesic starting at the identity and with tangent  $u_0 \in T_\r{\text{id}}M$ can be determined from
\be
f_t'( x) = \le( \cos{t}+{1\/2} \sin{t}\, u_0'( x) \ri)^2 \,.
\ee
In conclusion, for slowly driven CFTs  
there exists a mathematical framework relating trajectories to solutions
of Euler-Arnold equations of Sobolev $\dot{H}^2$ metrics in $\text{Dens}\,S^1$.
In the Cardy limit, where the $\dot{H}^2$ metric  reduces to the $\dot{H}^1$ metric, rigorous results and explicit solutions are available. 
An open problem for future work is
to extend the rigorous analysis from Sobolev $\dot{H}^1$ metric to Sobolev $\dot{H}^2$ metrics in $\text{Dens}\,S^1$, if possible. Solutions for the geodesics allow one to substitute them back to the driving Hamiltonian, and thus construct optimal\footnote{Optimal in the sense of minimizing dissipated work in a Type II process, and minimizing circuit complexity in a Type I process.} 
driving of a CFT in Type I and II processes. If such solutions would be available in the $\dot{H}^2$ case, optimal processes could be found for arbitrary finite temperature.

\subsection{Comments on Virasoro states on the infinite line}

In previous sections, we considered Virasoro states on the circle $ S^{1} $, but they can also be defined on the infinite line $ \mathbb{R} $. This corresponds to a CFT on Minkowski space $ \mathbb{R}^{1,1} $ which has $ \mathbb{R} $ as its space-like slices. Since $ \mathbb{R}^{1,1} $ can be embedded into the Lorentzian cylinder $ \mathbb{R}\times S^{1}$ by a conformal map \cite{besken_local_2020}, Virasoro states on $ \mathbb{R} $ behave in the same way as on $ S^{1} $. 

For simplicity, we will consider the flat metric on $ S^{1}\times \mathbb{R} $. Denoting light-ray coordinates on $ \mathbb{R}^{1,1} $ by $ X^{\pm} \in \mathbb{R} $, the embedding is given by $ x^{\pm} = D^{-1}(X^{\pm}) $ where $ D\colon (-\pi,\pi)\rightarrow\mathbb{R} $ is the map \cite{besken_local_2020}
\begin{equation}
D(x) = \tan{\frac{x}{2}}.
\label{Pmap}
\end{equation}
The Minkowski space stress tensor is $\Theta_{ab}$ whose components in $X^\pm$ coordinates are $\Theta_{--}(X^{-}) = \Theta(X^{-})\otimes \mathbf{1} $ and $\Theta_{++}(X^{+}) = \mathbf{1}\otimes\Theta(-X^{+}) $ where \cite{besken_local_2020}
\begin{equation}
\Theta(X) \equiv (D^{-1})'(X)^{2}\,T(D^{-1}(y))+\frac{c}{24\pi}\{D^{-1}(X),X\}
\label{stresstensorline}
\end{equation}
with $T(x)$ defined in \eqref{Tphi}. Virasoro states on the infinite line are then defined as
\begin{equation}
\sigma_h = \frac{e^{-\beta H_h}}{\tr{e^{-\beta H_h}}}, \quad H_h = \int_{-\infty}^{\infty}dX\,\frac
{\Theta(X)}{h'(X)}.
\end{equation}
where now $ h \in \text{Diff}_+\mathbb{R} $.

The projective representation $V_f$ of $ \widetilde{\text{Diff}}_+S^{1} $ can be lifted to a projective representation $V_h$ of $ \text{Diff}_+\mathbb{R} $ using the map \eqref{Pmap} as $ h = D\circ f \circ D^{-1} $. However, this is only valid when $ f \in \widetilde{\text{Diff}}_+S^{1}$ belongs to a subgroup of diffeomorphisms that keep a point on the circle fixed $ f(\pm \pi) = \pm \pi $ so that infinity on the infinite line is kept fixed $ h(\pm \infty) = \pm \infty $. There are further conditions which will be left for future work. The adjoint action on the stress tensor generalizes in exactly the same way \cite{fewster_quantum_2005,fewster_probability_2019}
\begin{equation}
V_h\,\Theta(X)\,V_h^{\dagger} = h'(X)^{2}\,\Theta(h(X)) - \frac{c}{24\pi}\{h(X),X\}.
\label{adjointinfiniteline}
\end{equation}
Virasoro unitaries $V_h$ are now exponentials of $\Theta(X)$ integrated over $ \mathbb{R} $ and the Virasoro orbit on $ \mathbb{R} $ is obtained as on the circle $ V_{h}\,\sigma_h\,V_{h}^{\dagger} = \sigma_\beta$. Because of this, the formula for relative entropy on the circle \eqref{ScurlyF} can be extended to $ \mathbb{R} $. The von Neumann entropy contribution to relative entropy vanishes again due to unitary invariance and the result is
\begin{equation}
S(\sigma_{h_2}\lVert \sigma_{h_1}) = \frac{c\beta}{24\pi} \int_{-\infty}^{\infty}dX\,\biggl[(\mathcal{F}'(X)^{2}-1)\,\frac{2\pi^{2}}{\beta^{2}} - \{\mathcal{F}(X),X\}\biggr]
\end{equation}
where $ \mathcal{F} = h_2\circ h_1^{-1} \in \text{Diff}_+\mathbb{R} $ and we used that on the infinite line the stress tensor expectation value in the thermal state is universal $ \langle \Theta(X)\rangle_{\sigma_\beta} = \frac{c\pi}{12\beta^{2}} $. We can further simplify the expression to
\begin{equation}
S(\sigma_{h_2}\lVert \sigma_{h_1}) = \frac{c\beta}{48\pi}\int_{-\infty}^{\infty} dX\,\biggl[\biggl(\frac{\mathcal{F}''(X)}{\mathcal{F}'(X)}\biggr)^{2}+\frac{4\pi^{2}}{\beta^{2}}\,(\mathcal{F}'(X)^{2}-1)\biggr]
\label{lineS}
\end{equation}
where we used $ \mathcal{F}''(\pm\infty) = 0 $ to cancel the total derivative term in the Schwarzian. 

These results for Virasoro states on $ \mathbb{R} $ are related to the relative entropy of reduced density matrices of the semi-infinite line  $ (0,\infty) \subset \mathbb{R} $ in global pure states as studied in \cite{hollands_relative_2020,panebianco_formula_2020}. They considered reduced density matrices
\begin{equation}
\rho = \tr_{(-\infty,0)}\lvert 0 \rangle \langle 0 \lvert, \quad \rho' = \tr_{(-\infty,0)}\lvert u \rangle \langle u \lvert, \quad \lvert u\rangle \equiv \exp{\biggl(i\int_{0}^{\infty}dy\,u(y)\,\Theta(y)\biggr)}\lvert 0 \rangle~,
\end{equation}
where $ u(y) $ is the component of a smooth vector field on $ \mathbb{R} $ that satisfies $ u(0) =0 $. The resulting relative entropy is given by \cite{hollands_relative_2020}
\begin{equation}
S(\rho'\lVert \rho) = \frac{c}{24}\int_{-\infty}^{\infty}dX\,\biggl[\biggl(\frac{\varphi''(X)}{\varphi'(X)}\biggr)^{2}+\varphi'(X)^{2}-1\biggr]
\label{HollandsS}
\end{equation}
where the diffeomorphism $ \varphi \in \text{Diff}_+\mathbb{R} $ is given by $ \varphi(X) = \log{\theta_1(e^{X})} $ where $ \theta_1(y) $ is the solution of the equation $ \dot{\theta}_s = u\circ \theta_s $ at $ s = 1 $ and $X =\log{y}\in \mathbb{R} $ when $y\in (0,\infty)$.

We see that \eqref{HollandsS} is equal to \eqref{lineS} after setting $ h_1 = \text{id} $, $ h_2 = \varphi $ and $ \beta = 2\pi $. Hence it appears that the reduced density matrices are equal to Virasoro states as $ \rho' = \sigma_h $ and $ \rho = \sigma_{\beta} $ when $ \beta = 2\pi $. Indeed, the thermal state $ \sigma_{\beta} $ with $ \beta = 2\pi $ on $ \mathbb{R} $ can be identified with the reduced density matrix of the semi-infinite line $ (0,\infty) \subset \mathbb{R} $ in the vacuum state $\lvert 0 \rangle$ \cite{bisognano_duality_1976}. A similar relation appears to hold for a generic Virasoro state: $ \sigma_h $ with $ \beta = 2\pi $ on $\mathbb{R}$ is the reduced density matrix of a global pure state $ \lvert u\rangle \equiv V_g \lvert 0 \rangle $ where $ V_g $ is a Virasoro unitary with $ g = \text{id} $ on $ (-\infty,0) $. This would give a state understanding for the modular Hamiltonians studied in \cite{Cardy:2016fqc}.

As in the case of the circle, non-negativity of \eqref{lineS} is not immediately clear and is sensitive to boundary conditions at $ \pm \infty $. A simple check is to expand \eqref{lineS} in $ \mathcal{F}(X) = X + \lambda u(X) $ which should not include a term linear in $ \lambda\rightarrow 0 $ when non-negativity is satisfied. It turns out that the expansion of \eqref{lineS} contains a total derivative term leading to
\begin{equation}
S(\sigma_{h_2}\lVert \sigma_{h_1}) = \frac{c\pi}{6\beta}\,\lambda\,(u(\infty) - u(-\infty)) + \mathcal{O}(\lambda^{2})~,
\label{Sexp}
\end{equation}
which implies a violation of non-negativity when $ u(\infty) -u(-\infty) \neq 0 $ (understood as a limit). We claim that the solution to this problem is that the change of von Neumann entropy $ S(\sigma_{h_2}) - S(\sigma_{h_1}) \neq 0 $ in \eqref{relentropydecomp} no longer vanishes when $ \mathcal{F}(\infty) -\mathcal{F}(-\infty) \neq 0 $. Hence we have to add the change in entropy to \eqref{lineS} which cancels the linear term in \eqref{Sexp} as expected by the entanglement first law. This of course requires that $ \sigma_{h_{1,2}} $ are no longer unitarily related. A more detailed investigation of these subtleties associated with boundary conditions is left for future work.

\section{Möbius processes and hyperbolic geometry }\label{sec:mobiusgeometry}

The $\text{SL}(2,\R)$ subgroup of $\text{Diff}_+ S^1$ gives a subset of the Virasoro states which we will refer to as Möbius states. They are given as Gibbs states of the form
\be\label{introMobiusdef}
\s_f= {e^{-\b H_f} \ov \r{Tr}\,e^{-\b H_f}},\qq H_f=  \r{cosh}\,\rho\,L_0+{1\/2} \,\r{sinh}\,\rho \, \le(e^{-i\chi}L_1 + e^{i\chi} L_{-1} \ri),
\ee
where $(\rho,\chi)$ are coordinates on the space of Möbius states that parametrize diffeomorphisms $f = f_{(\rho,\chi)}$ in $\r{SL}(2,\R)$ (or more precisely, its universal cover). The space of Möbius states is the unit disk $\r{SL}(2,\R)\slash \text{U}(1) = \mathbb{D} \subset \mathbb{C} $ with a complex coordinate $z=e^{i\chi}\,\r{tanh}(\tfrac12 \rho)$ in the parametrization \eqref{introMobiusdef}. In this section, we will show that the BKM geometry corresponds to a hyperbolic metric on this disk 
\be
\bD = \{z\in \C\mid |z|<1\},\qq ds^2_{\text{BKM}} = {4dz^2\/(1-|z|^2)^2} = d\rho^2 + \r{sinh}^2\rho\,d\chi^2,
\ee
where $ds^2_{\text{BKM}}$ is the BKM information metric on the space of Möbius states. The BKM geometry is the space of possible Möbius states $\s(z) = \sigma_{f_z}$ and it is in one-to-one correspondence with points $z\in \bD$ of the unit disk. As a result, Möbius trajectories are continuous curves on the disk.

In this section, we will focus on Type I Möbius processes that are obtained by starting with the thermal state and evolving unitarily with a time-dependent $\r{SL}(2,\R)$ Hamiltonian
\be\label{genHt}
H_t = c_0(t)\, L_0 + c_1(t)\, L_1 +c_{-1}(t)\,L_{-1}~.
\ee
Periodic driving\footnote{So far we have been discussing deforming the Hamiltonian $H_t$ with a one-parameter-family of diffeomorphisms. To make contact with work on the Floquet CFTs, we now consider a parametrization where the diffeomorphism that is applied to the Hamiltonian is alternated abruptly, for example to achieve a periodic exchange between two Hamiltonians $H_1,H_2$ that drive a system for constant times $T_1,T_2$. This stepwise parametrization could be smeared a bit to again achieve a smooth parametrization as in our previous discussion.} with such Hamiltonians  was considered in the context of Floquet CFT \cite{Wen:2018agb, Han:2020kwp,Wen:2020wee, Choo:2022lgm}. For such periodic processes, the dynamics is controlled by the iteration of a Möbius transformation on the disk and the asymptotic behavior of the trajectory under iteration gives the distinction between heating and non-heating phases. These two phases are illustrated in Figure \ref{Fig:KMphases} on the unit disk in a periodic two-step process. Using the BKM metric, we can define information-geometric quantities, such as a cost, state complexity and the BKM action. A summary of the quantities we consider is given in Table \ref{Fig:summarytable}.

The BKM geometry also provides a framework for investigations of ergodicity. It gives a quantum counterpart of the classical phase with a Möbius invariant metric. The resulting measure is suitable to study questions of ergodicity as it is invariant under the Möbius driving.  In the non-heating phase, ergodicity is determined by whether the process angle $\t$ is a rational multiple of $\pi$, see Figure \ref{Fig:ergodic}.

Two-dimensional  CFTs describe critical quantum many-body systems and it should be possible to realize or simulate driving Hamiltonians \eqref{genHt} in the future \cite{doi:10.1126/science.aal3837, Sch_fer_2020}. 
Periodic driving is one of the basic protocols to study non-equilibrium systems and leads to many interesting phenomena  \cite{PhysRevLett.106.220402,kitagawa_topological_2010,PhysRevLett.118.115301,PhysRevX.4.041048,PhysRevLett.114.140401,PhysRevLett.115.256803,PhysRevLett.117.147202,PhysRevLett.67.516}. The BKM geometry gives a powerful tool to solve and study these periodic processes.

\subsection{Information geometry of Möbius states}

The Möbius states are the Virasoro states obtained by acting with an $\r{SL}(2,\R)$ transformation on the thermal state
\be\label{mobiusunitary}
\s_f = V_F\, \s_\b V_F^\dg = {e^{-\b H_f}\over \r{Tr}\,e^{-\b H_f}},\qq H_f=\int_{0}^{2\pi}d x\,{T(x)\/f'(x)} ~,
\ee
where $V_F$ is a Virasoro unitary with $F=f^{-1}$ in the $\widetilde{\text{SL}}(2,\R)$ subgroup of $\widetilde{\text{Diff}}_+S^1$. The corresponding Möbius Hamiltonians are of the form
\be\label{HfMob}
H_f = b_0 L_0 + b_+ L_+ + b_{-}L_{-}
\ee
where we define $L_+ = {1\/2}(L_{1}+L_{-1}) $ and $L_- = {1\/2i}(L_{1}-L_{-1})$. Note that this could be easily generalized to other $\r{SL}(2,\R)$ subgroups generated by $L_n$, $L_{-n}$ and $L_0+c(n^2-1)/24$, see for example \cite{Liska:2022vrd}. The thermal state $\s_\beta$ corresponds to $H_\r{id} = L_0$. As the $\r{SL}(2,\R)$ action preserves the Casimir, we see that all Möbius states satisfy
\be
\cC^{(2)} = -b_0^2+b_+^2+b_-^2 =-1~.
\ee
As a result, we can parametrize
\be\label{bparameters}
b_0 = \r{cosh}\,\rho,\qq b_+  = \r{sinh}\,\rho\,\r{cos}\,\chi,\qq b_- = \r{sinh}\,\rho \,\r{sin}\,\chi~.
\ee
and we see that Möbius states (and Hamiltonians) are in one-to-one correspondence with a point $(\rho,\chi)$ such that $\rho >0$ and $\chi \sim \chi + 2\pi$. Introducing the complex coordinate $z = e^{i\chi}\,\r{tanh}(\tfrac12 \rho)$ we obtain the open unit disk in the complex plane
\be
\bD = \{ z\in \C \mid |z|<1\}~,
\ee
and the space of Möbius states is in one-to-one correspondence with $\mathbb{D}$. The diffeomorphism $f_z$ preparing a Möbius state from the thermal state $\s_\b = \s(0)$,
\be\label{mobiusstate}
\s(z) \equiv \s_{f_z} = V_{F_z}\s_\b V_{F_z^\dg}~,
\ee
corresponding to the point $z\in \bD$ takes the form
\be\label{diffMobius}
f_z( x) =  2 \,\r{arctan}\le( e^{\rho} \,\r{tan}(\tfrac12( x-\chi))\ri)~,
\ee
and $F_z = f_z^{-1}$. We can verify that this gives the Möbius Hamiltonian \eqref{HfMob} in the parametrization \eqref{bparameters} due to
\be
{1\/f_z'( x)} = \r{cosh}\,\rho - \r{cos}( x-\chi)\,\r{sinh}\,\rho~.
\ee

\paragraph{Unitary Möbius action.} The unitary operator representing an element $M\in \r{SU}(1,1)\cong \r{SL}(2,\R)$ is defined as
\be\label{mobiusaction}
U_M \equiv V_{F_M}
\ee
which is the Virasoro unitary for the diffeomorphism $F_M$ defined from
\be
e^{i F_M(\phi)}  =  {\a e^{i\phi}+\b\/\b^\ast e^{i\phi}+\a^\ast}~,\qq
M=\bpm \a & \b \\ \b^\ast & \a^\ast\epm\in \r{SU}(1,1)~.
\ee
Applying the Möbius action by the unitary \eqref{mobiusaction} to a generic Möbius state \eqref{mobiusstate} gives another Möbius state obtained by acting with $F_M\circ F_z$ on the thermal state $\sigma(0)$. As a result we have
\be\label{RepMobius}
U_M\, \s(z)\, U_M^\dg = \s(Mz)
\ee
which is the state corresponding to the Möbius-transformed point
\be\label{Mobiustransformation}
M  z = {\a z+ \b\over \b^\ast z+  \a^\ast}~.
\ee
The Type I Möbius process obtained by a time-dependent driving Hamiltonian will thus reduce to a succession of Möbius transformations on the disk $\mathbb{D}$.

\paragraph{Information metric.}

The BKM metric \eqref{BKMdiffS1} at a point $f_z \in \text{SL}(2,\mathbb{R}) \subset \widetilde{\text{Diff}}_+S^1$ takes the form
\begin{equation}
    ds^2_{\text{BKM}} = \mathcal{G}_{\text{id}}(dv_z,dv_z) = \frac{c\beta}{24\pi}\int_0^{2\pi} dx\,\Bigl(dv_z''(x)^2 + \gamma\,dv_z'(x)^2\Bigr),
\end{equation}
where $\g=48\pi \ln T\rn_\b/c$ and we have defined the 1-form
\begin{equation}
    dv_z = df_z\circ f_z^{-1} = (\partial_\rho f_z\circ f_z^{-1})\,d\rho+(\partial_\chi f_z\circ f_z^{-1})\,d\chi.
\end{equation}
Using the expression \eqref{diffMobius} for $f_z$, we obtain the metric
\be\label{hyperbolicplane}
ds^2_{\text{BKM}} = {c\b\/24} (\g+1) (d\rho^2 + \r{sinh}^2\rho\, d\chi^2)={c\b\over 24}(\g+1)\,{  4\,d z d\overbar{z}\/(1-|z|^2)^2}
\ee 
with $ |z|<1$. We see that up to a multiplicative prefactor, the BKM metric on Möbius states is the hyperbolic metric on the unit disk. For simplicity of the presentation, we will often ignore this prefactor in the computations below. The prefactor is easily reinstated by multiplying each distance by $\sqrt{{c\b\/24} (\g+1)}$ which can be viewed as the natural unit of length in which to measure BKM distances.

\subsection{Möbius processes from Floquet driving}

In this section, we consider Möbius trajectories of the form
\be
\sigma_t = \s(z_t) \equiv \sigma_{f_{z_t}}
\label{mobiusprocess}
\ee
where the time-dependent diffeomorphism $f_{z_t}$ is given by \eqref{diffMobius}. This trajectory corresponds to curve $t\mapsto z_t$ on the hyperbolic disk. 

We assume that the trajectory \eqref{mobiusprocess} is realized by unitary evolution of the initial state with a time-dependent Hamiltonian which is a Type I Möbius process. Our interest is on periodic Möbius processes where the driving Möbius Hamiltonian \eqref{genHt} is periodic in time (also known as Floquet driving). We focus on periodic two- and multi-step processes where the driving Hamiltonian is time-independent except at discrete points (end-points of the steps) where it changes instantaneously. Such multi-step processes give trajectories which have discontinuous first derivatives in time.

As explained in Section \ref{sec:hamiltonianderivation}, this is equivalent to putting the CFT on a cylinder with a time-dependent metric given, up to a Weyl factor, as
\be
ds^2 = (d\phi+\nu dt)(d\phi+\overbar{\nu} dt)
\ee
where, for Möbius trajectories $(\rho_t,\chi_t)$ and $(\overbar{\rho}_t,\overbar{\chi}_t)$ in the left and right-moving sector, we have explicitly using \eqref{nutrajectory} and \eqref{diffMobius}:
\be
\begin{split}
\nu(t,\phi) = \le(\r{cosh}\,\rho_t + \r{cos}\,\phi\,\r{sinh}\,\rho_t\ri) \dot\chi_t - \r{sin}\,\phi\,\dot\rho_t~,\\
\overbar{\nu}(t,\phi) = \le(\r{cosh}\,\overbar{\rho}_t + \r{cos}\,\phi\,\r{sinh}\,\overbar{\rho}_t\ri) \dot{\overbar{\chi}}_t - \r{sin}\,\phi\,\dot{\overbar{\rho}}_t~.
\end{split}
\ee
Multi-step processes correspond to discontinuous metrics where the discontinuities can be viewed as idealized quenches.

\subsubsection{Two-dimensional representation}

Consider a periodic process where each period corresponds to the application of a unitary $U_M$ where $M\in \r{SU}(1,1)$ is the Möbius transformation representing $U_M$. From the representation \eqref{RepMobius}, we see that the action of $U_M$ on $\s(z_0)$ is the Möbius action of $M$ on $z_0$. As a result, the system after $n$ periods is at the point 
\be\label{periodicstepprocess}
z_{nT} = M^n z_0
\ee
where $T$ is the duration of a period. Hence the process corresponds to iterations of the Möbius transformation $M$. When the matrix $M$ implementing one period takes the form  of a product $M = M_m\cdots M_1 $ with $M_i \in \r{SU}(1,1) $, the process is an $m$-step process.

The matrix $M\in \r{SU}(1,1)$ can be obtained from the operator $U_M$ using the two-dimensional representation of the $\r{SL}(2,\mathbb{R})$ algebra, where we represent
\be
L_0 = {1\/2}\bpm 1 & 0 \\ 0 & -1\epm,\qq L_{1} = \bpm 0 & 0 \\ 1 & 0 \epm,\qq L_{-1} = \bpm 0 & -1 \\ 0 & 0 \epm~.
\ee
which satisfies the usual commutation relations. This representation allows us to translate CFT unitary $\r{SL}(2,\R)$ operators in terms of two-dimensional matrices in $\r{SU}(1,1)$ which are the Möbius transformations \eqref{Mobiustransformation} that preserve the unit disk. The fact that disk preserving Möbius transformations must be in $\r{SU}(1,1)$ follows from the fact that they also preserve the unit circle $\partial \mathbb{D}$. Note that the group $\r{SU}(1,1)$ is isomorphic to $\r{SL}(2,\R)$.

For future reference, we record the expression of the Möbius Hamiltonian associated to the point $z = e^{i\chi}\,\r{tanh}(\tfrac12 \rho)\in \bD$:
\be\label{HMobzexpr}
H_z =  \r{cosh}\,\rho\,L_0+{1\/2} \,\r{sinh}\,\rho \, \le(e^{-i\chi}L_1 + e^{i\chi} L_{-1} \ri) = {1\/2(1-|z|^2)}\le( (1+|z|^2)\,L_0 + 2z L_{1}+2 z^\ast L_{-1}\ri)~.
\ee
Its two-dimensional representation as a $\r{SU}(1,1)$ matrix is given by
\be
H_z =   {1\/2(1-|z|^2)} \bpm 1+|z|^2 & -2z \\ 2 z^\ast & -1-|z|^2\epm = {1\/2}\bpm \r{cosh}\,\rho & - e^{i\chi}\,\r{sinh}\,\rho \\  e^{-i\chi}\,\r{sinh}\,\rho & -\r{cosh}\,\rho\epm,
\ee
which satisfies
\be
\r{Tr}\,H_z = 0,\qq \r{det}\,H_z = -{1\/4}
\ee
where the trace and determinant are taken in the two-dimensional matrix representation of $\r{SU}(1,1)$. It is also useful to note that the hyperbolic distance $d(z_1,z_2)$ between two points $z_1,z_2\in \bD$ takes the form \cite{Anderson:1164418}
\bea\label{hypdistance}
\r{cosh}\,d(z_1,z_2) \= \r{cosh}\,\rho_1\,\r{cosh}\,\rho_2 -\r{cos}(\chi_1-\chi_2)\,\r{sinh}\,\rho_1\,\r{sinh}\,\rho_2 \-
\=1 +{ 2 |z_1-z_2|^2 \ov (1-|z_1|^2(1-|z_2|^2)} ~, \\
\= 2 \,\tr{(H_{z_1}H_{z_2})}~.
\eea

\subsubsection{Floquet driving with a two-step process} 

\begin{figure}
	\centering
	\includegraphics[width=10cm]{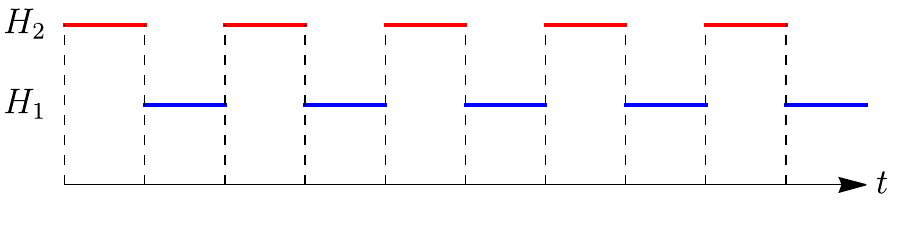}
	\caption{Two-step driving. We alternate between the Hamiltonian $H_1$ for a time $T_1$ and the Hamiltonian $H_2$ for a time $T_2$. }
\end{figure}

Throughout this section, we will illustrate our results using the simple example of a two-step process, which was also considered in \cite{PhysRevB.103.224303}. We take the Hamiltonian $H_1$ for a time $T_1$ and $H_2$ for a time $T_2$ where we define
\be\label{twostepH1h2}
H_1=  \r{cosh}\,\la_1\, L_0+ \r{sinh}\,\la_1\,L_+ ,\qq H_2= L_0.
\ee
The process then corresponds to the periodic iteration of $M = M_2M_1$ where $M_1=e^{i T_1 H_1}$ and $M_2= e^{i T_2 H_2}$. We have explicitly 
\bea
M_1 &=&  \left( \begin{array}{cc} \r{cos}(\tfrac12 T_1) + i\,\r{cosh}\,\la_1\,\r{sin}(\tfrac12 T_1) & - i \, \r{sinh}\,\la_1 \,\r{sin}(\tfrac12 T_1)\\ 
i \,\r{sinh}\,\la_1 \,\r{sin}(\tfrac12 T_1) & \r{cos}(\tfrac12 T_1) - i\,\r{cosh}\,\la_1\,\r{sin}(\tfrac12 T_1)  \end{array}  \right)\\
M_2 &=&\left(\begin{array}{cc}  e^{iT_2/2} & 0 \\ 0 & e^{-i T_2/2} \end{array}\right) \ ,
\eea
and the two-step Möbius transformation takes the form
\begin{align}
M &= M_2M_1\nonumber\\
&=\bpm e^{iT_2/2}\le(\r{cos}(\tfrac12 T_1) + i\,\r{cosh}\,\la_1\,\r{sin}(\tfrac12 T_1) \ri) & - i \, e^{i T_2/2}\,\r{sinh}\,\la_1 \,\r{sin}(\tfrac12 T_1)\\ 
i \,e^{-i T_2/2}\,\r{sinh}\,\la_1\,\r{sin}(\tfrac12 T_1) & e^{-iT_2/2}\le(\r{cos}(\tfrac12 T_1) - i\,\r{cosh}\,\la_1\,\r{sin}(\tfrac12 T_1) \ri)
\epm ~.
\end{align}
To any Möbius Hamiltonian $H_k$, we can associate a point $u_k\in \bD$. In hyperbolic coordinates  $(\la_k,\chi_k)$, we have $u_k=e^{i\chi_k}\r{tanh}(\tfrac12 \la_k) $ and the Hamiltonian is
\be
H_k =\r{cosh}\,\la_k\,L_0 +\r{sinh}\,\la_k\,(\r{cos}\,\chi_k \,L_+ + \r{sin}\,\chi_k\,L_-)~.
\ee
The points associated to \eqref{twostepH1h2} are respectively
\be
u_1 = \r{tanh}(\tfrac12 \la_1),\qq u_2 = 0~.
\ee
Each step is of the form $M_k = e^{i T_k H_k}$ so we see that $M_k$ is an anti-clockwise rotation by an angle $T_k$ about the point $u_k$. This is clear from $\r{SL}(2,\R)$ representation theory. The Hamiltonian $H_k$ is obtained by an $\r{SL}(2,\R)$ adjoint action on $L_0$ which corresponds to a Möbius transformation that maps the origin to the point $u_k$. Since $e^{i  T_k L_0}$ is a rotation around the origin by an angle $T_k$, we see that $e^{i T_k H_k}$ is a rotation around $u_k$ by the same angle.

\subsection{Phases of periodic Möbius processes}

\begin{figure}
	\centering
	\includegraphics[width=8cm]{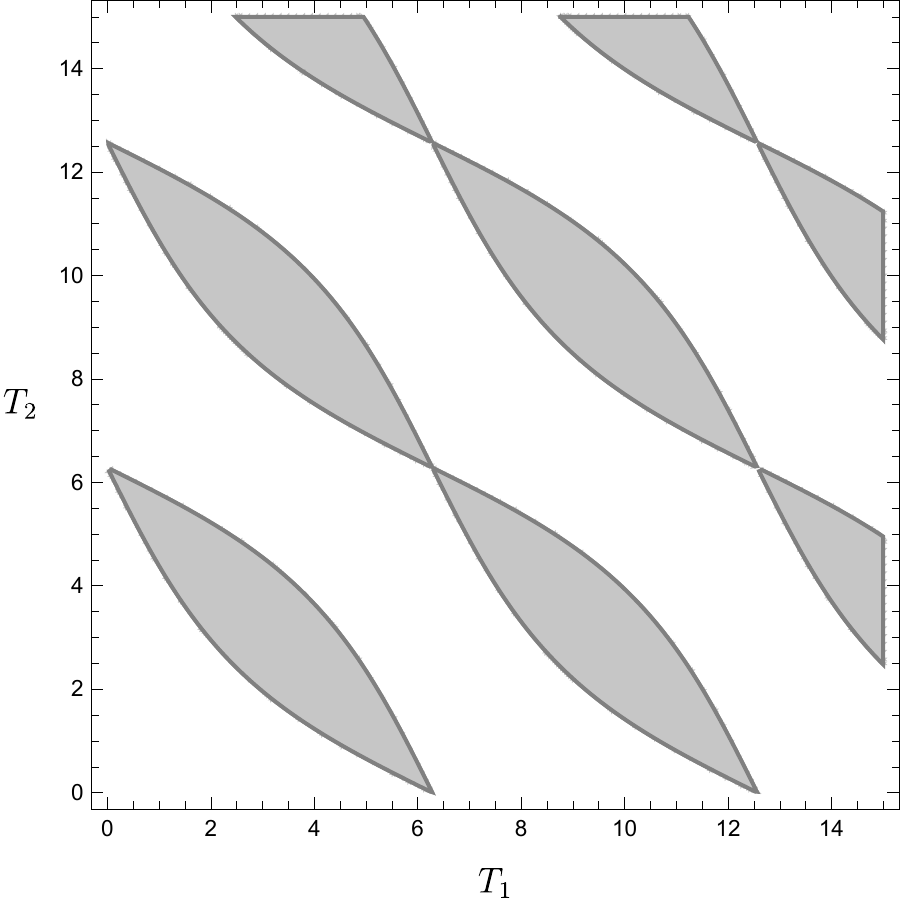}
	\caption{Phase diagram of a two-step process. The shaded region is the heating region where $|\Tr{M}|>2$.}\label{Fig:PhaseDiagram}
\end{figure}

We will now move on to study the late time $n\rightarrow \infty$ behaviour of the process \eqref{periodicstepprocess} obtained by iterating a Möbius transformation $M$. At late times, there are different possible behaviors depending on the type of Möbius transformation that we iterate, elliptic, parabolic or hyperbolic \cite{Anderson:1164418}. We then have two possible phases \cite{Wen:2018agb,Wen:2020wee, Choo:2022lgm}:
\bit
\item $|\r{Tr}\,M|<2$: $M$ is elliptic corresponding to the non-heating phase,
\item $|\r{Tr}\,M|> 2$: $M$ is hyperbolic   corresponding to the heating phase,
\item $|\r{Tr}\,M|=2$: $M$ is parabolic corresponding to the phase transition.
\eit
For the two-step example we have explicitly
\be
\r{Tr}\,M =2\, \r{cos}(\tfrac12 T_1)\,\r{cos}(\tfrac12 T_2)-2\,\r{cosh}\,\la_1\,\r{sin}(\tfrac12 T_1)\,\r{sin}(\tfrac12 T_2)~.
\ee
The phase diagram as a function of $T_1$ and $ T_2$ is depicted in Figure \ref{Fig:PhaseDiagram}. 

\begin{figure}
\begin{center}
	\begin{tabular}{cc}
		\subf{\includegraphics[width=6cm]{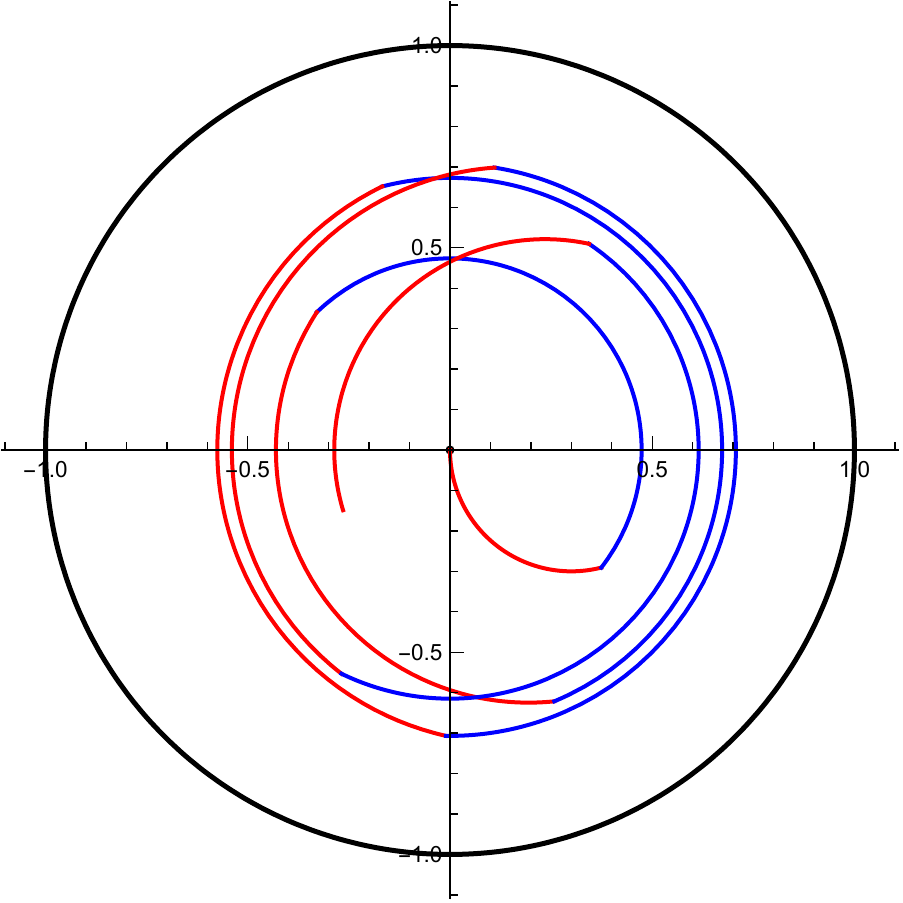}}{Non-heating phase} &
		\subf{\includegraphics[width=6cm]{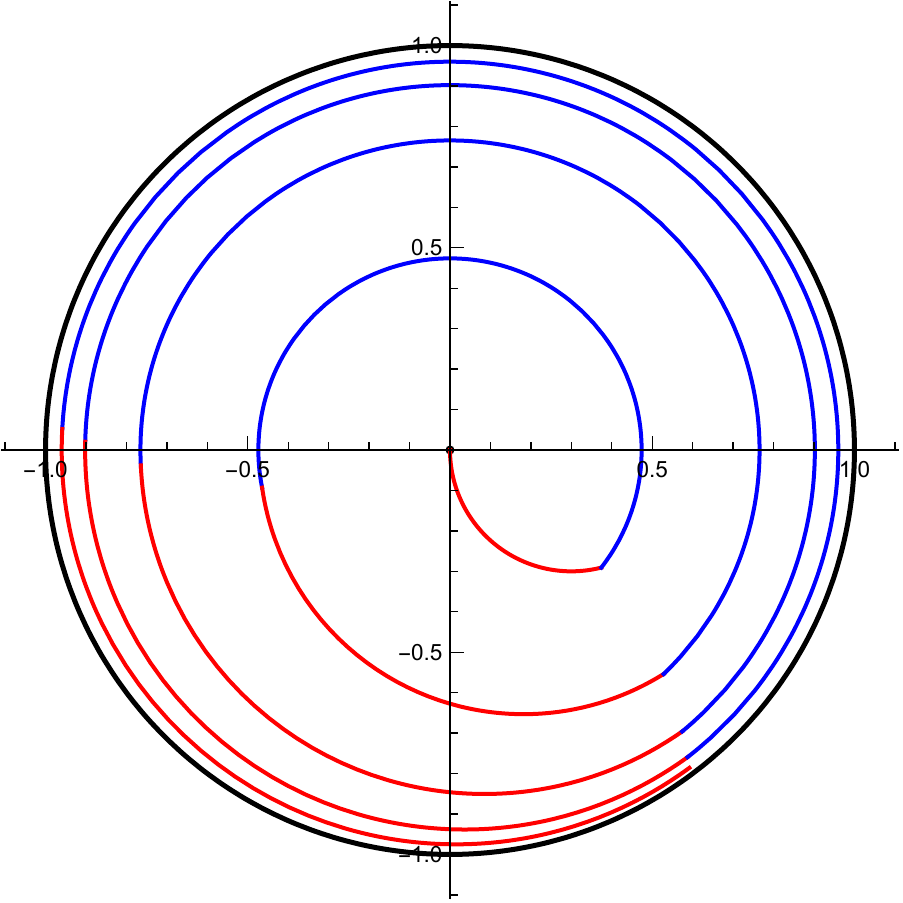}}{Heating phase}
	\end{tabular}
\end{center}
\caption{Representation of the two-step process in the BKM geometry after 10 steps. The trajectory on the hyperbolic disk is represented with alternating blue and red coloring corresponding to each step. In the non-heating phase, the process remains in a finite region, while in the heating phase, it diverges to the asymptotic boundary. The parameters chosen here are $\la=0.7, T_1=1.6$, $T_2=3$ for the non-heating phase and $T_2=4$ for the heating phase.}\label{Fig:KMphases}
\end{figure}

The heating phase corresponds to the regions where $|\r{Tr}\,M|>2$. This corresponds to a situation where the process diverges toward the asymptotic boundary $\p \bD$ of the hyperbolic disk at late time. The non-heating phase corresponds to the regions where $|\r{Tr}\,M|<2$ in which the process remains in a finite region of the disk. This is illustrated for the two-step process in Figure \ref{Fig:KMphases}. From this we see that the BKM geometry gives a simple understanding of the two possible phases of Floquet CFTs.

Moreover the trace of $M$ characterizes the late time behavior of the process. We can write
\be
\lvert\r{Tr}\,M\lvert = \begin{cases} 2\,\r{cosh}\,\la & \text{heating phase} \\ 2\,\lvert\cos{\theta}\lvert & \text{non-heating phase} \end{cases}
\ee
with $\lambda >0$ and $0<\theta<\pi$. In the heating phase, this defines the Lyapunov exponent $\la_L = {2\/T}\la$ which controls the exponential growth of physical quantities. In the non-heating phase, we obtain the \textit{period angle} $\t$ which determines the ergodic properties of the process, as will be discussed in more detailed below. We will now give a detailed analysis of the BKM dynamics in the heating, non-heating and transition regimes.

\subsubsection{Heating phase}

The heating phase corresponding to $M$ being a hyperbolic Möbius transformation so we can write\footnote{Here we could also have a minus sign but this is irrelevant as $M$ and $-M$ give the same Möbius transformation}
\be
\r{Tr}\,M = 2\,\r{cosh}\,\la~.
\ee
Without loss of generality we can represent the matrix $M$ as
\be\label{Mheating}
M = \bpm \r{cosh}\,\la - i\d & e^{i\chi}(\r{sinh}\,\la +i\d) \\ e^{-i\chi}(\r{sinh}\,\la -i\d) & \r{cosh}\,\la+i\d\epm \in \r{SU}(1,1)~.
\ee
After $n$ periods we then have
\be
z_{nT} = M^n z_0 = e^{i\chi} - { 2 e^{i\chi}\,\r{sinh}\,\la\/e^{2n\la}(\r{sinh}\,\la+i\d)+\r{sinh}\,\la - i\d}~.
\ee
At late time, we converge towards the boundary point
\be
z_\infty = e^{i\chi} \in \p \bD
\ee
and the distance from any fixed point $v\in \mathbb{D}$ is
\be
d(v,z_{nT}) = 2n \la + \mathcal{O}(1),\qq n\to \infty~.
\ee

\subsubsection{Phase transition}

The phase transition between heating and non-heating case corresponds to $M$ being parabolic
\be
\r{Tr}\,M=2~.
\ee
This can be studied by taking $\la=0$ in the parametrization \eqref{Mheating}.  We then have
\be
z_{nT} = M^n z_0= e^{i\chi} - {e^{i\chi}\/1+ i n\d}
\ee
We see that the process also reaches the boundary at the point
\be
\lim_{n\to\infty} z_{nT}=z_\infty = e^{i\chi}
\ee
but in this case the distance from a point $v\in \bD$ is
\be
d(v, z_{nT}) = 2 \log n +\mathcal{O}(1)
\ee
and grows logarithmically with $n$.

\subsubsection{Non-heating phase}

In the non-heating phase, $M$ is an elliptic Möbius transformation so we have
\be
\r{Tr}\,M = 2\,\r{cos}\,\t
\ee
which defines an angle $0<\t<\pi$ that we will call the \emph{period angle}. This angle  controls the ergodic properties of the process in the non-heating phase. The eigenvalues of the matrix $M$ are $e^{\pm i\t}$ so we see that iteration of $M$ corresponds to rotations on a circle by an angle $\t$. 

To make this more precise, we can write the matrix $M$ in the form
\be
M = \bpm \r{cos}\,\t +i\,\r{cosh}\,R\,\r{sin}\,\t & e^{i\chi} \,\r{sinh}\,R\,\r{sin}\,\t\\ e^{-i\chi}\,\r{sinh}\,R\,\r{sin}\,\t & \r{cos}\,\t-i\,\r{cosh}\,R\,\r{sin}\,\t \epm
\ee
and after $n$ periods we have
\be
z_{nT} = M^n z_0 = A(\z e^{2i n\t})
\ee
where we have written $ M^n z_0$ as the Möbius transformation by $A$ of a circle centered at the origin and of radius $|\z|$ where
\be
\z = e^{i\chi} \,\r{tanh}(\tfrac12 R)
\ee
and for $z_0 = 0$ we have
\be\label{defAnonheat}
A = \bpm  i\,\r{cosh}(\tfrac12 R) & -i e^{i\chi}\,\r{sinh}(\tfrac12 R) \\ i  e^{-i\chi}\,\r{sinh}(\tfrac12 R) & - i\,\r{cosh}(\tfrac12 R)\epm \in \r{SU}(1,1)~.
\ee
Note that $\z$ is the fixed point of $M$ that is inside $\mathbb{D}$.

\begin{figure}[t]
	
	\begin{center}
		\begin{tabular}{cc}
			\subf{\includegraphics[width=7cm]{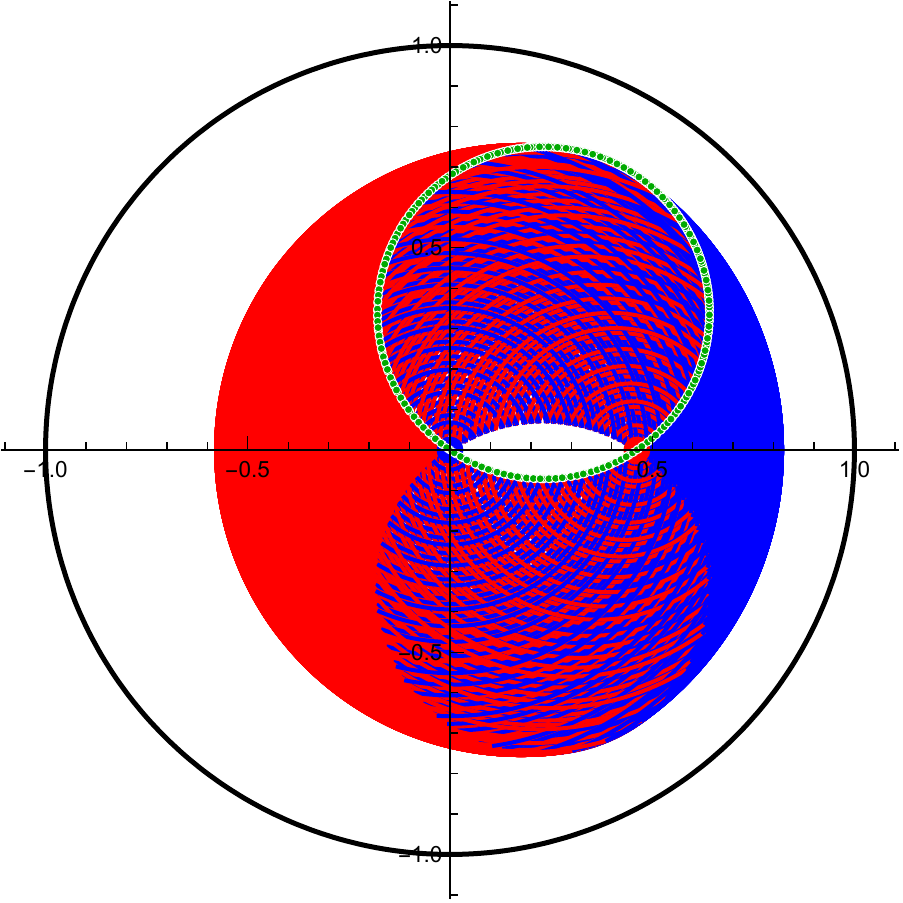}}{Ergodic: $\t=2.4$} &
			\subf{\includegraphics[width=7cm]{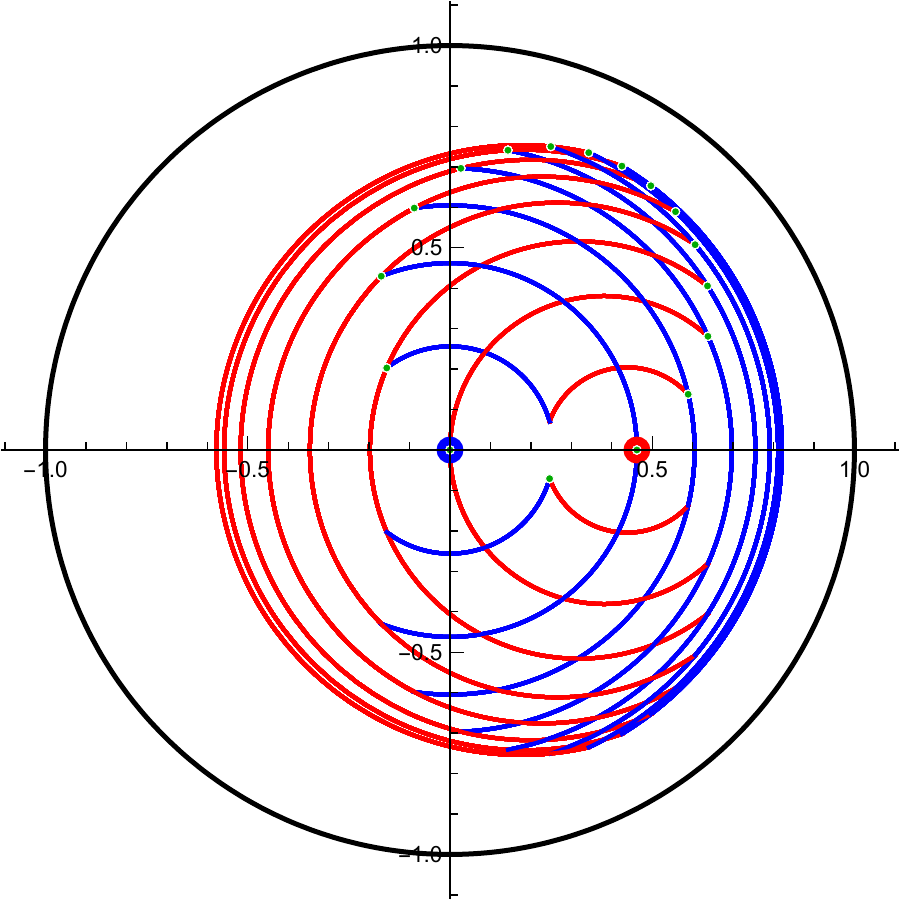}}{Non-ergodic: $\t={13\pi\/17}\approx 2.402$} 
		\end{tabular}
	\end{center}
	\caption{Two-step process in the non-heating phase. The green points correspond to the positions of the process after each period. They lie on the process circle. Ergodicity of the dynamics on this circle, characterized by an angle $\t$, leads to ergodicity of the process in a region $\cR$ of a disk. Each step is a rotation of angle $T_k$ around a center the point $u_k$ representing the Hamiltonian $H_k$. The red and blue points are $u_1$ and $u_2$ corresponding to $H_1$ and $H_2$. This picture corresponds to 1000 steps and $\la=1$. }\label{Fig:ergodic}
\end{figure}

The curve parametrized as $\vphi\mapsto A(\z e^{i\vphi})$ is a circle, indeed it is the image of a circle centered at the origin under the Möbius transformation $A$. After each period, the process comes back to this circle with a rotation by $\t$. We will call this circle the \emph{process circle} and it can be viewed in green in Figure \ref{Fig:ergodic}.

A period gives a rotation by the angle $\t$ on that circle. As a result the process is periodic if and only if $\t$ is a rational multiple of $\pi$. When the process is not periodic, it will be ergodic as it will fill densely the available domain $\mathcal{R}$ of the state space. (We will discuss the domain in more detail in Section \ref{sec:hyp-area}.) These two cases for a two-step process are represented in Figure \ref{Fig:ergodic}.

For the two-step process, the process angle satisfies the equation
\be
\r{cos}\,\t = {1\/2}\r{Tr}(M)=  \r{cos}(\tfrac12T_1)\,\r{cos}(\tfrac12T_2)-\r{cosh}\,\la_1 \,\r{sin}(\tfrac12T_1)\,\r{sin}(\tfrac12T_2)
\ee
The resulting value of $\t$ gives the condition for ergodicity, depending on whether it is a rational multiple of $\pi$. As we discussed above,  the evolution for a duration $T_k$ with the Hamiltonian $H_k$ is a rotation around the point $u_k\in \mathbb{D}$ associated to $H_k$ with angle $T_k$. After each period, we come back on the circle (shown in green in Figure \ref{Fig:ergodic}). This simple geometric description of the process shows that ergodicity on the circle implies ergodicity in the region covered by the process in the hyperbolic disk.

The BKM geometry gives a way to understand Floquet CFTs in the context of ergodic theory: the space of states has a measure given by the BKM metric and the dynamics (given by Möbius transformations) preserves the measure.

\subsection{Information quantities}

\begin{figure}

\begin{center}
	\begin{tabular}{cc}
		\subf{\includegraphics[width=5cm]{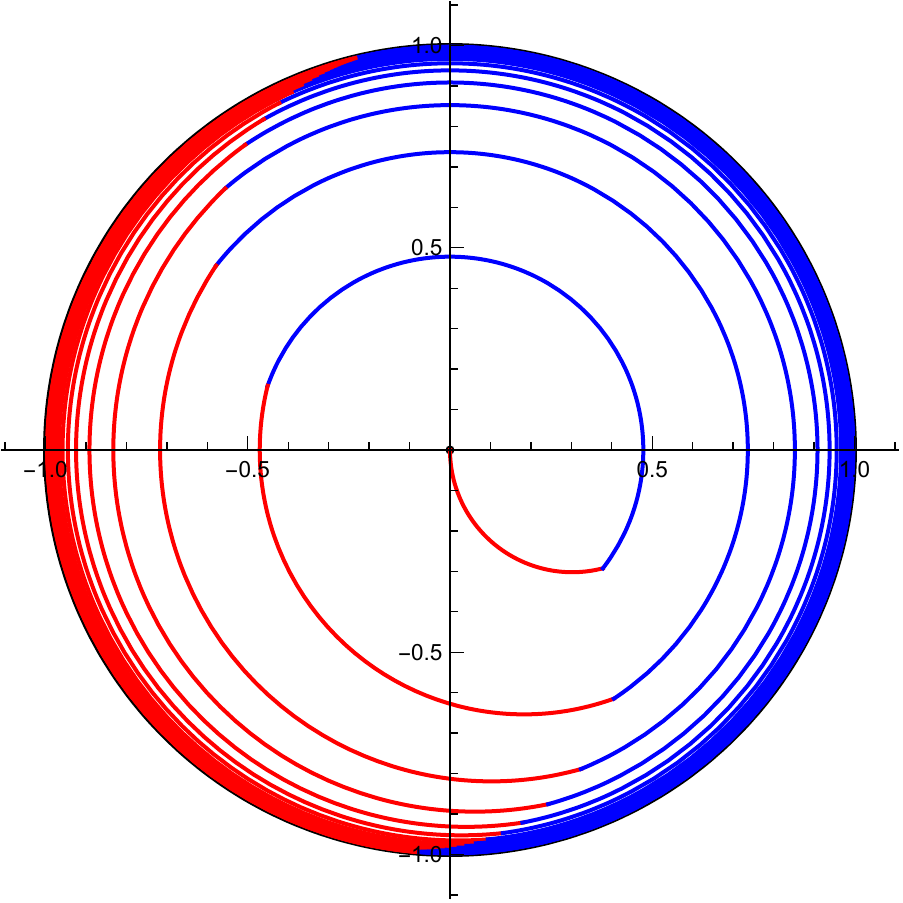}}{BKM representation} &
		\subf{\includegraphics[width=7cm]{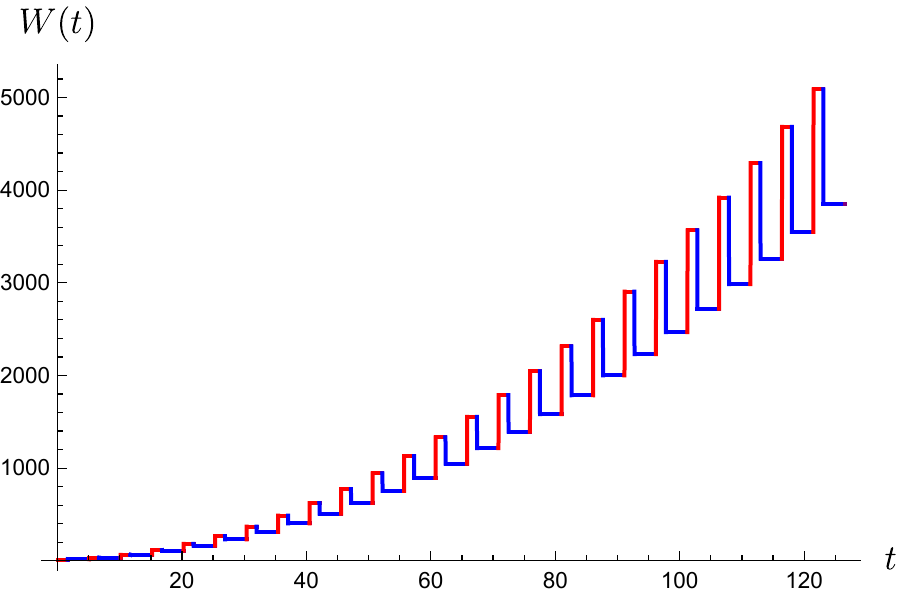}}{Work} \\
		\subf{\includegraphics[width=7cm]{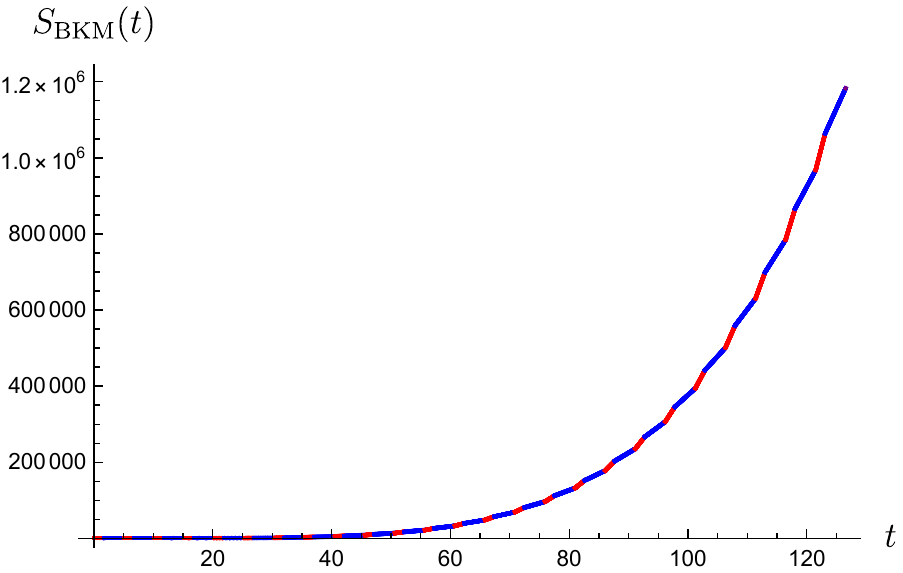}}{BKM action} &
		\subf{\includegraphics[width=7cm]{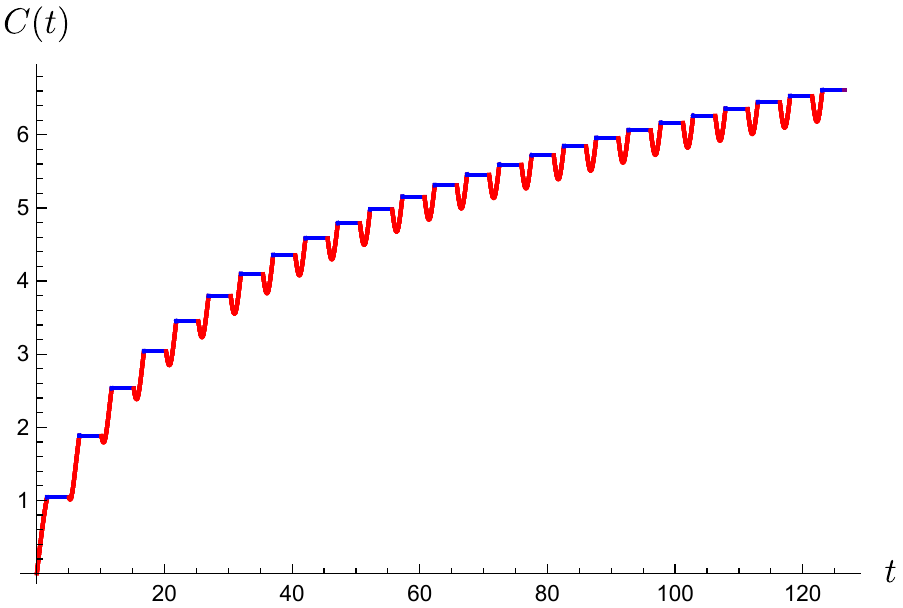}}{Complexity} 
	\end{tabular}
	
\end{center}
\caption{Transition between the heating and non-heating phase after 50 steps. The parameters are $g=0.7, T_1=1.6$ and $T_2\approx 3.462$ is chosen to be at the transition point $|\r{Tr}(M)|=2$. In this case, we observe power law behavior.}\label{Fig:transition}
\end{figure}

We will now consider various information quantities that can be used to characterize periodic two- and multi-step Möbius processes using BKM information geometry. In particular, we consider a measure of complexity of the state at each instant along the process and measures of work\slash energy from non-equilibrium thermodynamics. The summary of considered quantities and their late-time behaviors in the different phases are given in Table \ref{Fig:summarytable}. Plots of the quantities can be found in Figures \ref{Fig:heat}, \ref{Fig:nonheat} and \ref{Fig:transition}.

\subsubsection{BKM complexity}

We consider processes starting from the origin $z_0=0$ (from the initial thermal state) in the BKM geometry. After time $t$, the process is at the point $z_t$ with the Möbius state $\sigma (z_t)$. Following Nielsen's ideas \cite{nielsen_geometric_2005}, we can define a notion of circuit complexity by computing the cost of the trajectory (which we can alternatively view as resulting from a sequence of infinitesimal $\r{SL}(2,\R)$ unitary transformations or "gates"). The complexity of the output state $\sigma (z_t)$, relative to the initial state, is then the minimal circuit complexity, between the initial and final states.
In our case the natural cost function to use is simply the length of the trajectory using the BKM metric. The minimal circuit complexity is then the geodesic BKM distance between the initial and final states.  
Thus we define the complexity\footnote{Incidentally, the BKM distance relates also to the majorization \eqref{majorization}
of the associated classical probability densities: it could be viewed as a notion of what is the ``distance in majorization'' between two probability distributions. We thus have a curious relationship between complexity and (classical) majorization.}
\be
C_{\text{BKM}}(t) =d(z_0,z_t) ~,
\ee
where $d(\cdot, \cdot)$ is the hyperbolic (BKM) distance \eqref{hypdistance}.

\begin{table}[t]\arraycolsep=10pt\def\arraystretch{1.3}
	\centering
\begin{tabular}{c|c|c|c}
&	Heating phase	&  Transition & Non-heating phase \\\hline
	Complexity $C(t)$ & $\sim \la_L t$ & $\sim \log t$ & oscillatory \\\hline
		BKM action $S_\r{BKM}(t)$ &	 $\sim e^{2\la_Lt}$  &  $\sim t^5$ & $\sim \eta_\r{BKM} t$ \\\hline
BKM cost $L_\r{BKM}(t)$  &	 $\sim e^{\la_Lt}$  &  $\sim t^3$ & $\sim \k_\r{BKM} t$ \\\hline
Work $W(t)$  & $\sim e^{\la_L t}$ &  $\sim t^2$ & oscillatory
\end{tabular}
	
	\caption{Time-dependence of  the information quantities for the heating and non-heating phases and at the phase transition. }\label{Fig:summarytable}
\end{table}

\paragraph{Heating phase.}

In the heating phase, the complexity after each period is $C_\r{BKM}(nT) \sim 2\la n$ for $n\to\infty$. As a result we have
\be
C_\r{BKM}(t) \sim \la_L t,\qq t\to\infty
\ee
where the Lyapunov exponent is given by
\be
\la_L = {2\la\/T}~.
\ee
This is the same Lyapunov exponent as defined in \cite{Wen:2020wee}.

\paragraph{Phase transition.} In the parabolic case corresponding to the phase transition, the hyperbolic distance from the origin after $n$ periods is
\be
C_\r{BKM}(nT) = d(z_0,z_{nT})= 2 \,\log\le( n \d + \sqrt{n^2\d^2+1}\ri)
\ee
so that the complexity is
\be
C_\r{BKM}(t) \sim 2 \log\,t + \mathcal{O}(1)
\ee
which grows logarithmically with time.

\paragraph{Non-heating phase.} In the non-heating phase, the complexity remains bounded. This reflects the fact that the process is confined to a finite region of the hyperbolic disk. After $n$ periods, the complexity is 
\be
C_\r{BKM}(n T)= d(z_0,M^n z_0) = d( \z, \z e^{2in\t})
\ee
and we obtain the  formula
\be
C_\r{BKM}(nT) = 2\,\r{arcsinh}\le( \,\r{sinh}(\tfrac12 R)\,|\,\r{sin}(\tfrac12 n\t)|\ri)
\ee
where $R$ is the hyperbolic radius of the process associated to $\z$ via $|\z|=\r{tanh}(\tfrac12 R)$.

We can consider the averaged complexity $\overline{C}_\r{BKM}$ obtained by averaging over $N$. In the ergodic phase this is the same thing as integrating over the circle. As a result, the average complexity takes the form
\be
\overline{C}_\r{BKM} = {1\/\pi}\int_0^{2\pi}d\vphi\, \r{arcsinh}\le( \,\r{sinh}(\tfrac12 R)\,\r{sin}(\tfrac12 \vphi)\ri).
\ee
This integral can be evaluated numerically for any value of $R$. For large $R$, we can expand and integrate analytically to find
\be
\overline{C}_\r{BKM} =  R - \log{4} + \mathcal{O}(e^{-R/2}) \qq \qq R\gg 1.
\ee
In a small $R$ expansion, we can also compute the coefficients
\be
\overline{C}_\r{BKM} = {1\/2\pi}\le( 8 R + {4\/9}R^3 - {13\/225}R^5 + \mathcal{O}(R^{7}) \ri)\qq R\ll 1.
\ee
As expected the averaged complexity grows with the hyperbolic radius of the process. Note that at small $R$, the average complexity is linear but with the slope $4/\pi$ rather than $1$ for large $R$.

\subsubsection{BKM action}

We will now consider the \emph{BKM action} \eqref{BKMcost0} which is the classical action of the Möbius trajectory $z_t$, or equivalently $(\rho_t,\chi_t)$, viewed as a particle on the hyperbolic disk:
\be\label{BKMaction}
S_\r{BKM}(t) =\int_0^{t} ds\, { 4|\dot{z}_s|^2\/(1-|z_s|^2)^2}.
\ee
This should be multiplied by the prefactor ${c\b\/24}(\g+1) $ which we ignore for simplicity.

The motivation to consider this quantity comes from its relation to dissipated work in non-equilibrium quantum thermodynamics. Instead of considering unitary evolution in a closed system (Type I process), we could realize the Möbius trajectory in an open system as a Type II Möbius process. In this case, the BKM action computes the amount of dissipated work along the process, see Section \ref{subsec:dissipation}. However, note that in a Type I process, the BKM action does not have such an interpretation.

The BKM action can be computed explicitly for multi-step processes. Let's consider an $m$-step process where we first apply the Hamiltonian. There the iterated matrix corresponding to a period is
\be
M = M_m M_{m-1}\dots M_2 M_1,\qq M_k\equiv e^{i T_k H_k}~.
\ee
The Hamiltonians $H_k$ are associated to points $u_k\in \bD$ using \eqref{HMobzexpr}. Thus the data of the $m$-step process is the list $\{(u_k,T_k),k=1,\dots,m\}$ of Möbius Hamiltonians and durations. Note that the from step to step we change the driving Hamiltonian abruptly so this gives a trajectory whose derivatives are not continuous. It can be viewed as the idealization of a process where the change between Hamiltonians is quick but not instantaneous.

During the step where we apply $H_k$ for a duration $T_k$, the contribution to the cost is
\be
\int_{t_0}^{t_0+ T_k} ds\,  {4|\dot{z}_s|^2\/(1-|z_s|^2)^2} = T_k \,\r{sinh}^2 d(u_k,z(t_0)),\qq z_t = e^{i t H_k}z_{t_0}
\ee
which can be written in terms of the hyperbolic distance $d(u_k,z(t_0))$ between the point $u_k$ associated to $H_k$ and $t_0$ is beginning of the step we are considering. This can be understood from the fact that for the Hamiltonian $H_k = L_0$, corresponding to $u_k=0$, the evolution $e^{i t H_k}$ is a rotation. A general $H_k$ can be mapped to the origin using a Möbius transformation.

For an $m$-step process, the point after $n$ periods and $k$ steps is at the position
\be
z_{nm+k} = M_k \dots M_1 M^n z_0
\ee
As a result the BKM action after $N$ periods can be written as a sum over periods
\be
S_\r{BKM}(N T) = \sum_{n=0}^{N-1} s_n
\ee
where the contribution of the period $n$ is
\be
s_n=\sum_{k=1}^m T_k\,\r{sinh}^2 d(u_k, z_{nm+k-1})~,
\ee
and $T=\sum_{k=1}^m T_k$ is the period duration.  It is useful to use Möbius invariance of the distance to  rewrite this as
\be
s_n =\sum_{k=1}^m T_k\,\r{sinh}^2 d(v_k,M^n z_0)~,
\ee
where
\be
v_1 = u_1,\qq v_2 = M_1^{-1}u_2,\qq v_3 = (M_2M_1)^{-1}u_3,\qq \dots
\ee
As a result the  behavior of the BKM action follows from the behavior of the quantity $\r{sinh}^2 d(v,M^n z_0)$ as a function of $n$. We will write the points $v_k$ in hyperbolic coordinates
\be
v_k=e^{i\chi_k}\tanh{(\tfrac12 \rho_k)},\qq k=1,\dots,m~. 
\ee

\begin{figure}
\begin{center}
	\begin{tabular}{cc}
		\subf{\includegraphics[width=5cm]{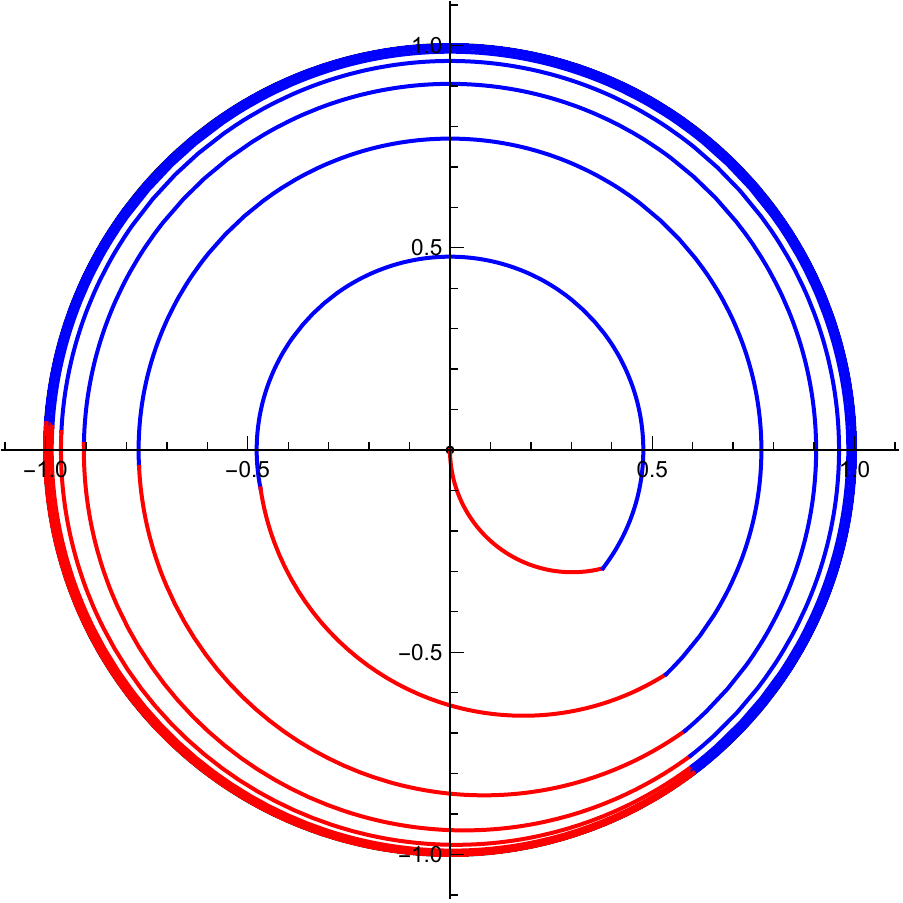}}{BKM representation} &
		\subf{\includegraphics[width=7cm]{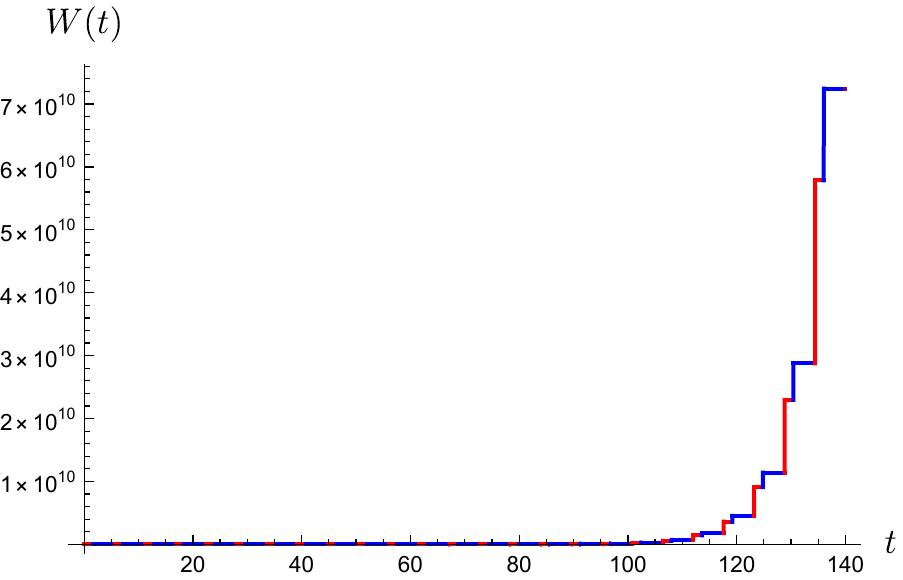}}{Work} \\
		\subf{\includegraphics[width=7cm]{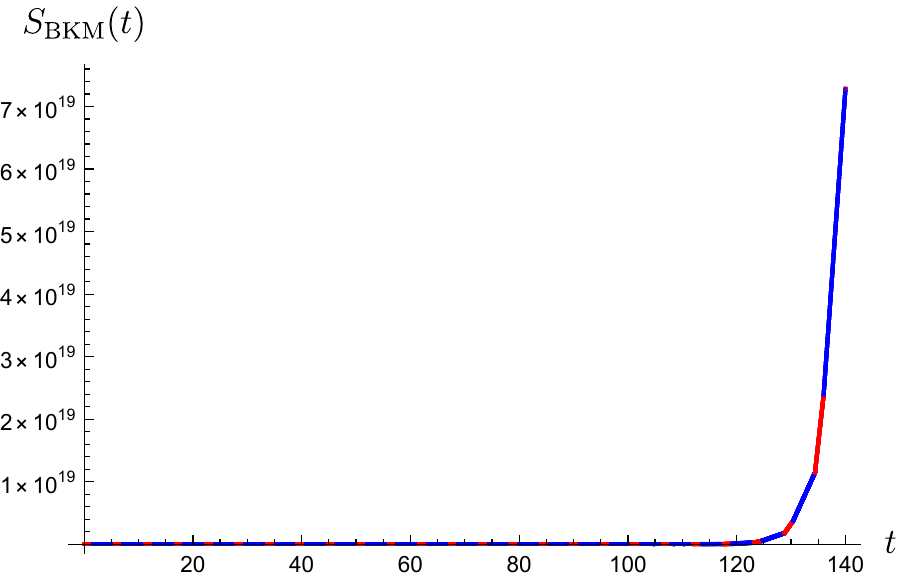}}{BKM action} &
		\subf{\includegraphics[width=7cm]{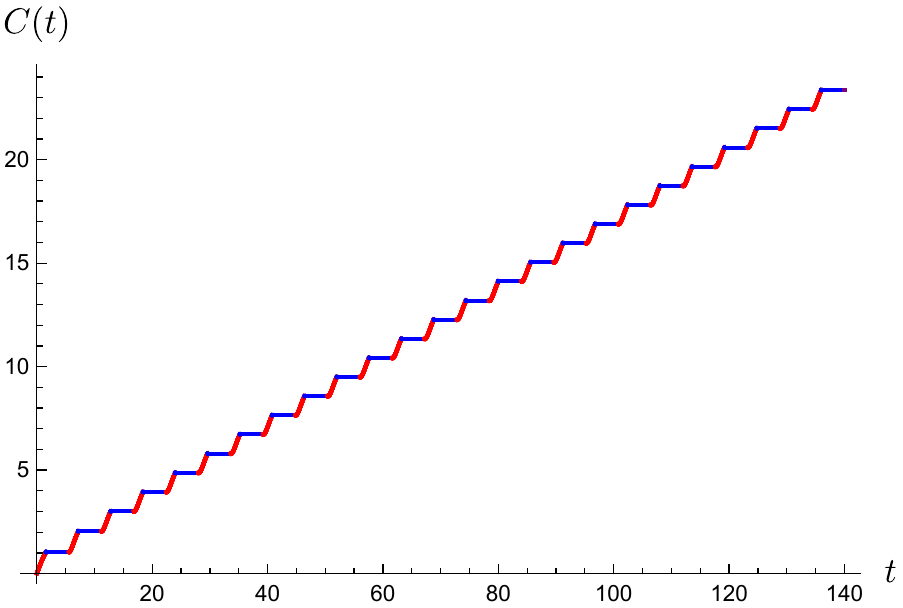}}{Complexity} 
	\end{tabular}
\end{center}
\caption{Heating phase after 50 steps. The parameters are $\la=0.7, T_1=1.6, T_2=4$.}\label{Fig:heat}
\end{figure}

\paragraph{Heating phase.}

In the heating phase, the contribution of a step at late time is
\be
\r{sinh}^2 d(v,M^n z_0) = D(v)\, e^{4n\la}+ \mathcal{O}(e^{2n\la})~, \qq n\to \infty~,
\ee
where  we have
\be
D(v_k) = {\d^2+\r{sinh}^2\la\/2\,\r{sinh}^2\la} \,(\r{cosh}\,\rho_k-\r{cos}(\chi-\chi_k)\,\r{sinh}\,\rho_k)~.
\ee
This shows that the BKM action at late time takes the form
\be
S_\r{BKM} =\sum_{k=1}^m T_k D(v_k)\,e^{2 \la_L t} + \mathcal{O}(e^{\la_L t}),\qq t\to\infty~.
\ee
We see that the BKM action grows exponentially in time with the Lyapunov exponent, see Figure \ref{Fig:heat}.

\paragraph{Phase transition.}

At the phase transition, we can take $M$ to be of the form \eqref{Mheating} with $\lambda = 0$. The contribution of a single step is
\be
\r{sinh}^2 d(v_k, M^{N}z_0) = 4 e^{-2\rho_k} \d^4 n^4 + 2(1+e^{-2\rho_k}) \d^2 n^2 + \r{sinh}^2\rho_k~.
\ee
As a result the contribution of a period is a quartic polynomial in $n$. At large times, we have
\be
s_n =  4 \d^4 \sum_{k=1}^m T_k \,e^{-2\rho_k}  n^4+O(n^2),\qq n\to\infty~,
\ee
The BKM action after $N$ periods is then 
\be
S_\r{BKM}(N T)  = {4 \d^4\/5 T^5} \sum_{k=1}^m T_k \,e^{-2\rho_k} t^5 + O(t^4),\qq t\to\infty~.
\ee

\begin{figure}

\begin{center}
	\begin{tabular}{cc}
		\subf{\includegraphics[width=5cm]{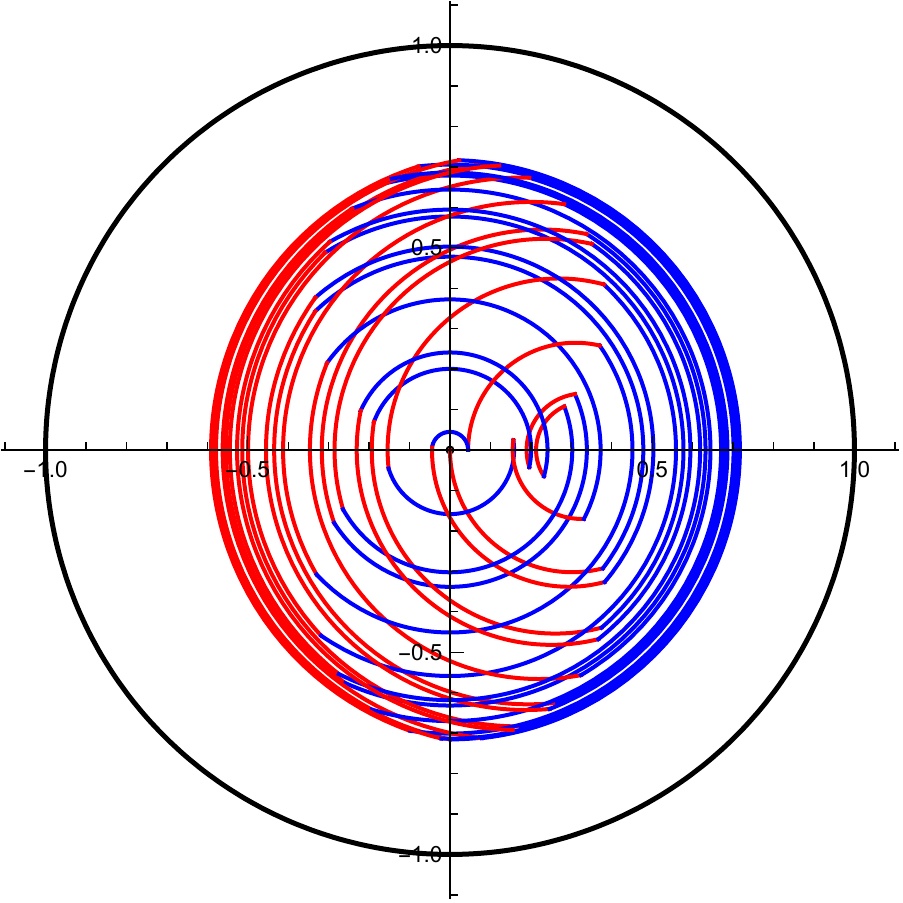}}{BKM representation} &
		\subf{\includegraphics[width=7cm]{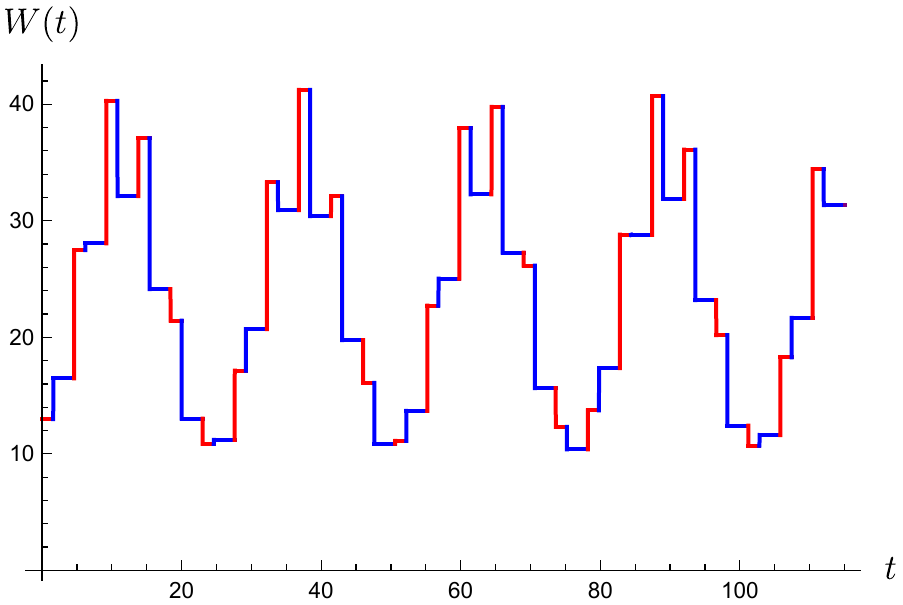}}{Work} \\
		\subf{\includegraphics[width=7cm]{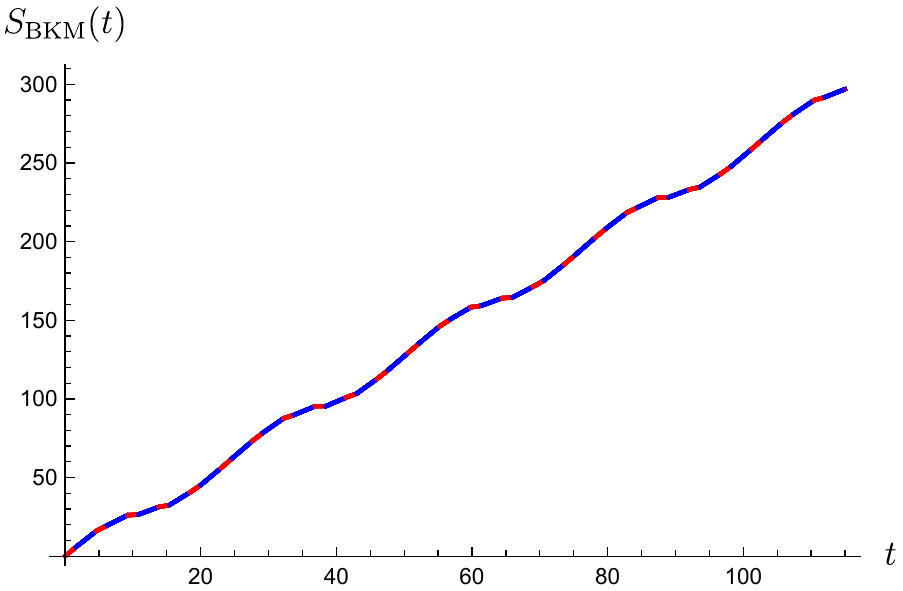}}{BKM action} &
		\subf{\includegraphics[width=7cm]{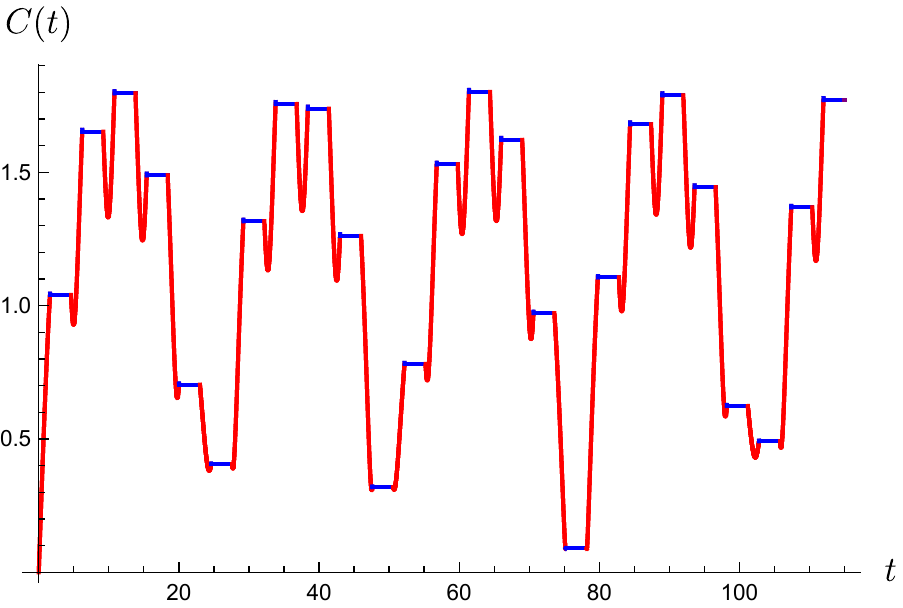}}{Complexity} 
	\end{tabular}
	
\end{center}
\caption{Non-heating phase after 50 steps. The parameters are $\la=0.7, T_1=1.6, T_2=3$. We see that the internal energy and complexity oscillate. In the BKM geometry, the system is restricted to a bounded region and the trajectory intersects itself many times.  We observe approximate periods in the physical quantities. This corresponds to the quasi-periodic motion discussed in \cite{PhysRevB.103.224303}. This is natural from the BKM geometry as the system is restricted to a finite region, so it is inevitable that quasi-periods will emerge.}\label{Fig:nonheat}
\end{figure}

\paragraph{Non-heating phase.}   In the non-heating phase, the BKM action grows linearly, see Figure \ref{Fig:nonheat}. This motivates the definition of the \emph{BKM rate}:
\be
\eta_\r{BKM} \equiv \lim_{t\to\infty} {S_\r{BKM}(t)\/t}~
\ee
which is a finite quantity in the non-heating phase characterizing the process.

We can use the fact the Möbius transformation $A$ defined in \eqref{defAnonheat} to write
\be
d(v_k, M^n z_0) = d(w_k,\z e^{2i n\t})
\ee
where $w_k =A v_k$ using that $A$ is an involution. This allows to write the contribution of a period to the BKM cost as
\be
s_n = \sum_{k=1}^m T_k\,\r{sinh}^2 d(w_k, \z e^{2in\t})~.
\ee
The contribution of a single step is
\be
\r{sinh}^2 d(w_k, \z e^{2in\t}) =\eta(w_k,\z) + (\L_1 (w_k,\z) e^{-2in\t}+\L_2(w_k,\z) e^{-4in\t}+\hc)
\ee
where we use hyperbolic coordinates for $w_k$:
\be\label{defrhok}
w_k = e^{i \phi_k}\tanh{(\tfrac12 d_k)} ~,
\ee
so that the coefficients functions can be written 
\bea\label{BKMcostHyp}
\eta(w_k,\z)  &\equiv& \r{sinh}^2 R +\r{sinh}^2 d_k +{3\/2}\sinh^2{R}\, \sinh^2{d_k}~,\\
\L_1(w_k,\z)  &\equiv& -{1\/4}e^{i (\phi_k-\chi)} \sinh{(2 R)}\,\sinh{(2 d_k)}~,\\
\L_2(w_k,\z)  &\equiv& {1\/4}e^{2i (\phi_k-\chi)} \sinh^2{R}\,\sinh^2{d_k}.
\eea
We then obtain the BKM action after $N$ periods 
\be\label{IBKM}
S_\r{BKM}(NT) =  \sum_{k=1}^m T_k \eta(w_k,\z) N  + \sum_{k=1}^m T_k( \L_1(w_k,\z) s_N(\t) +\L_2(w_k,\z) s_{N}(2 \t)+\hc)
\ee
so we see that the cost is linear in the number of steps and the fluctuations are controlled by the angle $\t$ in terms of the sum over phases
\be
s_N(\t)=  \sum_{n=0}^{N-1} e^{2in\t}~.
\ee
This sum over phases control fluctuations in the rate. In the ergodic case, we have ${s_N\ov N} \to 0$ as $N\to\infty$ so the rate tends to a constant which can be expressed as 
\be
\eta_\r{BKM} = {1\ov T}\sum_{k=1}^m T_k \eta(w_k,\z)   = {1\ov T} \sum_{k=1}^m T_k\le( \r{sinh}^2 R +\r{sinh}^2 d_k +{3\/2}\r{sinh}^2 R\, \r{sinh}^2 d_k\ri)~.
\ee
The rate in the ergodic case can also be obtained as an integral over the process circle:
\be
\eta_\r{BKM}= {1\/2\pi}\int_0^{2\pi} d\vphi \, \sum_{k=1}^m {T_k\/T} \,\r{sinh}^2 d(w_k, \z e^{i\vphi})~.
\ee
This is a reflection of ergodicity as instead of averaging over time, we can integrate over space. This is because for an irrational choice of $\t$, the set of points $\{n\t\mid n\in \bN\}$ becomes dense on the circle with the uniform measure.

\begin{figure}
	
	\begin{center}
		\begin{tabular}{cc}
			\subf{\includegraphics[width=6cm]{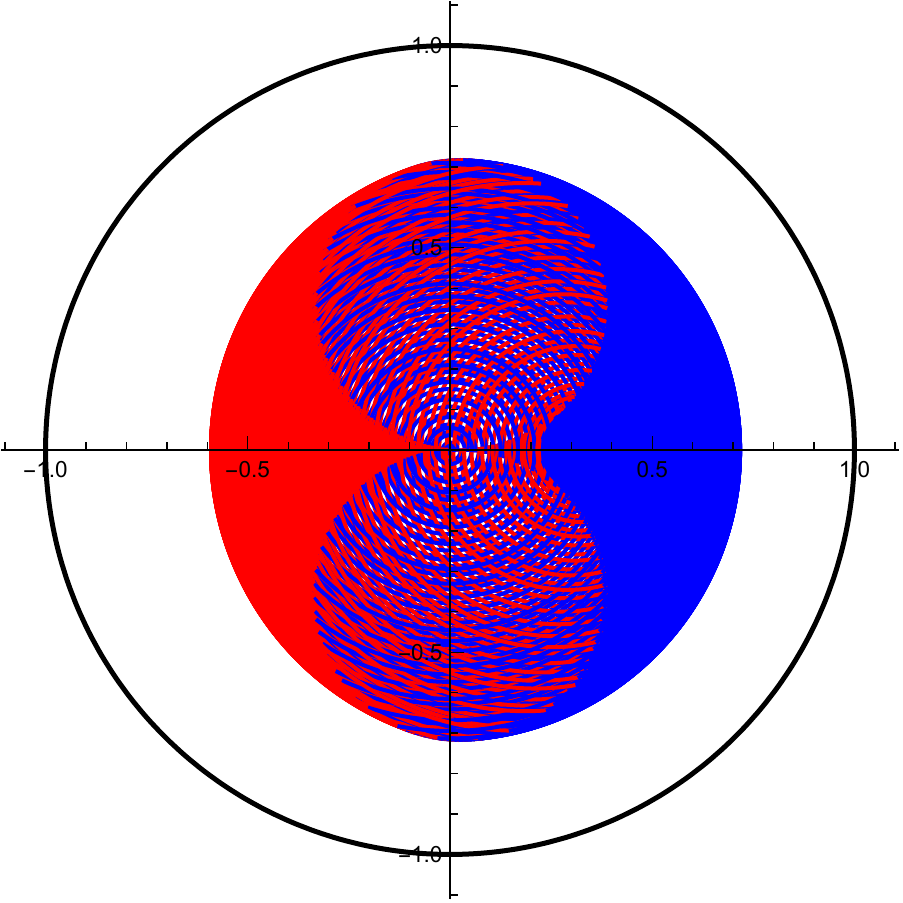}}{} &
			\subf{\includegraphics[width=9cm]{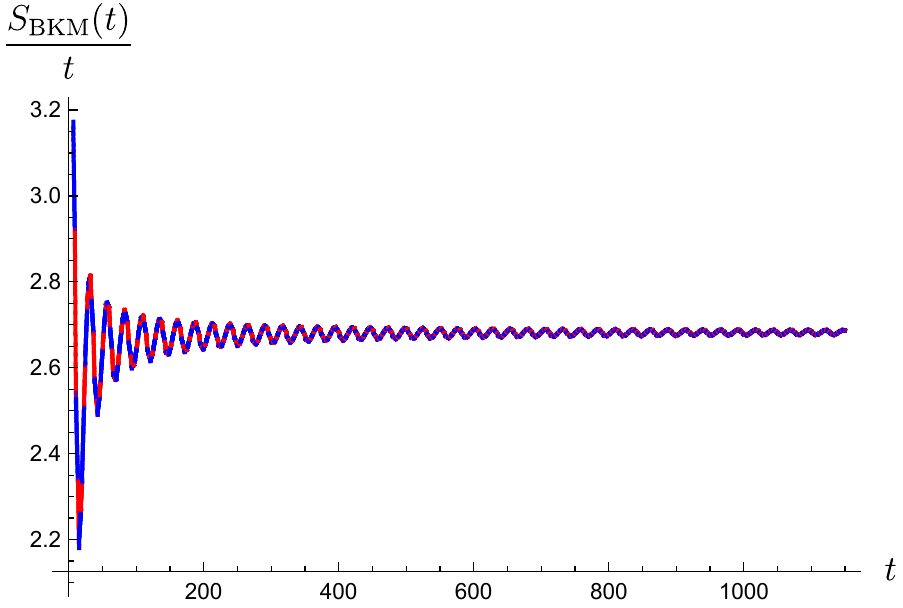}}{BKM rate} \\	\end{tabular}
		
	\end{center}
	\caption{Non-heating phase with parameters are $\la=0.7, T_1=1.6, T_2=3$ after 5000 steps. The system is confined in a region and has ergodic behavior. The BKM action grows linearly with a specific rate.}\label{Fig:rate}
\end{figure}

\subsubsection{BKM cost}

In the previous section, we studied the BKM action, which is the action of a particle representing the Möbius process on the hyperbolic disk. From the perspective of the BKM geometry, a related and more natural quantity is the BKM cost \eqref{thermolength} given by
\be\label{BKMcost}
L_\r{BKM}(t) =\int_0^{t} ds\, { 2|\dot{z}_s|\/\lvert 1-|z_s|^2\lvert}.
\ee
defined with a square root, so that this gives the integrated length of the trajectory.

For an $m$-step process, the BKM cost is
\be
L_\r{BKM}(N T) = \sum_{n=0}^{N-1}\sum_{k=1}^m T_k\,\r{sinh}\,d(v_k,M^n z_0)~,
\ee
In the heating phase, we have at late time
\be
\r{sinh}\,d(v_k,M^n z_0) \sim \la_L t,\qq t\to\infty 
\ee
so we see that the BKM cost grows exponentially
\be
L_\r{BKM}(t) \sim e^{\la_L t },\qq t\to\infty  ~.
\ee
At the phase transition, we have
\be
\r{sinh}\,d(v_k, M^N z_0 ) \sim N^2
\ee
and as a result the BKM cost is
\be
L_\r{BKM}(t) \sim t^3
\ee
In the non-heating phase, the BKM cost is also linear in time just as the BKM action and we can define the cost rate
\be
\k_\r{BKM} = \lim_{t\to\infty} {L_\r{BKM}(t)\/t}~.
\ee
We can give an expression for the cost rate by using that 
\be
\r{sinh}\, d(v,\z e^{i\vphi}) = \sqrt{\eta(v,\z) + (\L_1 (v,\z) e^{-i\vphi}+\L_2(v,\z) e^{-2i\vphi}+\hc)}~.
\ee
From this the cost rate is given by
\be
\k_\r{BKM} = {1\/T}\sum_{k=1}^m T_k \k(v_k,\z)
\ee
where $\k(v,\z)$ can be computed as the integral
\be
\k(v,\z ) = {1\/2\pi}\int_0^{2\pi} d\vphi \,\r{sinh}\, d(v,\z e^{i\vphi})~.
\ee

\subsubsection{Internal energy}

As defined in Section \ref{subsec:dissipation}, the amount of work performed on the CFT along the Möbius trajectory \eqref{mobiusprocess} after time $t$ is given by
\begin{equation}
    W(t) = \int_0^t ds\,\tr{(\sigma_s \dot{H}_s)},
\end{equation}
where $\sigma_t = \sigma(z_t) = \sigma_{f_{z_t}} $. In this section, the CFT is a closed system with unitary state evolution $\dot{\sigma}_t = i[H_t,\sigma_t]$ which implies that there is no heat exchange \eqref{heatQ} and all the work is transformed into internal energy \eqref{EandS} of the CFT:
\begin{equation}\label{cumulativework}
    W(t) = E(t)-E(0) = \tr{(\sigma_t H_t)} - \tr{(\sigma_0 H_0)}.
\end{equation}
To compute this, we can use the fact that the expectation value of the stress tensor in the Möbius state $ \sigma_t = \sigma_{f_{z_t}}$ is
\be
\r{Tr}(\s_t \,T_{--})= f_{z_t}'( x^-)^2 \ln T\rn_\b -{c\/24\pi}\{f_{z_t}( x^-), x^-\}~.
\ee
For the Möbius process $z_t$, we obtain using \eqref{diffMobius} 
\be\label{energydensity}
\r{Tr}(\s _t\, T_{--}) = -{c\over 48\pi} + {1\/(\r{cosh}\,\rho_t - \r{cos}(x^--\chi_t)\,\r{sinh}\,\rho_t)^2}\le( \ln T\rn_\b +{c\over 48\pi}\ri)
\ee
in terms of $z_t= e^{i\chi_t}\,\r{tanh}(\tfrac12 \rho_t)$ so that $(\rho_t,\chi_t)$ are the hyperbolic coordinates of the process.\footnote{In the heating phase, the second term in \eqref{energydensity} develops a sharpening peak (the position of which rotates around the
circle), leading to a heating spot surrounded by a cooling region. This process was called as conformal Floquet cooling, and discussed in \cite{Wen:2022pyj}.} The internal energy can then be computed as
\be
E(t) =c_0(t)\,\r{Tr}(\sigma_t\, L_0) + c_+(t)\,\r{Tr}(\sigma_t\, L_+ )+ c_-(t)\,\r{Tr}(\sigma_t\, L_-)
\ee
using the representation \eqref{genHt} for the driving Hamiltonian. We can then use that
\bea
\r{Tr}(\s_t \, L_0) \= -{c\/24}+ {c\/24}(1+\g)\,\r{cosh}\,\rho_t~,\\
\r{Tr}(\s_t\, L_+)\= {c\/24}(1+\g)\,\r{sinh}\,\rho_t\,\r{cos}\,\chi_t~,\\
\r{Tr}(\s_t\, L_-) \=  {c\/24}(1+\g)\,\r{sinh}\,\rho_t\,\r{sin}\,\chi_t~,
\eea
where we are using $\g = 48\pi \ln T\rn_\b /c$.

We can now give the late time behavior of the work \eqref{cumulativework}. In the heating phase, we have $\rho_t\sim \la_L t$ so we the work grows exponentially as
\be
W(t) \sim \r{exp}\le( \la_L t + o(t)\ri),\qq t\to\infty~.
\ee
At the phase transition, we have $\rho_t \sim 2 \log t + \mathcal{O}(1)$ which leads to
\be
W(t) \sim t^2~.
\ee
Finally, the work is oscillating in the non-heating phase: energy is then being exchanged back and forth between the CFT and the unspecified external system that creates the Möbius driving (the classical force).

\subsection{Discussion and extensions}

We have seen that the BKM geometry is a powerful tool to study the dynamics of Möbius processes. Information quantities such as the BKM action, cost and complexity can be defined and computed. In this subsection, we discuss some extensions of these ideas.

\subsubsection{Hyperbolic area}
\label{sec:hyp-area}

\begin{figure}
	\centering
	\includegraphics[width=8cm]{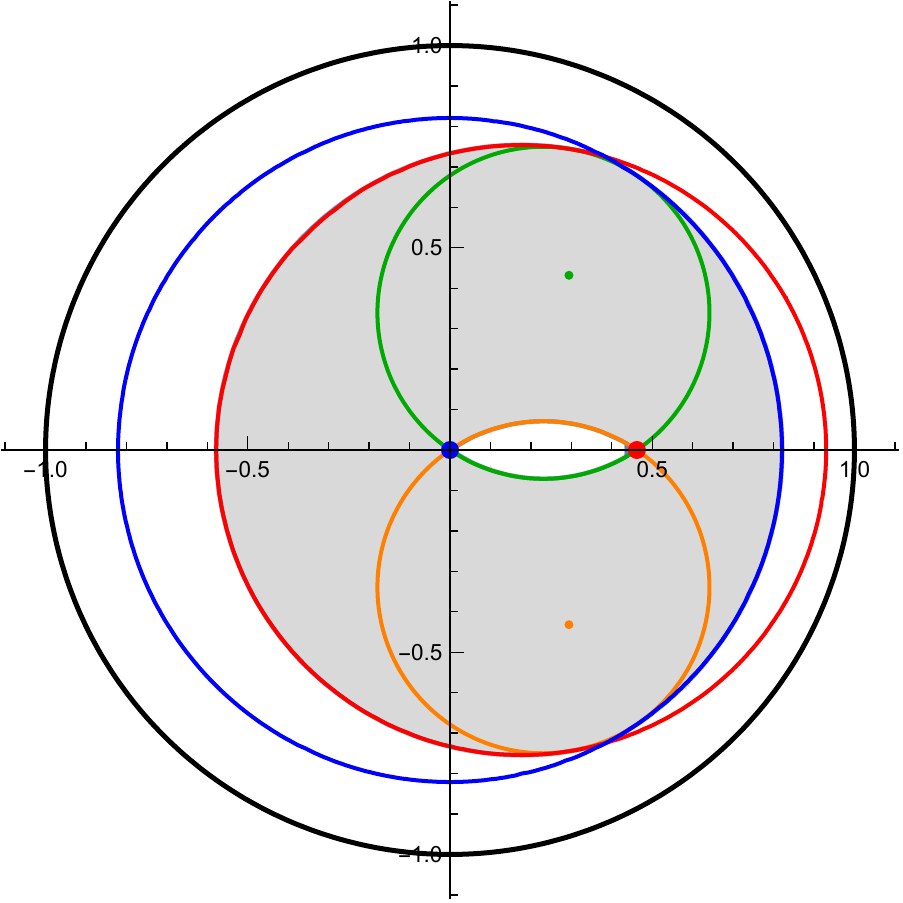}
	\caption{Geometric description of the domain $\cR$ covered by the two-step process of Figure \ref{Fig:ergodic} in the ergodic case. The process circle  with centered $\z$ and radius $R$ is represented in green. Its image under $M_1$ is in orange. The domain $\cR$ is the intersection region of the two blue and red disks centered at $u_1,u_2$ and with radii $R_1= R+\rho_1,R_2=R+\rho_1$, to which we exclude the intersection region of the green and orange disks.}\label{Fig:geoProcess}
\end{figure}

As we have seen, the process in the non-heating phase fills a subregion of the hyperbolic disk completely, as can be viewed from Figure \ref{Fig:ergodic}. The process circle, represented in green, is the circle on which the process returns after each period. During each period, the process rotates around $u_1$ for an angle $T_1$ and then rotates around $u_2$ for an angle $T_2$. 

In the ergodic case corresponding to $\t/\pi$ irrational, the process is ergodic on the circle so it is also ergodic in the domain $\cR$ in which it is confined. This domain can be described geometrically as it is bounded by four circles. Its hyperbolic area
\be
\r{Area}(\cR) = \int_{\cR} d^2 z\,\sqrt{\mathcal{G}_{\text{BKM}}},
\ee
where $\mathcal{G}_{\text{BKM}}$ denotes the determinant of the hyperbolic metric \eqref{hyperbolicplane}, is a natural information quantity in the BKM geometry. It gives a measure of the volume of phase space available to the process.

The area $\r{Area}(\cR)$ can be computed as the difference between the areas of two intersections of pairs of disks, as represented in Figure \ref{Fig:geoProcess}. The outermost intersection is given by the two disks $D_1$ and $D_2$ centered at $u_1$ and $u_2$ and with radii 
\be
R_1= R+\rho_1,\qq R_2=R+\rho_2~.
\ee
in terms of $|\z|=\r{tanh}(\tfrac12 R), |v_k|=\r{tanh}(\tfrac12 \rho_k)$. The innermost intersection corresponds to two disks centered at $\g$ and $M_1\g$ with radii $R$. The area is the difference between the areas of these two intersections,

\begin{figure}

\begin{center}
	\begin{tabular}{cc}
		\subf{\includegraphics[height=5cm]{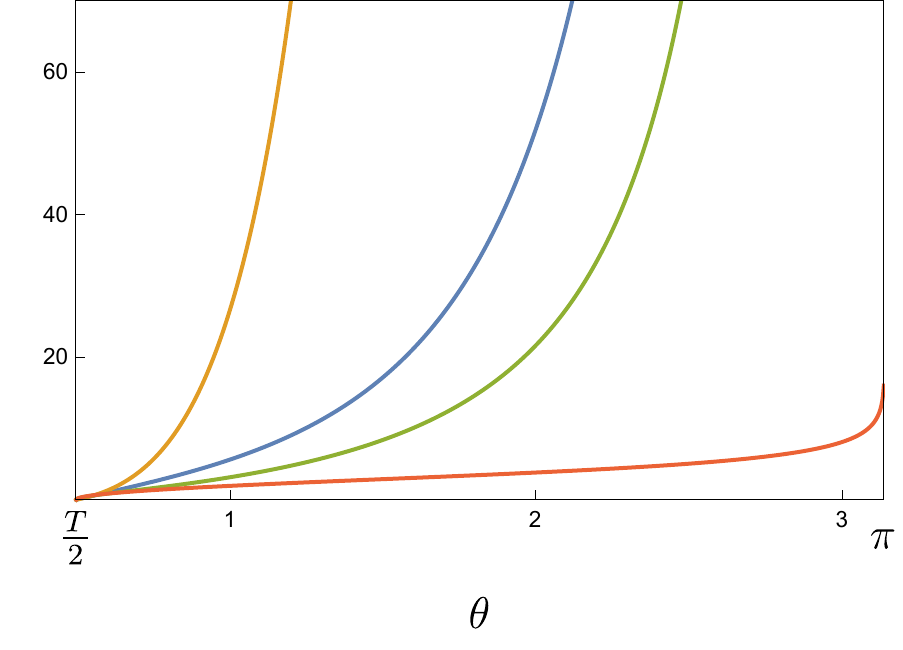}}{} &
		\subf{\includegraphics[height=5.1cm]{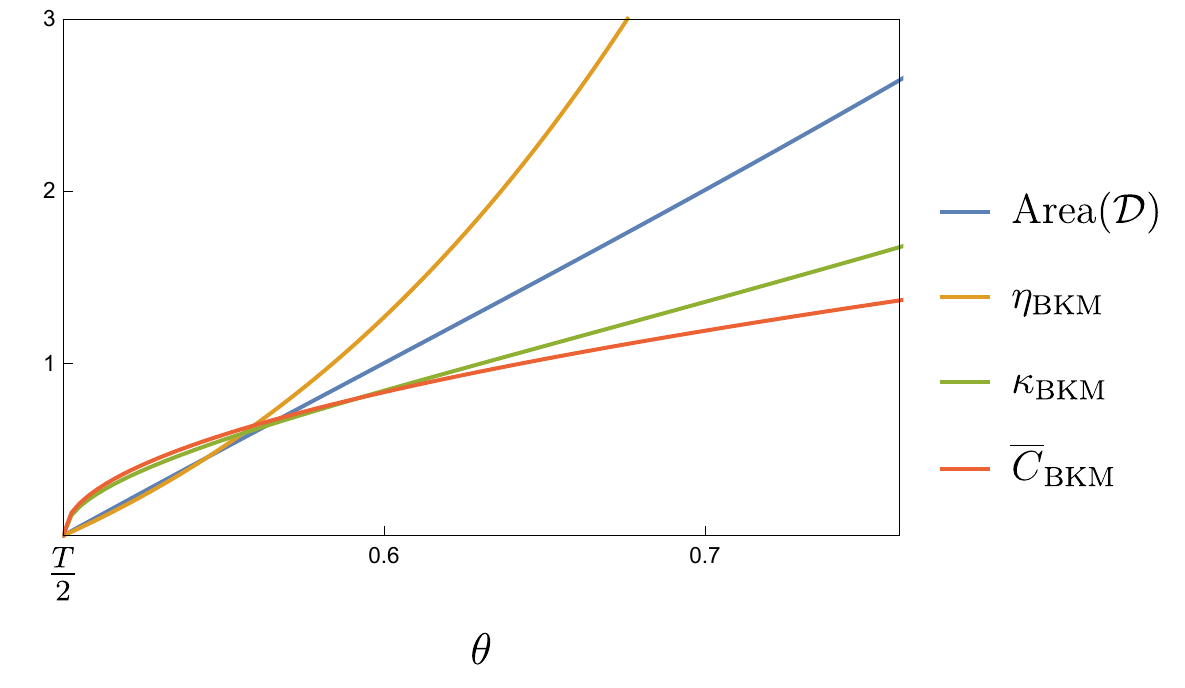}}{} 
	\end{tabular}
	
\end{center}

\caption{Comparison of the BKM area of the domain, the rates $\eta_\r{BKM}$ and $\k_\r{BKM}$ associated with the BKM action and cost, and the average complexity obtained from the BKM distance. At $\t= {T\/2}$, the process becomes trivial with the three quantities vanishing. At $\t=\pi$, the process becomes heating and the four quantities diverge. Although these quantities all describe the ``size'' of the process region, they differ quantitatively, as shown on the right plot which zooms over the region close to the trivial process. For these plots, we use the two-step process with $T= 1$.}\label{Fig:comparisonArea}
\end{figure}

For simplicity, we consider the case $T_1=T_2={T\/2}$ where the process can be parametrized by the period $T$ and the angle $\t$. In this case we have $\rho_1=\rho_2=R$ so we just have to consider symmetric intersections of disks. The intersection of two disks $D$ and $D'$ with the same radius $R_0$ and separated by a distance $2 d_0$ has the  hyperbolic area 
\be
A(D\cap D') =  2( 2\a_0 (\r{cosh}(R_0)-1) - (\pi - 2\a_0-2 \b_0) ) = 4\a_0 \,\r{cosh}\,R_0+4 \b_0- 2\pi
\ee
which is obtained as twice the difference of the area of a portion of the disk and the area of a hyperbolic triangle. The angles appearing in the formula are acute angles of a right triangle which can be expressed using hyperbolic trigonometry as \cite{Anderson:1164418}
\be
\r{cos}\,\a_0 = {\r{tanh}\,d_0\/\r{tanh}\,R_0},\qq \r{sin}\,\b_0 = {\r{sinh}\,d_0\/\r{sinh}\,R_0}~.
\ee
For our two-step process, the outermost intersection consists of disks centered at $u_1= 0$ and $u_2= \r{tanh}(\tfrac12 \la)$ with radii $R_1=R_2=2R$. The area of the first intersection is then
\be
\r{Area}(D_1\cap D_2) = 8 \a_1 \,{\r{tan}^2 ({\t\/2})\/\r{tan}^2({T\/4})}+4 (\b_1-\a_1)-2\pi
\ee
where the acute angles $\a_1$ and $\b_1$ are determined from
\be
\r{cos}\,\a_1 = \r{cos}(\tfrac14 T)\le( 1- { \r{tan}^2({T\/4})\/2\,\r{tan}^2({\t\/2}) }\ri),\qq \r{sin}\,\b_1 = {\r{cos}({\t\/2})\,\r{tan}({\t\/4})\/2\,\r{tan}({\t\/2})}~.
\ee
The innermost intersection consists of two disks $D_\g$ and $D_{M_1\g}$ centered at $\g$ and $M_1\g$ and with radius $R$. The area of their intersection is then
\be
\r{Area}(D_\g\cap D_{M_1\g\textbf{}}) = 4\a_2\,{\r{tan}({\t\/2})\/\r{tan}({T\/4})} + T-2\pi 
\ee
where $\b_2= {T\/4}$ and $\a_2$ is defined from
\be\r{cos}\,\a_2  = \r{cos}(\tfrac{\t}{2})\,\r{tan}(\tfrac{T}{4})
\ee
The area of the domain is then obtain by taking the difference and gives
\be
\r{Area}(\cR) =   {4\,\r{tan}({\t\/2})\/\r{tan}({T\/4}) }\le(  {2 \,\r{tan}({\t\/2})\/\r{tan}({T\/4})} \a_1 -\a_2\ri)+4 (\b_1-\a_1)-T~.
\ee
This hyperbolic area is a natural information quantity associated with the process. Like the BKM action and cost rates or the complexity, it measures the size of the region $\cR$ of which the domain is confined. See Figure \ref{Fig:comparisonArea} for a comparison of these three quantities in the two-step example. It would be interesting to find a physical interpretation for this quantity.

\subsubsection{Aperiodic processes}

We can also consider aperiodic  processes  \cite{Wen:2020wee, Wen:2021mlv}. In general we a multi-step process is given by a sequence $a_k\in \{1,2\}$ where at the step $k$ we apply the Hamiltonian $H_{a_k}$ for a time $T_{a_k}$. The two-step process we considered had an alternating sequence but we can take a more general sequence.

\paragraph{Quasi-periodic driving.}

\begin{figure}[t]
	\centering
	\includegraphics[width=10cm]{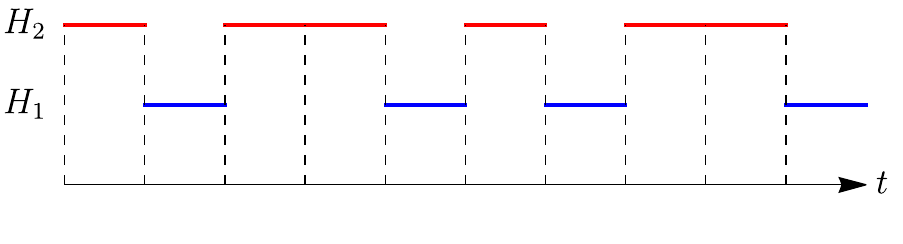}
	\caption{Fibonacci quasi-periodic driving. We alternate between the Hamiltonian $H_1$ and $H_2$ according to the Fibonacci sequence. }
\end{figure}

\begin{figure}

\begin{center}
	\begin{tabular}{cc}
		\subf{\includegraphics[width=5cm]{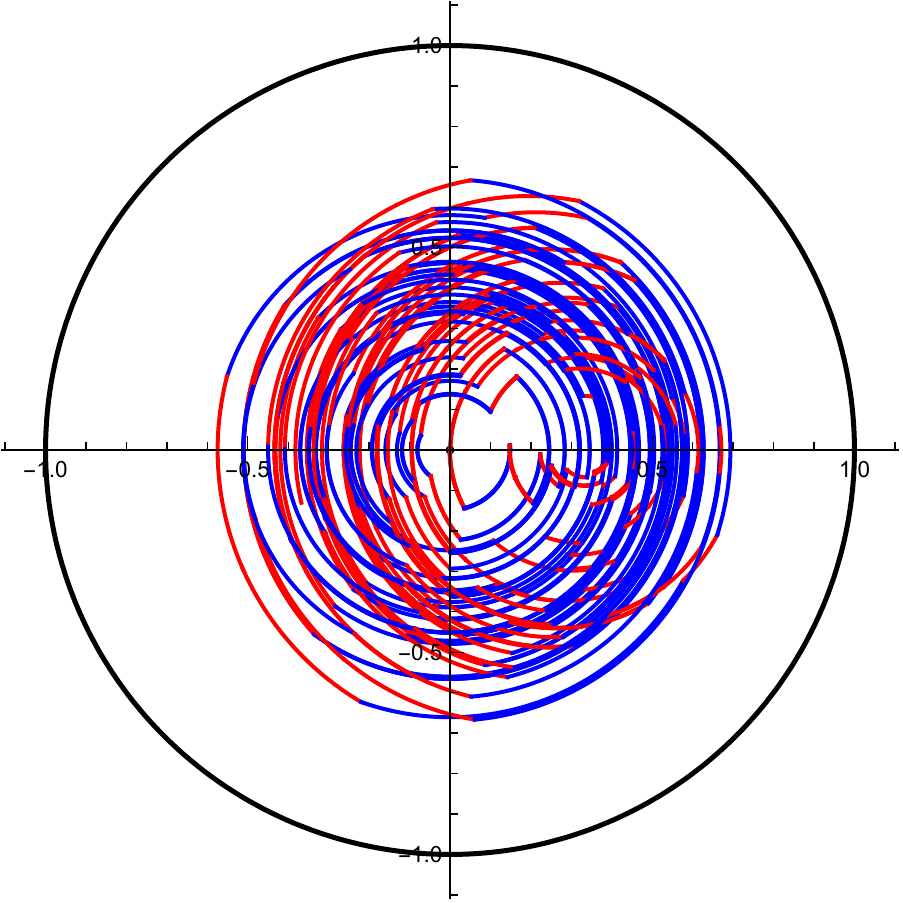}}{BKM representation} &
		\subf{\includegraphics[width=7cm]{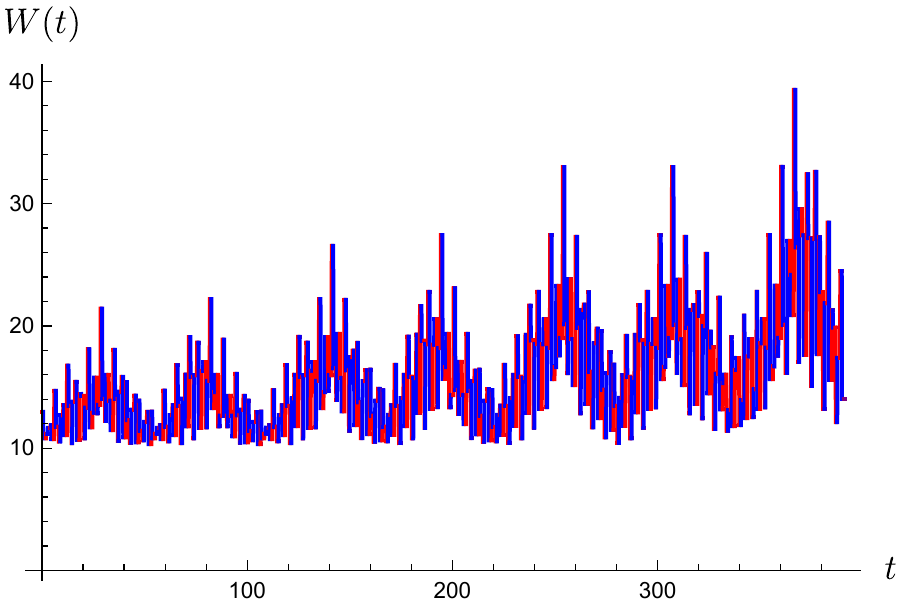}}{Work} \\
		\subf{\includegraphics[width=7cm]{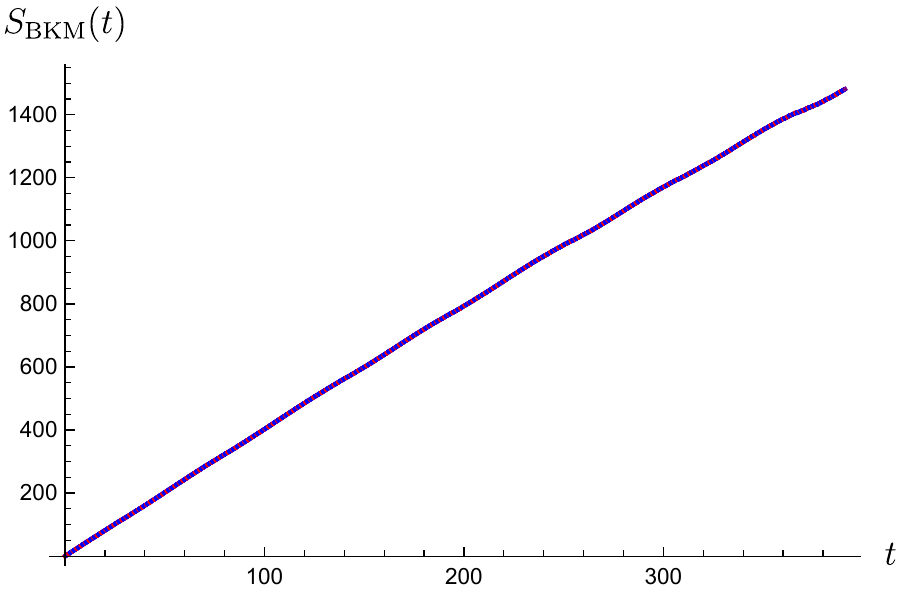}}{BKM action} &
		\subf{\includegraphics[width=7cm]{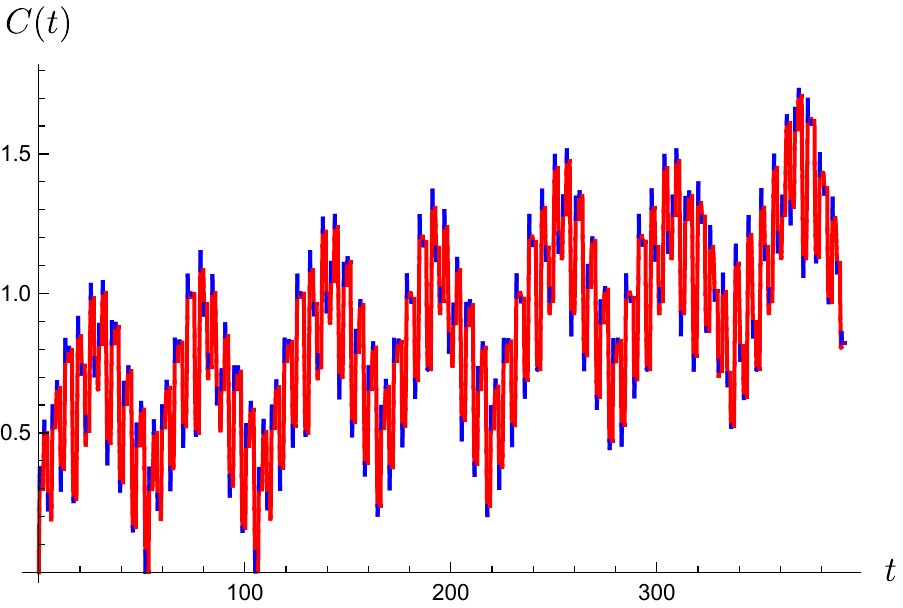}}{Complexity} 
	\end{tabular}
	
\end{center}

\caption{Fibonnaci driving for 500 steps. The parameters used are $\la=0.7,  T_1=0.4, T_2=1.4$. The energy and complexity oscillate and we see quasi-periodic behavior.}\label{Fig:Fib}
\end{figure}

An example of quasi-periodic driving is the Fibonacci driving \cite{Lapierre:2020roc,Wen:2020wee} which is closely related to the Fibonacci quasi-crystal \cite{PhysRevLett.50.1870,PhysRevLett.50.1873, FibInv}. In the context of \cite{Wen:2020wee}, it is a two-step process with driving times $T_1,T_2$ where we alternate the two Hamiltonians using the Fibonnaci sequence 
\be
a_k = 1,0,1,1,0,1,0,\dots
\ee
as a rule. 
(Here 0/1 corresponds to choosing the driving Hamiltonian to be $H_1/H_2$.) 
The Fibonacci sequence can be generated by a limiting process starting with $1$ and repeating the substitution $1\ra 10, 0\ra 1$, which gives
\be
1\ra 10 \ra 101 \ra 10110\ra 10110101 \ra\dots
\ee
The physical quantities for the Fibonacci process are represented in Figure \ref{Fig:Fib}. We see a quasi-periodic behavior.

Compared with periodic driving with the phase diagram Figure \ref{Fig:PhaseDiagram}, Fibonacci driven CFTs can have a complicated phase structure. Ref. \cite{Lapierre:2020roc} reported a fractal structure between heating and non-heating phases, with the latter restricted to ``Cantor lines'' in the phase diagram. An interesting problem for future study is to investigate ergodicity in the Cantor lines and see if non-ergodic regimes are at all possible.\footnote{We thank Bastien Lapierre for pointing this out to us.}

\begin{figure}

\begin{center}
	\begin{tabular}{cc}
		\subf{\includegraphics[width=5cm]{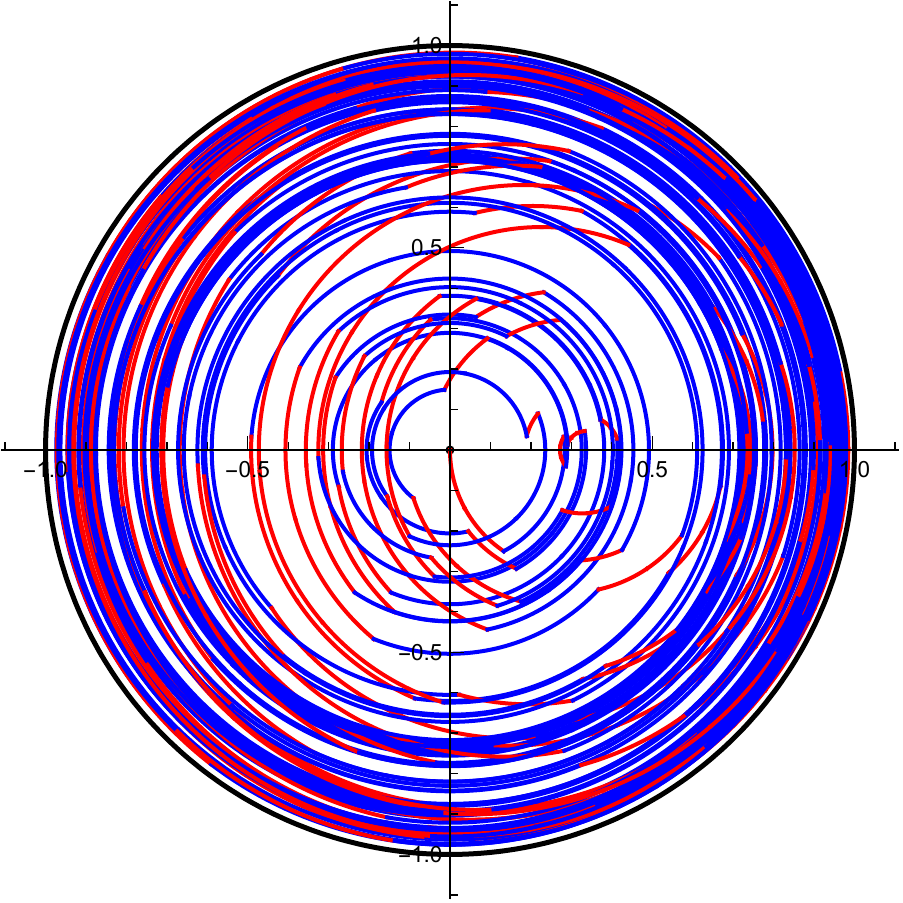}}{BKM representation} &
		\subf{\includegraphics[width=7cm]{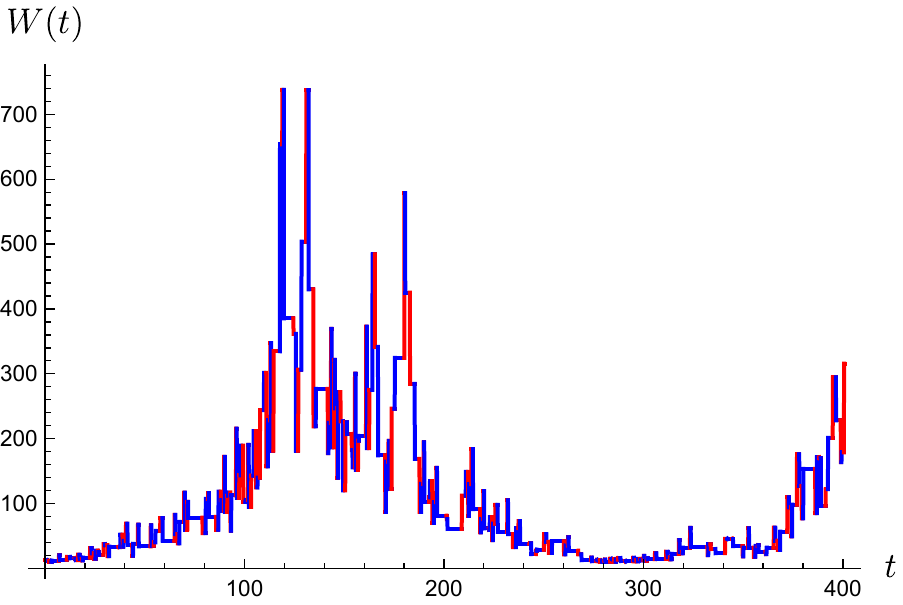}}{Work} \\
		\subf{\includegraphics[width=7cm]{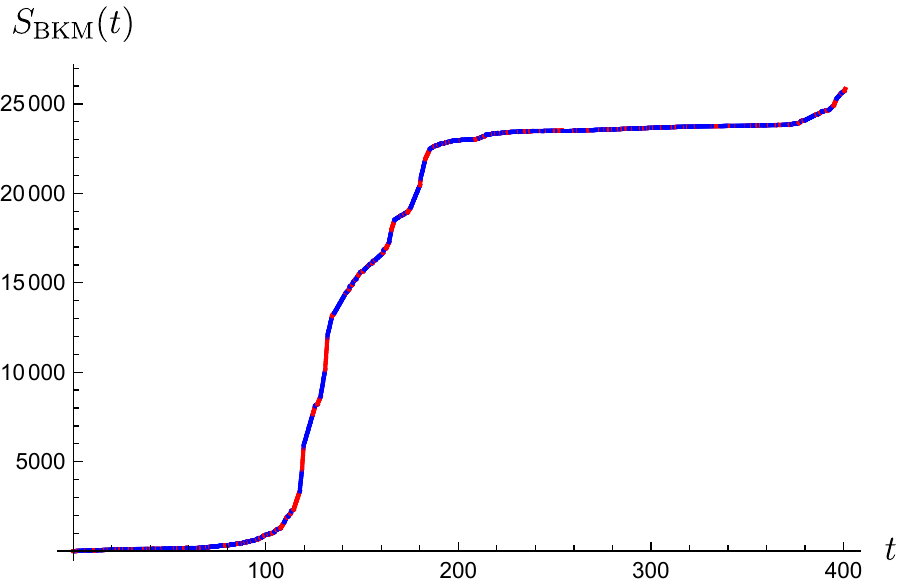}}{BKM action} &
		\subf{\includegraphics[width=7cm]{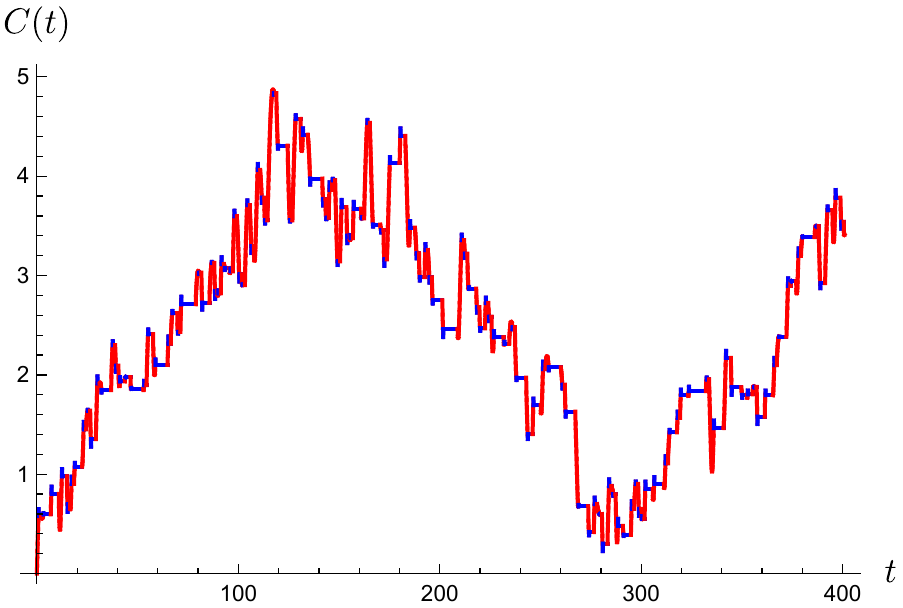}}{Complexity} 
	\end{tabular}
	
\end{center}
\caption{Random driving for 500 steps.  The parameters used are $\la=0.7, T_1=0.4, T_2=1.2$.  }\label{Fig:random}
\end{figure}

\paragraph{Random driving.} We can also consider a random sequence $a_k$ corresponding to applying $H_1$ or $H_2$ randomly with equal probability. The results are given in Figure \ref{Fig:random}.

For aperiodic processes, the phase diagram is more complicated. They can be thought of as random walks on the hyperbolic disk. In this case, we could ask whether the process will be in the heating phase, \ie will diverge towards the boundary, but this is difficult to answer in general, see for example \cite{mccarthy2019random}.

The analysis of ergodicity is more complicated. The process must be periodic if the subgroup of $\r{SL}(2,\R)$ generated by the steps is discrete and two-generator discrete subgroups of $\r{SL}(2,\R)$ have been classified  \cite{Rosenberger1986AllGP,Gilman1995TwogeneratorDS}.  This is related to Furstenberg's theorem \cite{Furstenberg1963NoncommutingRP} and was discussed in this context in \cite{Wen:2020wee}.
 
\subsubsection{Classical probability interpretation}
\label{subsec:probinterpretation}

We will now present connections of the above quantum Möbius states to the theory of classical probability.

\paragraph{Relation with statistical models.}
Recall from Section \ref{sec:ccop} that since $ f_z( x)$ defines a coordinate transformation on $S^1$, we have $\int_{0}^{2\pi}\frac{df_z}{2\pi}=\int_{0}^{2\pi}\frac{d x}{2\pi}\,f'_z( x)=1$ so that $f'_z( x)$ is a properly normalized probability density function on $S^1$ with respect to the measure $d x/(2\pi)$. Following the discussion in \cite{khesin_geometry_2011}, the family $\mathcal{S}=\{f'_z\}$ of probability distributions parametrized by $z\in\mathbb{D}$ is a two-dimensional 
{\em statistical model}. We already discussed that in the Cardy limit the BKM metric in the Virasoro orbit reduces to
a classical information metric, the spherical Hellinger distance on $\text{Dens}\,S^1$. As \cite{khesin_geometry_2011} point out, it further descends to the Fisher-Rao metric on a finite-dimensional statistical model. Note that the Cardy limit only affects the overall scale factor of the hyperbolic metric \eqref{hyperbolicplane}. This suggests that even without the limit it is the Fisher--Rao metric in the statistical model $\mathcal{S}$, defining a distance among classical probability distributions $f'_z$. In fact, up to the overall scale factor this can be verified by computing the components 
\be
G^{\r{FR}}_{ij}(s) = \int_{0}^{2\pi}\frac{d x}{2\pi}~p_{s}( x) \left(\frac{\partial \log p_s( x)}{\partial s_i}\right)\left(\frac{\partial \log p_s( x)}{\partial s_j}\right)
\ee
of the two-parameter probabilility distribution $p_{(s_1,s_2)}( x)\equiv f'_{(\rho ,\chi)}( x)$ arising from the diffeomorphism \eqref{diffMobius}, in agreement with the hyperbolic metric \eqref{hyperbolicplane}.

\paragraph{Continuous majorization.} We point out that the probability density functions associated with
the diffeomorphisms \eqref{diffMobius}
\be\label{pmobius}
p_z ( x) \equiv f'_z( x) = \frac{1}{2\pi}\frac {1}{\r{cosh}\,\rho - \r{cos}( x-\chi)\,\r{sinh}\,\rho}
\ee
satisfy a continuous majorization\footnote{Historical references for the concept of continuous majorization are \cite{HLPbook, Joe}.} order along the radial direction of the Poincaré disk\footnote{Here we adjusted the normalization such that $\int^{2\pi}_0d x~p_z( x) =1$.} Let us first quote
the definition, following the presentation in \cite{HJC}. Let $p,q$ be two probability distributions
defined on the same domain $\Omega\subset \mathbb{R}^n$. We say that $p$ majorizes $q$, and denote $p\succ q$, if 
\be
\int_{||x||\leq s}d^nx~p^{\downarrow}(x) \geq \int_{||x||\leq s}d^nx~q^{\downarrow}(x)
\ee
for all $s\geq 0$ with an equality as $s\rightarrow \infty$, where $p^\downarrow ,q^\downarrow$ are decreasing arrangements (see \cite{HJC}) of
the functions $p,q$. 

Now consider the distribution \eqref{pmobius}. It is symmetric about $x=\chi$ and monotonically decreasing
as a function of $|x-\chi|\in [0,\pi]$. Let us shift the origin of $x$ to $\chi$ ({\em i.e.,} set $\chi=0$) and consider the integral
\bea
I_s(\rho)\equiv\int_{|x|\leq s} dx~p^\downarrow_\rho (x) = 2\int^s_0 dx~p_\rho (x)
 =\frac{2}{\pi} \r{arctan}\le( e^{\rho} \,\r{tan}(s/2)\ri)~.
\eea
Since $\partial I_s(\rho) /\partial \rho >0$, we have continuous majorization order in the radial
direction of the Poincaré disk:
\be\label{majorization} 
 p_{\rho_1} \succ p_{\rho_2} \ \ \text{when}\ \rho_1 \geq \rho_2~.
\ee
At the origin we have the ``least ordered'' flat distribution $p_0 (x)=1/(2\pi)$, whereas
moving out radially the distribution deforms towards a delta function which is "most ordered". All distributions related by a rotation $x\to x-\chi$ are equivalent. 
A consequence of the majorization order is that every Schur convex (concave) functional of the 
distributions increases (decreases) in the radial direction. For example, the Schur concave Shannon
entropy $S[p_\rho] = -\int_{0}^{2\pi} dx~p_\rho (x)\log p_\rho (x)$ decreases as $\rho$ increases. 
We also note that the second term in \eqref{energydensity} can be written as $(2\pi p_{\rho_t})^2\le( \ln T\rn_\b +{c\over 48\pi}\ri)$, thus the conformal Floquet cooling in the heating phase leading to the 
hot spot and cooling region is connected with continuous majorization in time-evolution as the probability
distribution $p_{\rho_t}$ deforms towards a delta function. 

Finally, it is important to note that the majorization order is only a property of the classical probablility densities
associated with the diffemorphisms. The Möbius
states are related to the thermal state by the unitary conjugation \eqref{mobiusunitary} and
thus have identical spectra, and thus as quantum states do not have any non-trivial majorization.

\section{Holographic dual of Virasoro processes}\label{sec:holographicdual}

The AdS/CFT duality \cite{Maldacena:1997re,Aharony:1999ti} implies that a special class of two-dimensional CFTs  are expected to have a gravitational dual with AdS$_3$ asymptotics. The thermal state of a holographic CFT$_2$ is dual to the BTZ black hole \cite{Banados:1992wn} and Virasoro states correspond to pure gravity excitations of the geometry \cite{banados_three-dimensional_1999}.

In a holographic CFT, a Type I Virasoro process is dual to an asymptotically AdS$_3$ spacetime with a time-dependent boundary metric, as discussed in  \cite{MacCormack:2018rwq,Flory:2020dja,Erdmenger:2021wzc, Caputa:2020mgb,Das:2022pez}. The existence of a gravity dual gives additional tools to study the system as, for example,  the Ryu-Takayanagi formula for entanglement entropy \cite{Ryu:2006bv,Ryu:2006ef} which was applied in a related context in \cite{Liska:2022vrd, Caputa:2022zsr}. As we will see, the dual spacetime of a Virasoro process can be obtained by applying an appropriate diffeomorphism to the BTZ black hole and can be realized as a quotient of AdS$_3$. The spacetime dual can then be interpreted as  a BTZ black hole with an evolving horizon.

\subsection{Three-dimensional bulk metric}

In Section \ref{sec:hamiltonianderivation}, we showed that Virasoro trajectories $ \sigma_t = \sigma_{f_t,\overbar{f}_t} $ on the space of Virasoro states are in one-to-one correspondence with conformal classes of two-dimensional Lorentzian metrics $ g $ on the Lorentzian cylinder. Such a conformal class is obtained as the boundary limit of a three-dimensional asymptotically locally AdS$ _3 $ metric. Hence to every Virasoro trajectory there is a three-dimensional solution of Einstein's equations whose form is determined by the expectation value of the stress tensor operator $ \tr{(\sigma_t\,T_{ab})} $ and the initial state. This gives a holographic realization of the Type I Virasoro process. We will now show how to do this holographic mapping in detail, clarifying and extending results of \cite{erdmenger_complexity_2022}.

A Virasoro trajectory corresponds to the two-dimensional metric
\begin{equation}
ds^2 = g_{ab}\,dx^{a}dx^{b} = e^{\omega}(d\phi+\nu\,dt)(d\phi+\overbar{\nu}\,dt)
\label{holobdymetric}
\end{equation}
where $ \omega $ is arbitrary and $\nu,\overbar{\nu}$ are fixed by the trajectory. The three-dimensional asymptotically locally AdS$ _3 $ bulk metric is given by
\begin{equation}
ds^{2} = \frac{dr^{2}}{r^{2}} + \gamma_{ab}\,dx^{a}dx^{b}
\label{generalbulkmetric}
\end{equation}
such that \eqref{holobdymetric} lives on the conformal boundary:
\begin{equation}
	g_{ab}(x) = \lim_{r\rightarrow \infty} r^{-2}\,\gamma_{ab}(r,x).
\end{equation}
The explicit bulk metric can be written in the Fefferham--Graham expansion
\begin{equation}
\gamma_{ab}(r,x) = r^{2} g_{ab}(x)+ g_{ab}^{(2)}(x)+\frac{1}{r^{2}}\, g_{ab}^{(4)}(x),
\end{equation}
which truncates at finite order three dimensions. The fact that the metric is a solution of Einstein's equations gives the relations
\be
\n^a T_{ab}^\r{hol} = 0,\qq g^{ab}\, T_{ab}^\r{hol} ={c\/24\pi}R,\qq g_{ab}^{(4)} = {1\ov 4} g_{ab}^{(2)} g^{cd} g_{db}^{(2)},
\ee
where we have defined the holographic stress-tensor
\be
T_{ab}^\r{hol}\equiv {c\/12\pi} \le(g_{ab}^{(2)} +{1\/2}R g_{ab}\ri).
\ee
By the holographic dictionary, the holographic stress tensor is identified with the one-point function of the dual CFT in the boundary metric $g$ as
\begin{equation}
   T_{ab}^\r{hol} = \langle T_{ab}\rangle_g~.
   \label{holdic}
\end{equation}
Hence a locally AdS$ _3 $ bulk metric is determined by the input of the boundary metric $ g $ and of the one-point function $ \langle T_{ab}\rangle_g $,  determined in our case by the Virasoro trajectory $ \sigma_t = \sigma_{f_t,\overbar{f}_t} $. 

Using the form \eqref{CFTginitial} for the metric $g$, the full stress tensor one-point function can be written as a Liouville stress tensor (see Section \ref{sec:Polyakov})
\begin{equation}
\langle T_{ab}\rangle_g={c\/24\pi}\left[{1\/2}\,\p_a\varphi\,\p_b\varphi +\n_a\n_b\varphi + g_{ab} \le(\square\varphi - {1\/4} g^{ab}\, \p_a\varphi\, \p_b\varphi\ri)\right]~,
\label{liouvillerecall}
\end{equation}
where the Liouville field is
\begin{equation}
    \varphi =\hat\omega + \log{\mathcal{F}'_t(x^-)} + \log{\overbar{\mathcal{F}}'_t( x^{+})}
    \label{liouvillevarphirecall}
\end{equation}
and the diffeomorphisms are
\begin{equation}
    \mathcal{F}_t(x^-) = \tanh{\biggl(\sqrt{\frac{12\pi\langle T\rangle_{\beta}}{c}}\,f_t(x^-)\biggr)},\quad \overbar{\mathcal{F}}_t(x^+) = \tanh{\biggl(\sqrt{\frac{12\pi\langle T\rangle_{\smash[b]{\overbar{\beta}}}}{c}}\,\overbar{f}_t(x^+)\biggr)}.
\end{equation}
Assuming that the CFT is holographic, the stress tensor one-point function takes the universal Cardy value \eqref{thermal1pointvalues} also at finite values of the temperature \cite{hartman_universal_2014}
\begin{equation}
    \langle T\rangle_{\beta} = \frac{c\pi}{12\beta^2}, \quad \beta < 2\pi.
    \label{holographicTbeta}
\end{equation}
Hence assuming $\beta < 2\pi$ in our holographic setup, we have explicitly\footnote{In a holographic CFT, the condition $\beta < 2\pi$ amounts to being in the phase where the bulk geometry contains a BTZ black hole.}
\begin{equation}
    \mathcal{F}_t(x^-) = \tanh{\Bigl(\frac{\pi}{\beta}\,f_{t}(x^-)\Bigr)},\quad \overbar{\mathcal{F}}_t(x^+) = \tanh{\Bigl(\frac{\pi}{\overbar{\beta}}\,\overbar{f}_{t}(x^+)\Bigr)}~.
    \label{holographiccurlyF}
\end{equation}
On a slice of constant $t$, the state $ \sigma_t= \sigma_{f_t,\overbar{f}_t} $ is the Virasoro state associated with $(f_t,\bar{f}_t)$ and the stress tensor expectation value $ \tr{(\sigma_t\,T_{ab})} $ is
\be
\tr{(\sigma_t\,T_{--})}  = \frac{c}{24\pi}\,\{\mathcal{F}_t(x^-),x^-\},\quad \tr{(\sigma_t\,T_{++})}  = \frac{c}{24\pi}\,\{\overbar{\mathcal{F}}_t(x^+),x^+\}
    \ee
where the Schwarzian derivatives do not act on the $t$ label. The stress tensor one-point function is
\be
\ln T_{ab} \rn_g = \tr(\s_t T_{ab})+\text{counterterms},
\ee
where as explained in Section \ref{sec:Polyakov}, the counterterms are necessary to make the stress tensor conserved with the correct anomalous trace.

In holography, a Virasoro state is dual to a Bañados geometry:  a configuration of soft hair on a BTZ black hole. In the process, we evolve along a trajectory in the space of Virasoro states so we have an evolving configuration of soft hair. Equivalently, this can be viewed as the BTZ geometry with an evolving horizon from the action of large diffeomorphisms.

\subsection{Evolution of the horizon in the slow-driving limit}

In this section, we will consider the holographic dual of a closed CFT with unitary time evolution, a Type I Virasoro process, and we will determine the evolution of the position of the BTZ black hole horizon in the bulk. While one could in principle determine the horizon location exactly, we will for simplicity consider a quasi-static situation where the time evolution is slow and the background metric of the CFT at any given time is well approximated by a flat metric. This also implies that the bulk metric at each instant is well approximated by a particular BTZ geometry. One can then read off the horizon location with respect to the conformal boundary in a given time slice. This will be sufficient to connect the different driving regimes to the qualitative dynamics of black hole horizons. 

Geometrically, the slow-driving regime corresponds to having an approximately flat boundary metric \eqref{holobdymetric} which corresponds to
\be
\nu|_{t=t_\ast} = -1 + \mathcal{O}(\sde),\qq\overbar\nu|_{t=t_\ast}  = 1+ \mathcal{O}(\sde)~,
\ee
where we focus on the timeslice $t=t_\ast$. As in Section \ref{subsec:dissipation}, we can write $\sde = 1/t_\r{dr}$ where $t_\r{dr}$ is viewed as the variation timescale of the process.

In the slow-driving approximation around the slice $t=t_\ast$, the diffeomorphisms \eqref{holographiccurlyF} are approximately chiral diffeomorphisms
\be
\cF_t(x^-) = \cF_{t_\ast}(x^-)+\cO(\sde),\qq \overbar{\cF}_t(x^+) = \overbar{\cF}_{t_\ast}(x^+)+\cO(\sde)~.
\ee
As a result, the stress tensor one-point function reduces to its expectation value
\begin{equation}
    \langle T_{--}\rangle_g = \frac{c}{24\pi}\,\{\mathcal{F}_{t_\ast}(x^-),x^-\}+ \mathcal{O}(\sde),\quad \langle T_{++}\rangle_g = \frac{c}{24\pi}\,\{\overbar{\mathcal{F}}_{t_\ast}(x^+),x^+\}+ \mathcal{O}(\sde)
    ,\quad \ln T_{+-}\rn_g = \cO(\sde).
    \label{slowdrivingT}
\end{equation}
In particular the counterterms needed to make the stress tensor conserved and with the correct trace all disappear in the slow-driving limit.

The bulk metric around the slice $t=t_\ast$ is then the Bañados metric \cite{banados_three-dimensional_1999}
\be
ds^2 = {dr^2\/r^2} + r^2 dx^+ dx^- + {12\pi\ov c}\,\ln T_{ab}\rn_g\, dx^a dx^b +  {(12\pi)^2 \/c^2 r^2}\ln T_{++}\rn_g\ln T_{--}\rn_g\, dx^+ dx^- + \mathcal{O}(\e)
\label{banadosmetric}
\ee
To find the location of the horizon, we use the fact that the Bañados metric can be written as a quotient of the Poincaré AdS$_3$ metric 
\begin{equation}
ds^{2} = {-d T^2 + dX^2 + dY^2 \over Y^2},\qq Y >0~.
\label{PoincareAdS3}
\end{equation}
In terms of $X^\pm = X\pm T$, the diffeomorphism can be written \cite{roberts_time_2012,anand_exact_2017}
\begin{eqnarray}
X^- \= \cF_{t_\ast}(x^-) - {2  \cF_{t_\ast}'(x^-)^2 \overbar{\cF}_{t_\ast}''(x^+)\ov 4 r^2  \cF_{t_\ast}'(x^-)\overbar{\cF}'(x^+) + \cF_{t_\ast}''(x^-) \overbar{\cF}_{t_\ast}''(x^+)}~,\\
X^+\=\overbar{\cF}_{t_\ast}(x^+) - {2  \overbar{\cF}_{t_\ast}'(x^-)^2 \cF_{t_\ast}''(x^-)\ov 4 r^2  \cF'(x^-)\overbar{\cF}_{t_\ast}'(x^+) +  \cF''(x^-) \overbar{\cF}_{t_\ast}''(x^+)}~,\\
Y \= {  4 r (\cF_{t_\ast}'(x^-) \overbar{\cF}_{t_\ast}'(x^+))^{3/2}\over   4r^2 \cF_{t_\ast}'(x^-) \overbar{\cF}_{t_\ast}'(x^+)+  \cF_{t_\ast}''(x^-) \overbar{\cF}_{t_\ast}''(x^+)}~.
\label{poincarediffeo}
\end{eqnarray}
Noting that the angle $\phi$ on the slice $t=t_\ast$ can be written in two ways $\phi = f_{t_\ast}(x^-) = \overbar{f}_{t_\ast}(x^+)$, and assuming further that $\beta = \overbar{\beta}$, we get
\begin{equation}\label{phitwoexpr}
    \mathcal{F}_{t_\ast}(x^-) = \overbar{\mathcal{F}}_{t_\ast}(x^+)~.
\end{equation}
As a result the slice  $t=t_\ast$ is mapped to the slice $T=0$ in Poincaré AdS$_3$. The Bañados horizon is then a portion of the Rindler horizon which is the semi-circle of equation
\begin{equation}
	X^2+ Y^2 = 1, \quad T = 0~.
	\label{rindlerhorizon}
\end{equation}
Using the mapping \eqref{poincarediffeo}, we obtain the horizon location in $x^\pm$ coordinates  in the slow-driving limit. On the slice $t=t_\ast$, this is the curve parametrized by
\be\label{horizoncurve}
r(x^-)  = {\pi f_{t_\ast}'(x^-)\/\b} \sqrt{1-{\b^2\/4\pi^2} {f_{t_\ast}''(x^-)^2\/f_{t_\ast}'(x^-)^4}} ~,
\ee
where $x^- = F_{t_\ast}(\phi)$ with $\phi\in [0,2\pi)$ can be used to parametrize the circle.

The bulk metric is written using $x^\pm = x\pm \tau$ coordinates so to obtain a simple representation, we further demand that the constant $t=t_\ast$ slice corresponds to a constant $\tau=\tau_\ast$ slice.  The normal covector of the $t=t_\ast$ slice is given by
\be
dt = {dx^+\/ 2\overbar{F}_{t_\ast}'(\phi)}- {dx^-\/ 2 F_{t_\ast}'(\phi)} + \mathcal{O}(\e)~
\ee
using that in the slow-driving limit we have
\be
dx^- = F_{t}'(\phi)\,(d\phi+\nu dt) = F_{t}'(\phi)\, d\phi^- + \mathcal{O}(\e),\quad dx^+ = \overbar{F}_{t}'(\phi)\,(d\phi+\overbar\nu dt) = \overbar{F}_{t}'(\phi)\, d\phi^+ + \mathcal{O}(\e)~.
\ee
For the slice $t=t_\ast$ to be mapped to a slice $\tau=\tau_\ast$, we must then impose
\be\label{sameFprimes}
F_{t_\ast}'(\phi) = \overbar{F}_{t_\ast}'(\phi)
\ee
as this implies that the normals of the $t=t_\ast$ and $\tau=\tau_\ast$ slices are parallel $dt = d\tau/F_{t_\ast}'(\phi)$. The relation \eqref{sameFprimes} also implies that $F_{t_\ast}'(\phi)$ and $\overbar{F}_{t_\ast}'(\phi)$ differ only by a $\phi$-independent constant of integration which is identified with $\tau_*$:
\begin{equation}
    \tau_\ast = \frac{1}{2}(\overbar{F}_{t_\ast}-F_{t_\ast}).
\end{equation}
It follows that on the $t= t_*$ slice, we have $x^- = x-\tau_\ast$ is just a constant shift of $x$. As a result the horizon \eqref{horizoncurve} is now a curve on the $\tau = \tau_*$ slice of the Bañados spacetime \eqref{banadosmetric} whose equation is
\be
r(x)  = {\pi f_{t_\ast}'(x)\/\b} \sqrt{1-{\b^2\/4\pi^2} {f_{t_\ast}''(x)^2\/f_{t_\ast}'(x)^4}}
\ee
where $x\in[0,2\pi)$ parametrizes the circle. If there is a range of values of $t_\ast$ where the slow-driving approximation is valid and \eqref{sameFprimes} is satisfied, the constant $t$ slices in that range are also contant-$\tau$ slices and the dynamical horizon for each $t$ is the curve
\be
r_t(x)  = {\pi f_{t}'(x)\/\b} \sqrt{1-{\b^2\/4\pi^2} {f_{t}''(x)^2\/f_{t}'(x)^4}}~,\qq x\in[0,2\pi)~.
\label{horizonfinal}
\ee
Note that the slow-driving regime can always be achieved with a time reparametrization of the process. Given a Virasoro trajectory $(f_t,\overbar{f}_t)$, we can define a new trajectory  $(h_t,\overbar{h}_t)$ satisfying
\be
h_t' =f_{\e t}' ,\qq \overbar{h}_t' = \overbar{f}_{\e t}'~.
\ee
If the original trajectory has bounded time derivatives, the new trajectory is in the slow-driving regime for sufficiently small $\sde$.

\subsection{Heating phase in holography}

We can obtain the horizon location of Möbius processes by plugging in the diffeomorphism 
\be\label{ftxdiffeo}
f_{t}(x) = 2 \,\r{arctan}\le( e^{\rho_t} \,\r{tan}(\tfrac12(x-\chi_t))\ri)~,
\ee
We will assume that there is a range of values of $t_\ast$ for which the slow-driving approximation is valid. The position of the horizon is then obtained by substituting \eqref{ftxdiffeo} to \eqref{horizonfinal} which gives
\be\label{mobiushorizon}
r_t(x)= {\pi\/\b} {\sqrt{ 1-{\b^2\/4\pi^2}\r{sin}^2(x-\chi_{t})\,\r{sinh}^2\rho_{t}}\/\r{cosh}\,\rho_{t} -\r{cos}(x-\chi_{t})\,\r{sinh}\,\rho_{t}}~.
\ee
The horizon has an elliptical shape and there are two extremal points corresponding to $x=\chi_t$ and $\phi=\chi_t+\pi$ corresponding to the radial position
\be\label{rminmax}
\max_{x\in S^1} r_t(x) = {\pi\/\b}\, e^{\rho_{t}},\qq \min_{x\in S^1} r_t(x) = {\pi\/\b}\, e^{-\rho_{t}}~.
\ee
We see that the point $(\rho_t,\chi_t)$ in the BKM geometry gives essentially the position of the black hole on the Cauchy slice of the dual spacetime (whose geometry is also the hyperbolic disk). The process in the BKM geometry can then be viewed, at a qualitative level, as the trajectory of the black hole in spacetime.

We can measure the radial distance between the horizon and the boundary
\be
d_\text{hor}(t)  = \min_{x\in S^1} \int_{r_t(x)}^{r_\r{UV}} {dr\/r}
\ee
where $r_\r{UV}$ is a cutoff and we minimize the radial distance over the circle. For a Möbius process this is 
\be
d_\text{hor}(t)  = d_\text{hor}(0)- \rho_t~
\ee
where $(\rho_t,\chi_t)$ are the hyperbolic coordinates the system. In the heating phase, we have $\rho_t \sim \la_L t$. The holographic dual statement is that the black hole horizon comes closer to the boundary at a linear rate, given by the Lyapunov exponent.

It is interesting to note that $\rho_t$ is also the BKM complexity $C_\r{BKM}(t)$, the circuit complexity for a unitary mapping the thermal state to $\s_t$. As a result we see that the distance to the horizon $d_\text{hor}$ is equal to the ``uncomplexity'' \cite{Susskind:2018pmk}.

\begin{figure}

\begin{center}
	\begin{tabular}{cc}
		\subf{\includegraphics[height=7cm]{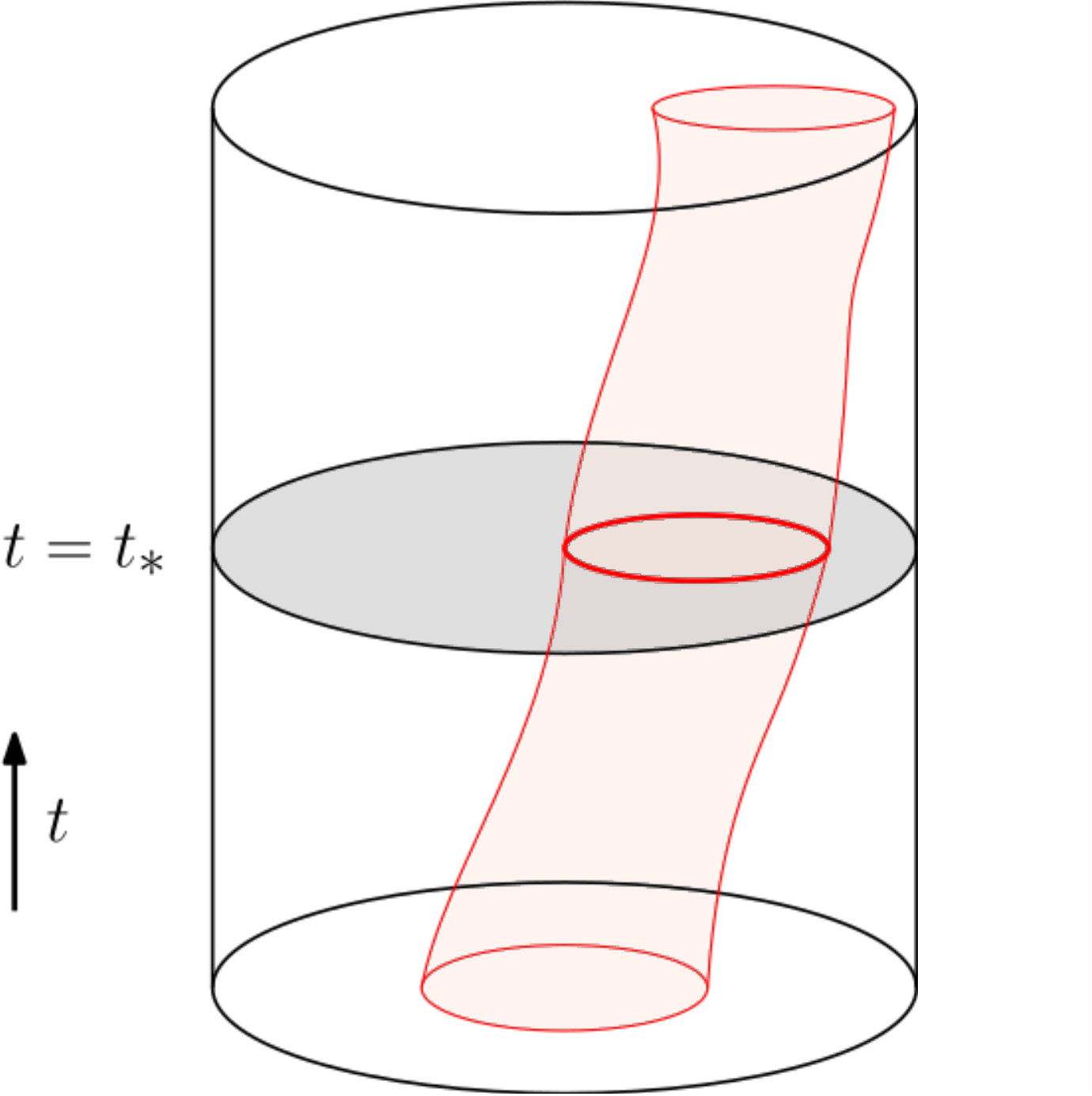}}{Spacetime dual to a Virasoro process} &
		\subf{\includegraphics[height=7cm]{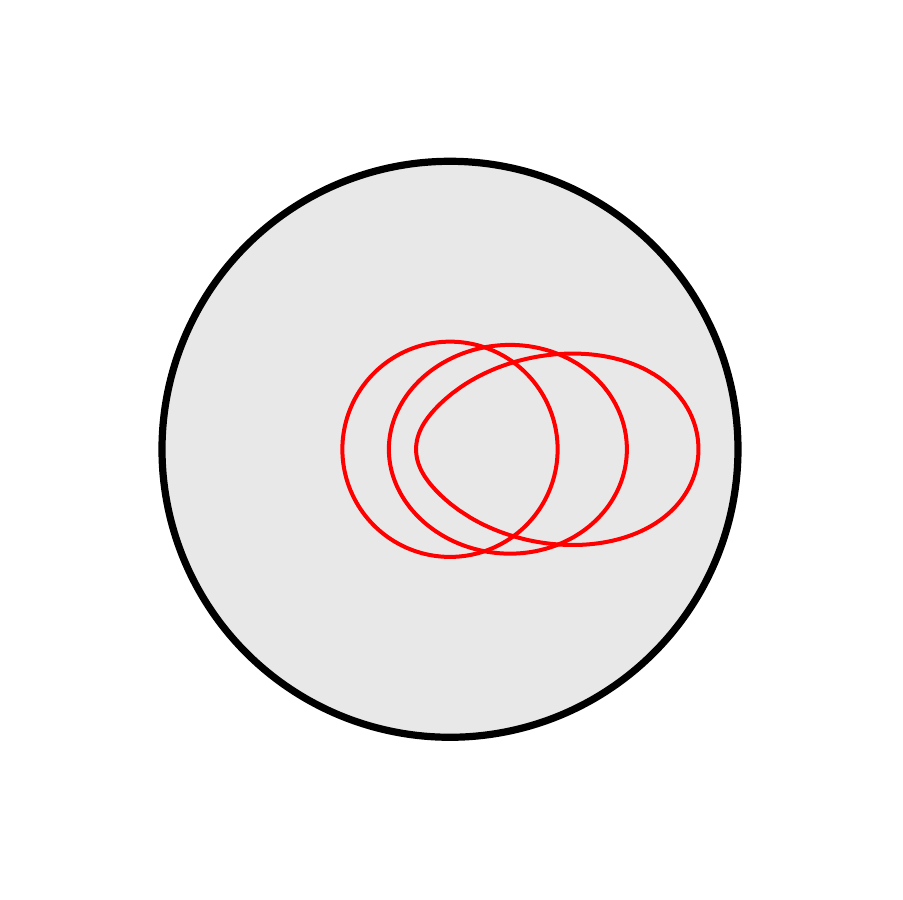}}{Horizon shapes on $t=\r{const}$ slices.} \\	\end{tabular}
	
\end{center}
\caption{The spacetime dual to the Virasoro process can be viewed, in the slow-driving regime, as a succession of Bañados geometries. The shape of the black hole horizon represented in red reflects the dynamics of the process. In the heating phase, the horizon approaches a point of the boundary.}\label{Fig:processSpacetime}
\end{figure}

A notion of heating phase can be defined for general Virasoro processes. A general diffeomorphism $f_{t}(x)$ corresponds to a probability distribution $ f_{t}'(x)$ on the circle. Indeed it is positive and satisfies 
\be
{1\/2\pi}\int_0^{2\pi} dx\,f_t'(x)  =1~.
\ee
The radius $r_t(x)$ has a maximum at $x_\ast$. This corresponds to a maximum of $f_{t}'(x)$ so we have $f_{t}''(x_\ast) = 0$. Hence the value of the radius there is
\be
r_t(x_\ast) = {\pi\/\b}\, f_{t}'(x_\ast)~.
\ee
The heating phase corresponds to $r_t(x_\ast) \to\infty $ so that the black hole horizon approaches the boundary point $x_\ast$. The heating phase then corresponds to 
\be
\max_{x\in S^1}{f_t'(x)} \ra \infty~.
\ee
In the case where $f_t'(x)$ has a single maxima at $x_*$ going to infinity, we have
\be
f_t'(x) = 2\pi \d_t(x-x_*)
\ee
where $\d_t(x)$ is a smooth approximation of the delta function on the circle so that
\begin{equation}
    \lim_{t\to\infty}\d_t(x)=\d_{S^1}(x)\equiv \sum_{n=-\infty}^{\infty}e^{inx}.
\end{equation}
The Möbius processes achieve this with the expression
\be
f_t'(x) = {1\/\r{cosh}\,\rho_t-\r{cos}(x-x_*)\,\r{sinh}\,\rho_t}~.
\ee
where $\lim_{t\to\infty } \rho_t =\infty $. This discussion mirrors the radial continuous majorization of the probability distributions $p_t(x)=f'_t(x)/(2\pi)$ \eqref{pmobius} in Section \ref{subsec:probinterpretation}. The holographic picture is then that the black hole horizon  moves  toward the point $x=x_*$ of the conformal boundary as depicted in Figure \ref{Fig:processSpacetime}. We could also imagine more complicated limits where $f_t'$ has multiple maxima going to infinity. In this case, the holographic picture involves a black hole horizon with a more complicated shape, approaching the boundary at various points.

The point of the circle to which the black hole tends becomes a heating point. Its antipode on the circle is a cooling point. We can define an effective temperature $T_\r{eff}=1/\b_\r{eff}$ using the definition of the Virasoro state 
\be
\s_{t} \propto \r{exp}\le[ -\int_0^{2\pi} dx\,\b_\r{eff}(x) \,T(x) \ri] ~,
\ee
so we have $\b_\r{eff}(x)= \b/f_t'(x)$ and the effective temperature is
\be
T_\r{eff}(x) = T f_{t}'(x) 
\ee
where $T=1/\b$. In the heating phase, we see that there is a point on the circle at which the temperature becomes infinite. Note that the average temperature is conserved
\be
{1\/2\pi}\int_0^{2\pi}T_\r{eff}(x) = T~.
\ee
So if a point becomes infinitely hot, the rest of the circle must cool down. This gives a holographic picture for the Floquet refrigerator \cite{Wen:2022pyj}. The emergence of a hot spot and its holographic interpretation with a deforming black hole horizon was observed in the context of time evolution after a SSD quench in  \cite{Goto:2021sqx}.

\section{Outlook and discussion}\label{sec:discussion}

In this paper we have developed spacetime geometry and quantum information geometry of driven two-dimensional CFTs in the universal sector of Virasoro states.  The BKM information geometry of Virasoro states gives new insight on complexity and ergodicity of driven CFTs. A particular truncation to the $\r{SL}(2,\R)$ subgroup illuminates the rich dynamics of Möbius processes realized in Floquet CFTs. In this section we discuss some possible extensions for future work. It should be relatively straightforward to generalize our results to other $\r{SL}(2,\R)$ subgroups generated by $L_n$, $L_{-n}$ and $L_0+c\,(n^2-1)/24$ (see e.g. \cite{Liska:2022vrd}) but we will leave this to future work.

\paragraph{Ergodic theory and Möbius driving.} As we have seen, the dynamics of a thermal CFT under Möbius driving reduces to measure-preserving dynamics of the two-dimensional hyperbolic disk, the BKM geometry of Möbius states. This makes contact with a much studied branch of ergodicity -- dynamics in hyperbolic spaces (see for example \cite{McMullenLectures} and references therein). In this work we have investigated only some particular models with some first observations about ergodic and non-ergodic regimes
and potential measures of interest. We anticipate that there is much more to explore, with more general driving protocols 
and more details. For example, we did not investigate whether in the ergodic regime the dynamics may also be mixing. There are also many contexts leading to consider various driven CFTs, where to apply and possibly generalize our approach. For example, driven Tomonaga--Luttinger liquids motivate a  compact boson CFT with driving through alternating the compactification radius \cite{Datta:2022fwd}. See also other models discussed therein.

\paragraph{Modeling the dynamics of open systems.} As one application of BKM geometry, we considered work dissipation in slowly driven open CFTs in contact with a heat bath. The driving resulted from a classical background metric which produced a time-dependent Hamiltonian. The treatment of dissipation under slow driving was based on the general framework developed in \cite{scandi_thermodynamic_2019,abiuso_geometric_2020},  applicable to any open quantum system whose Hamiltonian is time-dependent due to explicit time-dependence in its parameters (arising from driving by classical time-dependent background fields). A simple example is a two-level system coupled to a harmonic potential where the width of the potential well can be varied.

In this setting, the classical background field injects energy into the system since the Hamiltonian eigenvalues depend on time. This leads to changes in the energy level spacing that affects the internal energy. A simplification in these models is that they hide the detailed origin of this energy by taking a limit of the master equation. There is a clear classical to quantum boundary where the system under consideration is quantum while the background fields are treated classically. Hence the internal energy can grow without bounds (which is what happens in the heating phase for example), but in practice, such growth cannot continue forever. The limit of the master equation also hides the details of how the environment exchanges heat $Q$ with the system and where the dissipated work is absorbed.

A complete description involves a composite system-environment Hilbert space with a unitary composite Hamiltonian time evolution, from which open-system dynamics is obtained by tracing out the environment. In the case of finite-dimensional systems, there is extensive literature on such models. It would be interesting  to develop analogous 
detailed models of driven CFTs interacting with an environment. It would also be interesting to develop a holographic understanding of such open quantum systems.
Such a holographic description will presumably require matter fields, as without matter fields we can only consider diffeomorphisms in AdS which necessarily leave {\em e.g.} the entropy invariant. To include matter fields and their backreaction will introduce many new conceptual and technical complications beyond the scope of the present paper. It may also be possible to model open systems in a more universal way incorporating changes in energies and entropy using singular diffeomorphisms representing shockwaves. Diffeomorphisms of the real line which change the temperature of the thermal states appear to be of this form when we embed the Poincaré patch in global AdS. It would be interesting to explore this possibility in more detail.

\paragraph{Physical interpretation of geodesics.} The geodesics in the information
geometry give the shortest path from an initial state to a target space. From a geodesic
one can compute a particular driving protocol, which minimizes dissipated work in a slowly driven open CFT. For unitary driving it suggests an optimal way
to synthesize the target state (an optimal quantum circuit minimizing a thermodynamically motivated notion of circuit complexity). 

\paragraph{Higher work moments.} In this paper, we only considered the expected amount of work $ W $, the integral of the expectation value $ \tr{(\sigma_t\dot{H}_t)} $ where $H_t$ is the driving Hamiltonian. As a further step, one could consider integrals of higher moments of $ \dot{H}_t $, in particular the variance of performed work
\begin{equation}
\text{Var}\,W \equiv \int_{0}^{t_{\hspace{0.5pt}\text{f}}} dt\,\bigl[\tr{(\sigma_t\dot{H}_t^{2})} - (\tr{(\sigma_t\dot{H}_t)})^{2}\bigr]
\label{cumulativevar}
\end{equation}
measuring the expected amount of cumulative work fluctuations during the driving protocol. In general, one can define full work statistics by the so called two-projective-measurement (TPM) scheme, where work is measured by two projective measurements before and after the driving protocol \cite{tasaki_jarzynski_2000, kurchan_quantum_2001, talkner_fluctuation_2007, talkner_fluctuation_2009}. In this way one can define a moment generating function, and take derivatives to compute moments of the work distribution. However, only in discrete processes these moments agree with \eqref{cumulativevar} (see Appendix of \cite{miller_work_2019}).

Quantum information geometry can also be used to optimize work fluctuations \cite{miller_work_2019}: there exists a metric whose geodesics minimize \eqref{cumulativevar}. In fact, the integrand of \eqref{cumulativevar} is related to the variance of relative entropy \cite{de_boer_quantum_2021} 
\begin{equation}
V(\rho\lVert \sigma) = \tr{[\rho\,(\log{\rho}-\log{\sigma})^{2}]} - S(\rho\lVert \sigma)^{2},
\end{equation}
when the process is a step-equilibration process \cite{scandi_quantum_2020}. We define the variance metric as (which is well defined due to $ V(\rho\lVert \sigma)\geq 0 $)
\begin{equation}
\mathcal{V}_{\sigma}(\delta \sigma,\delta \rho) \equiv -\frac{\partial^{2}}{\partial \lambda_1\partial \lambda_2}V(\sigma +\lambda_1 \delta \sigma\lVert \sigma +\lambda_2\delta \rho)\bigg\lvert_{\lambda_1=\lambda_2 = 0},
\end{equation}
and by using results of Section \ref{subsec:dissipation}, we observe that for a step-equilibration process (see also \cite{scandi_quantum_2020})
\begin{equation}
\text{Var}\,W = \frac{1}{2}\int_{0}^{t_{\hspace{0.5pt}\text{f}}} dt\,\mathcal{V}_{\pi_t}(\dot{\pi}_t,\dot{\pi}_t) + \mathcal{O}(\lambda^{2}).
\label{varW}
\end{equation}
As a result, \eqref{varW} has its minimum value along geodesics of $ \mathcal{V}_{\sigma}(\delta \rho,\delta \sigma) $. As with the relative entropy variance here, one could use other information quantities to define many different information metrics on the BKM geometry.  We leave a more detailed analysis of higher moments to future work.

\paragraph{Other information metrics.}

In this work, we focused on the BKM quantum information metric which is a metric on mixed states and which is given by the Hessian of the relative entropy. More generally, the Hessian of any information divergence produces a metric on mixed states whose form is fixed by the perturbative expansion of the divergence. The variance metric presented in the previous paragraph is an example of this (see \cite{petz_riemannian_1996} for more examples). If the information divergence is unitary invariant, the corresponding metric should be right-invariant when restricted to Virasoro states like the BKM metric. For metrics restricted to Möbius states that can be identified with points of the unit disk $ \mathbb{D} $, this would translate to right-invariance under the action of $ \text{SL}(2,\mathbb{R}) $. Because the hyperbolic metric is the unique $ \text{SL}(2,\mathbb{R}) $-invariant metric on $ \mathbb{D} $ (up to a constant), all these metrics are expected to be hyperbolic, but have different curvature radii. This might have implications for Möbius driving. We leave a more detailed study of these metrics to future work.

Other information metrics can also be defined on the space of pure states, instead of mixed states, such as the Fubini--Study metric \cite{Erdmenger:2021wzc,erdmenger_complexity_2022}. In this case, the relevant state space is a Virasoro orbit of some initial pure state (such as the vacuum state). The space of pure states cannot be endowed with the BKM metric directly, because the relative entropy is ill-defined for states that are not full rank. However, it might still be possible to define a BKM metric on the vacuum orbit by taking a double scaling limit, where the temperature of the thermal state is taken to zero $ \beta \rightarrow \infty $ while simultaneously scaling the metric appropriately. It would be interesting to investigate the Fubini--Study geometry of Möbius states in Möbius driven CFTs and its relationship to the limit of the BKM geometry.

\paragraph{Higher dimensions.} The results of this paper can be generalized by considering different groups beyond $\r{SL}(2,\R)$ or Virasoro. For example, one could consider conformal processes where one drives a thermal state in a higher-dimensional CFT using their conformal groups. For holographic CFTs, this would correspond to driving the AdS-Schwarzschild black hole using a time-dependent conformal generator. It would be interesting to study the dynamics of such processes and their holographic dual description.

\bigskip

\noindent
{\large \bf Acknowledgments}

\medskip

\noindent  We thank Johanna Erdmenger, Marius Gerbershagen, Antti Kupiainen, Diego Liska, R. Loganayagam and Anna-Lena Weigel for useful discussions during the completion of this work. JdB is supported by the European Research Council under the European Unions Seventh Framework Programme (FP7/2007-2013), ERC Grant agreement ADG 834878. VG acknowledges the postdoctoral program at ICTS for funding support through the Department of Atomic Energy, Government of India, under project no. RTI4001. JK is supported by the Osk. Huttunen foundation and by the Magnus Ehrnrooth foundation. EKV's research has been conducted within the framework of InstituteQ - the Finnish Quantum Institute.

\addtocontents{toc}{\protect\setcounter{tocdepth}{1}}

\appendix 

\section{Virasoro trajectories and diffeomorphisms of the circle}\label{app:virasoro}

In this Appendix, we review projective representations of $\text{Diff}_+S^1$ which are representations of the Virasoro group: the central extension of the universal covering group $ \widetilde{\text{Diff}}_+S^{1} $. For other references on the topic with similar content, see \cite{fewster_quantum_2005,oblak_bms_2017,moosavi_inhomogeneous_2019}.

\subsection{Unitary representations of the Virasoro group}

The group of orientation-preserving diffeomorphisms of the circle is defined as
\begin{equation}
\text{Diff}_+S^{1} = \{h\colon S^{1} \rightarrow S^{1} \,\lvert\,h\text{ is smooth and orientation-preserving}\}
\end{equation}
whose group operations are composition and inversion of diffeomorphisms. The group manifold $ \text{Diff}_+S^{1} $ is not simply connected and hence globally different\footnote{The covering is infinite-to-one, because the fundamental group of $ \text{Diff}_+S^{1} $ is $ \mathbb{Z} $.} from its universal cover $ \widetilde{\text{Diff}}_+S^{1} $ which is given by
\begin{equation}
\widetilde{\text{Diff}}_+S^{1} =\{f\colon \mathbb{R}\rightarrow \mathbb{R} \,\lvert\,f( x+2\pi) = f( x) +2\pi,\, f'( x)>0,f\text{ is smooth}\}.
\label{universalcoverapp}
\end{equation}
The tangent space $ T_{f_t}\text{Diff}_+S^{1} $ of $ \widetilde{\text{Diff}}_+S^{1} $ at the point $ f_t $ is spanned by tangent vectors $ u_t $ obtained as derivatives
\begin{equation}
u_t( x) = \partial_t f_t( x)
\label{ut}
\end{equation}
with $ u_t( x+2\pi) = u_t( x) $. In this appendix, we assume that $ f_0 = \text{id} $ is the identity and tangent vectors at the identity are denoted as
\begin{equation}
u( x) \equiv \partial_t f_t( x)\lvert_{t=0}.
\label{uid}
\end{equation}
On the Hilbert space, a diffeomorphism $ f \in \widetilde{\text{Diff}}_+S^{1} $ is represented by unitary operators $ V_f $, and a tangent vector at the identity $ u \in T_{\text{id}}\text{Diff}_+S^{1} $ is represented by a Hermitian operator $ T_u $ which is defined as the derivative of $ V_{f_t} $. Given a trajectory $ f_t $ such that \eqref{uid} holds, we get
\begin{equation}
f_t \mapsto V_{f_t}, \quad u\mapsto T_u \equiv i\,\partial_t V_{f_t}\lvert_{t=0}.
\label{representation}
\end{equation}
The unitaries represent the group multiplication of $\widetilde{\text{Diff}}_+S^{1}$ projectively up to a phase (see \cite{oblak_bms_2017} for a review)
\begin{equation}
V_{f}V_{h} = e^{icB(f,h)}\,V_{f\circ h}
\label{composition}
\end{equation}
where the Thurston--Bott 2-cocycle
\begin{equation}
B(f,h) = \frac{1}{48\pi}\int_{0}^{2\pi}d x\,\frac{h''( x)}{h'( x)}\,\log{f'(h( x))}.
\label{2cocycle}
\end{equation}
The projective multiplication law \eqref{composition} is the group multiplication law of the \textit{Virasoro group} which is the central extension of $\widetilde{\text{Diff}}_+S^{1}$. Hence projective representations of $\widetilde{\text{Diff}}_+S^{1}$ correspond to ordinary representations of the Virasoro group.

In addition to the above representation, we can obtain an alternative one $\overbar{V}_f$ by taking the complex conjugate which satisfies
\begin{equation}
\overbar{V}_{f}\overbar{V}_{h} = e^{-icB(f,h)}\,\overbar{V}_{f\circ h}.
\end{equation}

\subsection{The Virasoro algebra and the stress tensor}

The Lie algebra $ T_{\text{id}}\text{Diff}_+S^{1} $ of $ \widetilde{\text{Diff}}_+S^{1} $ can be identified with the algebra $ \text{Vect}\,S^{1} $ of real valued vector fields on $ S^{1} $ via $ u = u( x)\partial_{ x} $. The Lie bracket of two vector fields is the standard one
\begin{equation}
[u,v] = (u( x)\,v'( x) - v( x)\,u'( x))\,\partial_{ x}
\label{vectorfieldcomm}
\end{equation}
which differs from the Lie bracket of $ T_{\text{id}}\text{Diff}_+S^{1} $ by a well-known minus sign \cite{modin_geometric_2019}. This sign will not play a role for us and we will write all expressions in terms of \eqref{vectorfieldcomm}.

The generators of $ \text{Vect}\,S^{1} $ are\footnote{More precisely, these are generators of the complexification $ \text{Vect}_{\mathbb{C}}\,S^{1} $.}
\begin{equation}
\ell_n( x) = e^{-in x}
\label{vectgenerators}
\end{equation}
and any vector field can be expanded in a Fourier series
\begin{equation}
u( x) = \sum_{n=-\infty}^{\infty}u_{n}\,\ell_n( x).
\end{equation}
For the generators, the bracket \eqref{vectorfieldcomm} is
\begin{equation}
[\ell_n,\ell_m] = i\,(n-m)\,\ell_{n+m}.
\label{vectgens}
\end{equation}
On the Hilbert space, a vector field is represented by the Hermitian operator $ T_u $ \eqref{representation} which by linearity has the expansion
\begin{equation}
 T_u = \sum_{n=-\infty}^{\infty}u_{n}L_n.
\label{Tu}
\end{equation}
where we have defined $ L_{n} \equiv T_{\ell_n} $ with $ L_n^{\dagger} = L_{-n} $. The stress tensor $ T( x) $ is defined as
\begin{equation}
T( x_1) \equiv T_{\delta_{2\pi}( x_1- x)\partial_x}
\label{stresstensordelta}
\end{equation}
where the $ 2\pi $-periodic delta function
\begin{equation}
\delta_{2\pi}( x) = \frac{1}{2\pi}\sum_{n=-\infty}^{\infty}e^{in x}.
\end{equation}
By linearity of the representation, it thus follows that
\begin{equation}
T( x) = \frac{1}{2\pi}\sum_{n=-\infty}^\infty L_n\,e^{i n  x}.
\end{equation}
and \eqref{Tu} can also be written as
\begin{equation}
T_u = \int_0^{2\pi}d x\,u( x)\,T( x).
\end{equation}
The operators $ T_u $ satisfy the central extension of the algebra $ \text{Vect}\,S^{1} $ which is the Virasoro algebra.\footnote{More precisely, the Virasoro algebra is the central extension of $ \text{Vect}_{\mathbb{C}}\,S^{1} $.} The Lie bracket between elements $ T_u $ can be derived using \eqref{representation} as follows.

Using \eqref{representation}, the adjoint action of a Virasoro unitary on an element of the Virasoro algebra is given by
\begin{equation}
V_h\,T_u\,V_h^{\dagger} \equiv i\,\partial_t( V_{h}V_{f_{t}}V_h^{\dagger})\big\lvert_{t=0}.
\end{equation}
By using the composition formula \eqref{composition} of Virasoro unitaries and taking derivatives of the 2-cocycle \eqref{2cocycle}, we get
\begin{equation}
V_h\,T_u\,V_h^{\dagger} = T_{\text{Ad}_hu} - \frac{c}{24\pi}\int_{0}^{2\pi}d x\,u( x)\,\{h( x), x\}
\label{adjointactionstresstensor}
\end{equation}
where the adjoint action of $ \widetilde{\text{Diff}}_+S^{1} $ on $ \text{Vect}\,S^{1} $ is given by
\begin{equation}
(\text{Ad}_hu)( x) = \partial_{t}(h\circ f_{t}\circ h^{-1})( x)\lvert_{t=0}\, = h'(h^{-1}( x))\,u(h^{-1}( x)), 
\label{Adfu}
\end{equation}
and the Schwarzian derivative is given by
\begin{equation}
\{h( x), x\} = \biggl(\frac{h''( x)}{h'( x)}\biggr)'-\frac{1}{2}\biggl(\frac{h''( x)}{h'( x)}\biggr)^{2}.
\end{equation}
Using \eqref{representation}, the commutator of two elements of the Virasoro algebra is obtained from the adjoint action as
\begin{equation}
[T_u,T_v] = -i\,\partial_t(V_{h_t}\,T_u\,V_{h_t}^{\dagger})\big\lvert_{t=0}.
\end{equation}
Denoting $ v( x) \equiv \partial_t h_t( x)\lvert_{t=0} $, we have
\begin{equation}
\partial_tT_{\text{Ad}_{h_t}u}\big\lvert_{t=0}\, = T_{[u,v]}, \quad \partial_t\{h_t( x), x\}\big\lvert_{t=0}\, = v'''( x),
\end{equation}
so that
\begin{equation}
[T_u,T_v] = -i\,T_{[u,v]} + \frac{ic}{24\pi}\int_{0}^{2\pi}d x\,u( x)\,v'''( x).
\label{TuTvcomm}
\end{equation}
For the generators \eqref{vectgenerators}, this amounts to the standard form of the Virasoro algebra
\begin{equation}
[L_n,L_m] = (n-m)\,L_{n+m} + \frac{c}{12}n^{3}\delta_{n,-m}
\label{appvirasoro}
\end{equation}
where we used \eqref{vectgens} in the first term. Similarly for the stress tensor \eqref{stresstensordelta}, we get
\begin{equation}
[T( x_1),T( x_2)] = -i\,\bigl(T( x_1)+T( x_2)\bigr)\,\delta'_{2\pi}( x_1- x_2)+\frac{ic}{24\pi}\delta'''_{2\pi}( x_1- x_2).
\end{equation}
The adjoint action \eqref{adjointactionstresstensor} on the stress tensor \eqref{stresstensordelta} amounts to the standard transformation law
\begin{equation}
V_{f}\,T( x)\,V_{f}^{\dagger} = f'( x)^{2}\,T(f( x)) -\frac{c}{24\pi}\{f( x), x\}.
\label{stresstransform}
\end{equation}

\paragraph{Complex conjugate representation.} The expressions for the complex conjugate represention of the Virasoro algebra is obtained by replacing $ i\rightarrow -i $ in the above formulae. The stress tensor of the conjugate representation is
\begin{equation}
    \overbar{T}( x) = \sum_{n=-\infty}^{\infty}L_n\,e^{-in x}= T(- x)
    \label{barTapp}
\end{equation}
and algebra elements are $ \overbar{T}_{u} = \int_{0}^{2\pi} d x\,u( x)\,\overbar{T}( x) $. In a CFT, one forms a tensor product of the original and the complex conjugate representations. The conjugate representation is used for the quantization of the left-moving component of the two-dimensional stress tensor, namely, we have
\begin{equation}
    T_{--}(x^-) = T(x^-) \otimes \mathbf{1},\quad T_{++}(x^+) = \mathbf{1}\otimes \overbar{T}(x^+)
\end{equation}
as in \eqref{TtildeT}. These conventions and signs match with \cite{moosavi_inhomogeneous_2019}.

\paragraph{Virasoro Hamiltonian.} The Virasoro Hamiltonian is defined as
\begin{equation}
H_f = V_{F}\,L_0\,V_{F}^{\dagger} = V_{F}\,T_{\ell_0}\,V_{F}^{\dagger}
\label{Hfasadjoint}
\end{equation}
up to a constant which cancels in the definition of the Virasoro state $\sigma_f$. Combined with \eqref{Adfu}, we get the expression for the Virasoro Hamiltonian in the main text
\begin{equation}
H_f = T_{\text{Ad}_{F}\ell_0} =\int_{0}^{2\pi}d x\,\frac{T( x)}{f'( x)} + \ldots,
\end{equation}
where we used \eqref{Adfu} and that $ \ell_0( x) = 1 $. Notice that the equality is up to a constant involving the Schwarzian derivative which cancels in the definition of a Virasoro state.

\subsection{Virasoro unitary as a time-ordered exponential}\label{app:virasorotimeordered}

The unitary operator can be written as a time-ordered exponential
\begin{equation}
	V_{f_t} = \mathcal{T}\exp{\biggl( -i\int_{0}^{t}ds\,G_s\biggr)}.
	\label{projectiveunitary}
\end{equation}
where $ G_t $ is a Hermitian operator and time decreases from the left to the right. It follows that the generator is given by
\begin{equation}
	G_t = i\,\partial_t V_{f_t}\,V^{\dagger}_{f_t} = i\,\partial_s(V_{f_s}V_{f_t^{-1}})\big\lvert_{s=t}.
	\label{processgenerator2}
\end{equation}
where we used $ B(f,f^{-1}) = 0 $. Using the composition rule \eqref{composition}, computing the derivative with respect to $ s $ and using \eqref{representation} gives
\begin{equation}
	G_t = T_{\dot{f}_t\circ f_t^{-1}} -\frac{c}{48\pi}\int_{0}^{2\pi} d x\,\frac{\dot{F}_t( x)}{F'_t( x)}\,\biggl(\frac{F''_{t}( x)}{F'_{t}( x)}\biggr)',
	\label{virasororightaction}
\end{equation}
where $F_t = f_t^{-1}$ and it matches with formulae in \cite{Oblak:2017ect,erdmenger_complexity_2020}. Hence the unitary \eqref{projectiveunitary} is explicitly
\begin{equation}
	V_{f_t} = e^{i c\alpha(f_t)}\;\mathcal{T}\exp{\biggl(-i\int_{0}^{t} ds\int_0^{2\pi}d x\,(\dot{f}_{s}\circ f^{-1}_{s})( x)\,T( x)\biggr)}
	\label{pathordered}
\end{equation}
where the phase is given by
\begin{equation}
	\alpha(f_t) = \frac{1}{48\pi}\int_{0}^{t} ds\int_0^{2\pi}d x\,\frac{\dot{F}_s( x)}{F'_s( x)}\,\biggl(\frac{F''_{s}( x)}{F'_{s}( x)}\biggr)'.
 \label{Virasorophase}
\end{equation}
Denote $ f \equiv f_1 $ for the diffeomorphism at $ t= 1 $ and consider the straight line $ \ddot{f}_t( x) = 0 $ from the identity to $ f $ whose tangent vector is independent of $ t $. In this case, the unitary \eqref{pathordered} reduces to
\begin{equation}
	V_f = e^{i c\alpha(f)}\,\exp{\biggl(-i\int_0^{2\pi}d x\,v( x)\,T( x)\biggr)},
	\label{Vftimeind}
\end{equation}
where $ v =\dot{f}_{t}\circ f^{-1}_{t} $ is the time-independent tangent vector and there is no more time-ordering. This is the formula presented in \cite{fewster_quantum_2005,fewster_probability_2019}. The argument can also be reversed: each tangent vector $ v $ at the identity defines a straight line such that \eqref{Vftimeind} represents the diffeomorphism $ f = f_1 $ obtained as the solution of the equation $ \dot{f}_t = v\circ f_t $ with $ f_0 = \text{id} $.

Lastly, for the complex conjugate representation $ \overbar{V}_{f_t} $, the expression is given by \eqref{pathordered} by replacing $ i \rightarrow -i $ and $T( x)\rightarrow \overbar{T}( x) = T(- x)$.

\section{Decomposition of the Polyakov action}\label{app:polyakovdecomposition}

In this Appendix, we will derive the decomposition of the Polyakov action when the metric is parametrized in terms of $ \omega,\nu,\overbar{\nu} $.

\subsection{Necessary identities}

We will first give an overview of identities. We consider Lorentzian cylinder with light-ray coordinates $\phi^\pm = \phi \pm t$ and a two-dimensional diffeomorphism $ F^{\pm}(t,\phi) $ and define
\begin{equation}
\nu = \frac{\partial_t F^{-}}{\partial_\phi F^{-}}, \quad \overbar{\nu} = \frac{\partial_t F^{+}}{\partial_\phi F^{+}}, \quad \mu = \frac{\partial_+ F^{-}}{\partial_- F^{-}}, \quad \overbar{\mu} = \frac{\partial_- F^{+}}{\partial_+ F^{+}},
\label{nusmusdefinition}
\end{equation}
where the derivative
\begin{equation}
    \partial_{\pm} \equiv \frac{\partial}{\partial \phi^{\pm}} = \frac{1}{2}\,(\partial_\phi \pm \partial_t).
\end{equation}
The ratios $ \mu,\overbar{\mu} $ are known as Beltrami differentials and we have
\begin{equation}
\nu = -\frac{1-\mu}{1+\mu},  \quad \overbar{\nu} = \frac{1-\overbar{\mu}}{1+\overbar{\mu}}, \quad \mu = \frac{1+\nu}{1-\nu}, \quad \overbar{\mu} = \frac{1-\overbar{\nu}}{1+\overbar{\nu}}.
\end{equation}
Notice that \eqref{nusmusdefinition} is equivalent to
\begin{equation}
(\partial_+ - \mu \partial_-)\,F^{-} = 0, \quad 	(\partial_- - \overbar{\mu}\partial_+)\,F^{+} = 0, \quad (\partial_t - \nu \partial_\phi)\,F^{-} = 0, \quad (\partial_t - \overbar{\nu} \partial_\phi)\,F^{+} = 0.
\label{firstequations}
\end{equation}
These equations are first order differential equations that fix $ F^{\pm} $ as non-local functionals of $ \nu,\overbar{\nu} $ or $ \mu,\overbar{\mu} $. In particular, the last two equations in \eqref{firstequations} are the Beltrami equations.

Using \eqref{firstequations} one can show that
\begin{align}
(\partial_t - \nu \partial_\phi)\,\log{(\partial_\phi F^{-})} &= \partial_\phi \nu\nonumber\\
(\partial_t - \overbar{\nu} \partial_\phi)\,\log{(\partial_\phi F^{-})} &= \partial_\phi \nu - (\overbar{\nu}-\nu)\,\partial_\phi\log{(\partial_\phi F^{-})}\nonumber\\
(\partial_t - \overbar{\nu} \partial_\phi)\,\log{(\partial_\phi F^{+})} &= \partial_\phi \overbar{\nu}\nonumber\\
(\partial_t - \nu \partial_\phi)\,\log{(\partial_\phi F^{+})} &= \partial_\phi \overbar{\nu} + (\overbar{\nu}-\nu)\,\partial_\phi\log{(\partial_\phi F^{+})}.
\label{nuidentities}
\end{align}
and similarly
\begin{align}
(\partial_+ - \mu \partial_-)\,\log{(\partial_- F^{-})} &= \partial_- \mu\nonumber\\
(\partial_- - \overbar{\mu} \partial_+)\,\log{(\partial_- F^{-})} &= -\overbar{\mu}(\partial_-\mu)+(1-\mu\overbar{\mu})\,\partial_-\log{(\partial_-F^{-})}\nonumber\\
(\partial_- - \overbar{\mu} \partial_+)\,\log{(\partial_+ F^{+})} &= \partial_+ \overbar{\mu}\nonumber\\
(\partial_+ - \mu \partial_-)\,\log{(\partial_+ F^{+})} &= -\mu(\partial_+\overbar{\mu})+(1-\mu\overbar{\mu})\,\partial_+\log{(\partial_+F^{+})}.
\label{muidentities}
\end{align}
Now a general two-dimensional metric can be written as
\begin{equation}
ds^{2} = e^{\omega}\,(d\phi + \nu dt)(d\phi + \overbar{\nu} dt) = e^{\tilde{\omega}}\,(d\phi^{-} + \mu\, d\phi^{+})(d\phi^{+} + \overbar{\mu}\, d\phi^{-}) = e^{\varphi}\,dx^-dx^+
\label{general2D}
\end{equation}
where we have defined the coordinate $  x^{\pm} = F^\pm(\phi^-,\phi^+) $ whose differentials are given by
\begin{align}
d x^{-} &=( \partial_{\phi}F^{-})\,(d\phi + \nu dt) = (\partial_-F^{-})(d\phi^{-}+\mu d\phi^{+})\nonumber\\
d x^{+} &= ( \partial_{\phi}F^{+})\,(d\phi + \overbar{\nu} dt)= (\partial_+F^{+})(d\phi^{+}+\overbar{\mu} d\phi^{-}).
\label{tildexcoordinates}
\end{align}
Hence the Weyl factors are related as
\begin{equation}
    \varphi = \omega -\log{( \partial_{\phi}F^{-})}-\log{( \partial_{\phi}F^{+})} = \tilde{\omega} -\log{( \partial_{-}F^{-})}-\log{( \partial_{+}F^{+})}.
\end{equation}
It follows that derivatives with respect to $  x^{\pm} $ are given by
\begin{align}
\widetilde{\partial}_{-}\equiv \frac{\partial}{\partial x^-} &=-\frac{1}{(\partial_\phi F^{-})}\frac{\partial_t - \overbar{\nu} \partial_\phi}{\overbar{\nu}-\nu}= \frac{1}{(\partial_-F^{-})}\frac{\partial_- - \overbar{\mu}\partial_+}{1 - \mu\overbar{\mu}} \nonumber\\
\widetilde{\partial}_{+}\equiv \frac{\partial}{\partial x^+} &= \frac{1}{(\partial_\phi F^{+})}\frac{\partial_t - \nu \partial_\phi}{\overbar{\nu}-\nu}= \frac{1}{(\partial_+F^{+})}\frac{\partial_+ - \mu\partial_-}{1 - \mu\overbar{\mu}}
\label{tildepartial}
\end{align}
These identities are sufficient to derive the decomposition of the Polyakov action.

\subsection{Decomposition in the $ \nu,\overbar{\nu} $ parametrization}

We consider the Polyakov action is
\begin{equation}
W[g] = -\frac{c}{192\pi}\int d^{2}x\sqrt{-g}\,R\,\frac{1}{\nabla^{2}}R
\label{polyakovapp}
\end{equation}
with the metric \eqref{general2D} given by
\begin{equation}
ds^2 = g_{AB}\,d\phi^{A}d\phi^{B} = e^{\omega}\,\hat{g}_{AB}\,d\phi^{A}d\phi^{B} = e^{\omega}\,(d\phi + \nu dt)(d\phi + \overbar{\nu} dt).
\end{equation}
Now the Polyakov action can be written as \cite{cappelli_stress_1989}
\begin{equation}
W[g] = I_{\text{L}}[\omega,\hat{g}] + W[\hat{g}]
\label{polyakovtransformation}
\end{equation}
where the Liouville action is given by
\begin{equation}
I_{\text{L}}[\omega,\hat{g}] = \frac{c}{96\pi}\int d^{2}x\sqrt{-\hat{g}}\,\left(\frac{1}{2}\,\hat{g}^{ab}\,\hat{\nabla}_{a} \omega\hat{\nabla}_{b}\omega+\omega \hat{R}\right).
\label{liouvilleapp}
\end{equation}
Here $\hat{\nabla}$ is the covariant derivative compatible with $\hat{g}$ and $\hat{R}$ is the corresponding Ricci scalar.

What remains is to decompose $ W[\hat{g}] $. In coordinates $  x^{\pm} $ defined in \eqref{tildexcoordinates}, the metric $\hat{g}$ takes the form
\begin{equation}
\hat{g}_{AB}\,d\phi^{A}d\phi^{B}  =e^{\hat{\varphi}}\,d x^{-}d x^{+}
\label{tildex}
\end{equation}
where the Weyl factor
\begin{equation}
\hat{\varphi} = \hat{\varphi}_- + \hat{\varphi}_+, \quad \hat{\varphi}_{\pm} \equiv -\log{(\partial_\phi F^{\pm})}.
\label{varphihatnu}
\end{equation}
Using again \eqref{polyakovtransformation}, the Polyakov action $ W[\hat{g}] $ is simply the Liouville action in these coordinates:
\begin{equation}
W[\hat{g}] = \frac{c}{48\pi}\int d x^{-} d x^{+}\,\widetilde{\partial}_+\hat{\varphi}\,\widetilde{\partial}_-\hat{\varphi}.
\end{equation}
Expanding the product and integrating the cross term $ \widetilde{\partial}_+\hat{\varphi}_+\,\widetilde{\partial}_-\hat{\varphi}_- $ by parts gives
\begin{equation}
W[\hat{g}] = \frac{c}{48\pi}\int d\phi dt\,(\partial_\phi F^{-})(\partial_\phi F^{+})\,(\overbar{\nu}-\nu)\,[\widetilde{\partial}_+\hat{\varphi}_-\,\widetilde{\partial}_-\hat{\varphi}_- + \widetilde{\partial}_+\hat{\varphi}_+\,\widetilde{\partial}_-\hat{\varphi}_+ + 2\,\widetilde{\partial}_-\hat{\varphi}_+\,\widetilde{\partial}_+\hat{\varphi}_-],
\label{startingdecomposition}
\end{equation}
where we also used that the integration measures are related as
\begin{equation}
d x^{-}\wedge d x^{+} =  d\phi\wedge  dt\,(\partial_\phi F^{-})(\partial_\phi F^{+})\,(\overbar{\nu}-\nu).
\end{equation}
By using \eqref{tildepartial} combined with the identities \eqref{nuidentities}, we get
\begin{align}
(\partial_\phi F^{-})(\partial_\phi F^{+})\,(\overbar{\nu}-\nu)\,\widetilde{\partial}_+\hat{\varphi}_-\,\widetilde{\partial}_-\hat{\varphi}_-&= -(\partial_\phi \nu)\biggl(\frac{\partial_\phi \nu}{\overbar{\nu}-\nu}-\partial_\phi\log{(\partial_\phi F^{-})}\biggr)\nonumber\\
(\partial_\phi F^{-})(\partial_\phi F^{+})\,(\overbar{\nu}-\nu)\,\widetilde{\partial}_+\hat{\varphi}_+\,\widetilde{\partial}_-\hat{\varphi}_+&= -(\partial_\phi \overbar{\nu})\biggl(\frac{\partial_\phi \overbar{\nu}}{\overbar{\nu}-\nu}+\partial_\phi\log{(\partial_\phi F^{+})}\biggr)\nonumber\\
(\partial_\phi F^{-})(\partial_\phi F^{+})\,(\overbar{\nu}-\nu)\,\widetilde{\partial}_-\hat{\varphi}_+\,\widetilde{\partial}_+\hat{\varphi}_-&= -\frac{(\partial_\phi \nu)(\partial_\phi \overbar{\nu})}{\overbar{\nu}-\nu}.
\label{derivativeproducts}
\end{align}
Substituting to \eqref{startingdecomposition}, rearranging and integrating by parts the terms involving $ \log{(\partial_\phi F^{\pm})} $ gives
\begin{align}
W[\hat{g}] &= -\frac{c}{48\pi}\int d\phi dt\,[ \nu\,\partial_\phi^{2}\log{(\partial_\phi F^{-})} - \overbar{\nu}\,\partial_\phi^{2}\log{(\partial_\phi F^{+})}] \\
&- \frac{c}{48\pi}\int d\phi dt\,\frac{1}{\overbar{\nu}-\nu}\,[(\partial_\phi \nu)+(\partial_\phi \overbar{\nu})]^{2}.
\end{align}
Hence we can write the full Polyakov action of the metric \eqref{general2D} as
\begin{equation}
W[g] = \Gamma[\nu] + \overbar{\Gamma}[\overbar{\nu}] + K[\nu,\overbar{\nu}] + I_{\text{L}}[\omega,\nu,\overbar{\nu}]
\end{equation}
where $ I_{\text{L}}[\omega,\nu,\overbar{\nu}] $ denotes the Liouville action \eqref{liouvilleapp} and
\begin{gather}
\Gamma[\nu]  = -\frac{c}{48\pi}\int d\phi dt\, \nu\,\partial_\phi^{2}\log{(\partial_\phi F^{-})} , \quad \overbar{\Gamma}[\overbar{\nu}] = \frac{c}{48\pi}\int d\phi dt\, \overbar{\nu}\,\partial_\phi^{2}\log{(\partial_\phi F^{+})}\\ K[\nu,\overbar{\nu}] = -\frac{c}{48\pi}\int d\phi dt\,\frac{1}{\overbar{\nu}-\nu}\,[(\partial_\phi \nu)+(\partial_\phi \overbar{\nu})]^{2}.
\label{appdecomposition}
\end{gather}

\subsubsection{Including a background Weyl factor}

We can also include a ``background'' in the Weyl factor $ \omega $ which will not be subtracted away as in Section \ref{sec:Polyakov}. In other words, we write the metric \eqref{general2D} as
\begin{equation}
ds^2 = g_{AB}\,d\phi^{A}d\phi^{B} = e^{\hat{\omega}}\,\hat{g}_{AB}\,d\phi^{A}d\phi^{B}
\end{equation}
where $\hat{\omega} = \omega-\chi-\overbar{\chi}$ so that this time
\begin{equation}
\hat{g}_{AB}\,d\phi^{A}d\phi^{B} = e^{\hat{\varphi}}e^{\chi+\overbar{\chi}}\,d x^{-}d x^{+}
\label{newbackground}
\end{equation}
includes an additional Weyl factor $ \chi+\overbar{\chi} $. To obtain $ W[\hat{g}] $, we have to simply replace
\begin{equation}
\hat{\varphi}_{-} \rightarrow \hat{\varphi}_{-} + \chi, \quad \hat{\varphi}_+ \rightarrow \hat{\varphi}_+ + \overbar{\chi}
\end{equation}
in equation \eqref{startingdecomposition} and compute each term again. The result of the calculation is the same as \eqref{derivativeproducts}, but with replacements
\begin{align}
\partial_\phi \nu &\rightarrow \mathcal{D}_{\phi} \nu - \partial_t \chi\nonumber\\
\partial_\phi \overbar{\nu} &\rightarrow \overbar{\mathcal{D}}_{\phi} \overbar{\nu} - \partial_t \overbar{\chi}\nonumber\\
\partial_\phi\log{\partial_\phi F^{-}} &\rightarrow\partial_\phi\log{\partial_\phi F^{-}} - \partial_\phi \chi\nonumber\\
\partial_\phi\log{\partial_\phi F^{+}} &\rightarrow\partial_\phi\log{\partial_\phi F^{+}} - \partial_\phi \overbar{\chi}
\label{replacementsnu}
\end{align}
where the covariant derivatives are defined as
\begin{equation}
\mathcal{D}_\phi\nu = \partial_\phi\nu + (\partial_\phi\chi)\,\nu, \quad \overbar{\mathcal{D}}_{\phi} \overbar{\nu} = \partial_\phi\overbar{\nu} + (\partial_\phi \overbar{\chi})\,\overbar{\nu}.
\end{equation}
We will now assume that the Weyl factors are of the form
\begin{equation}
\chi = \log{\partial_\phi D(\phi)}, \quad \overbar{\chi} = \log{\partial_\phi \overbar{D}(\phi)},
\end{equation}
that are independent of time $ \partial_t\chi = \partial_t \overbar{\chi} = 0 $. Taking this into account in \eqref{replacementsnu}, we can do the replacements to get that the Polyakov action of the metric \eqref{general2D} is given by
\begin{equation}
W[g] = \Gamma[\nu,\chi] + \overbar{\Gamma}[\overbar{\nu},\overbar{\chi}] + K[\nu,\overbar{\nu},\chi,\overbar{\chi}] + I_{\text{L}}[\hat{\omega};\nu,\overbar{\nu},\chi,\overbar{\chi}]
\end{equation}
where $ I_{\text{L}}[\hat{\omega};\nu,\overbar{\nu},\chi,\overbar{\chi}] $ is the Liouville action of $ \hat{\omega} = \omega- \chi-\overbar{\chi} $ in the background metric \eqref{newbackground} and
\begin{align}
\Gamma[\nu,\chi]  &= \frac{c}{48\pi}\int d\phi dt\, (\mathcal{D}_{\phi}\nu)\,(\partial_\phi\log{(\partial_\phi F^{-})} - \partial_\phi\chi)\nonumber\\
\overbar{\Gamma}[\overbar{\nu},\overbar{\chi}] &= -\frac{c}{48\pi}\int d\phi dt\, (\overbar{\mathcal{D}}_{\phi}\overbar{\nu})\,(\partial_\phi\log{(\partial_\phi F^{+})}- \partial_\phi\overbar{\chi})\nonumber\\ K[\nu,\overbar{\nu},\chi,\overbar{\chi}] &= -\frac{c}{48\pi}\int d\phi dt\,\frac{1}{\overbar{\nu}-\nu}\,[(\mathcal{D}_{\phi} \nu)+(\overbar{\mathcal{D}}_{\phi} \overbar{\nu})]^{2}.
\label{gengammaK}
\end{align}
By using the identity \eqref{nuidentities} and integrating by parts multiple times, we get
\begin{align}
\Gamma[\nu,\chi] &= \frac{c}{24\pi}\int d\phi dt\,\nu\,\biggl(-\frac{1}{2}\,\partial_\phi^{2}\log{(\partial_\phi F^{-})}+\{D(\phi),\phi\}\biggr)\nonumber\\
\overbar{\Gamma}[\overbar{\nu},\overbar{\chi}] &= \frac{c}{24\pi}\int d\phi dt\,\overbar{\nu}\,\biggl(\frac{1}{2}\,\partial_\phi^{2}\log{(\partial_\phi F^{+})}-\{\overbar{D}(\phi),\phi\}\biggr),
\label{genGammasimple}
\end{align}
where the Schwarzians are given by
\begin{equation}
\{D(\phi),\phi\} = \partial_{\phi}^{2}\chi - \frac{1}{2}\,(\partial_{\phi} \chi)^{2}, \quad \{\overbar{D}(\phi),\phi\} = \partial_{\phi}^{2}\overbar{\chi} - \frac{1}{2}\,(\partial_{\phi} \overbar{\chi})^{2}.
\end{equation}
We note that in the derivation of \eqref{genGammasimple} from \eqref{gengammaK}, terms proportional to $ \partial_\phi\partial_t \chi $ and $\partial_\phi\partial_t \overbar{\chi} $ appear that vanish by $ \partial_t\chi = \partial_t \overbar{\chi} = 0 $.

\subsection{Decomposition in the $ \mu,\overbar{\mu} $ parametrization}

For completeness, we will also derive the decomposition of the Polyakov action in the $ \mu,\overbar{\mu} $ parametrization written in \cite{Verlinde:1989hv,verlinde_conformal_1990,nguyen_holographic_2021}. As far as we are aware, an explicit derivation is not available in the literature.

The idea is the same as above. We consider the Polyakov action \eqref{polyakovapp} of the metric $g$ \eqref{general2D} given by
\begin{equation}
ds^2 = g_{AB}\,d\phi^{A}d\phi^{B} = e^{\tilde{\omega}}\,\hat{g}_{AB}\,d\phi^{A}d\phi^{B} = e^{\tilde{\omega}}\,(d\phi^{-} + \mu d\phi^{+})(d\phi^{+} + \overbar{\mu} d\phi^{-})
\label{mumug}
\end{equation}
where $\hat{g}$ should not be confused with \eqref{tildex} of the previous section. The Polyakov action decomposes as
\begin{equation}
    W[g] = I_{\text{L}}[\tilde{\omega},\hat{g}]+W[\hat{g}].
\end{equation}
In coordinates $  x^{\pm} $ defined in \eqref{tildexcoordinates}, the metric $\hat{g}$ in \eqref{mumug} takes the form
\begin{equation}
\hat{g}_{AB}\,d\phi^{A}d\phi^{B} = e^{\hat{\varphi}}\,d x^{-}d x^{+}
\label{tildexmu}
\end{equation}
where the Weyl factor this time is
\begin{equation}
\hat{\varphi} = \hat{\varphi}_- + \hat{\varphi}_+, \quad \hat{\varphi}_{\pm} \equiv -\log{(\partial_\pm F^{\pm})}
\label{hatvarphimu}
\end{equation}
which should not be confused with \eqref{varphihatnu} of the previous section. The remaining part of the Polyakov action is
\begin{equation}
W[\hat{g}] = \frac{c}{48\pi}\int d\phi^{-}d\phi^{+}\,(\partial_- F^{-})(\partial_+ F^{+})\,(1-\mu\overbar{\mu})\,[\widetilde{\partial}_+\hat{\varphi}_-\,\widetilde{\partial}_-\hat{\varphi}_- + \widetilde{\partial}_+\hat{\varphi}_+\,\widetilde{\partial}_-\hat{\varphi}_+ + 2\,\widetilde{\partial}_-\hat{\varphi}_+\,\widetilde{\partial}_+\hat{\varphi}_- ]
\label{startingdecompositionmu}
\end{equation}
where we used
\begin{equation}
d x^{-}\wedge d x^{+} = d\phi^{-}\wedge d\phi^{+}\,(\partial_- F^{-})(\partial_+ F^{+})\,(1-\mu\overbar{\mu}).
\end{equation}
Again by using \eqref{tildepartial} combined with the identities \eqref{muidentities}, we get
\begin{align}
(\partial_- F^{-})(\partial_+ F^{+})\,(1-\mu\overbar{\mu})\,\widetilde{\partial}_+\hat{\varphi}_-\,\widetilde{\partial}_-\hat{\varphi}_-&= (\partial_- \mu)\biggl(-\frac{\overbar{\mu}(\partial_-\mu)}{1-\mu\overbar{\mu}}+\partial_-\log{(\partial_-F^{-})}\biggr)\nonumber\\
(\partial_- F^{-})(\partial_+ F^{+})\,(1-\mu\overbar{\mu})\,\widetilde{\partial}_+\hat{\varphi}_+\,\widetilde{\partial}_-\hat{\varphi}_+ &=(\partial_+ \overbar{\mu})\biggl(-\frac{\mu(\partial_+\overbar{\mu})}{1-\mu\overbar{\mu}}+\partial_+\log{(\partial_+F^{+})}\biggr) \nonumber\\
(\partial_- F^{-})(\partial_+ F^{+})\,(1-\mu\overbar{\mu})\,\widetilde{\partial}_-\hat{\varphi}_+\,\widetilde{\partial}_+\hat{\varphi}_-&= \frac{(\partial_- \mu)(\partial_+\overbar{\mu})}{1-\mu\overbar{\mu}}.
\label{muidentitiescomputed}
\end{align}
Substituting to \eqref{startingdecompositionmu} and integrating by parts the terms involving $ \log{(\partial_\pm F^{\pm})} $ gives
\begin{align}
W[\hat{g}] &= -\frac{c}{48\pi}\int d\phi^{-}d\phi^{+}\,[\mu\,\partial_-^{2}\log{(\partial_-F^{-})} + \overbar{\mu}\,\partial_+^{2}\log{(\partial_+F^{+})}] \nonumber\\
&- \frac{c}{24\pi}\int d\phi^{-}d\phi^{+}\,\frac{1}{1-\mu\overbar{\mu}}\,\biggl[(\partial_- \mu)(\partial_+\overbar{\mu})-\frac{1}{2}\,\overbar{\mu}(\partial_-\mu)^{2}-\frac{1}{2}\,\mu(\partial_+\overbar{\mu})^{2}\biggr].
\end{align}
Hence we can write the full Polyakov action of the metric \eqref{general2D} as
\begin{equation}
W[g] = \widetilde{\Gamma}[\mu] + \widetilde{\overbar{\Gamma}}[\overbar{\mu}] + \widetilde{K}[\mu,\overbar{\mu}] + I_{\text{L}}[\tilde{\omega},\mu,\overbar{\mu}]
\end{equation}
where
\begin{gather}
\widetilde{\Gamma}[\mu]  = -\frac{c}{48\pi}\int d\phi^{-}d\phi^{+}\, \mu\,\partial_-^{2}\log{(\partial_-F^{-})}, \quad \widetilde{\overbar{\Gamma}}[\overbar{\mu}]  = -\frac{c}{48\pi}\int d\phi^{-}d\phi^{+}\, \overbar{\mu}\,\partial_+^{2}\log{(\partial_+F^{+})}\\
\widetilde{K}[\mu,\overbar{\mu}] = -\frac{c}{24\pi}\int d\phi^{-}d\phi^{+}\,\frac{1}{1-\mu\overbar{\mu}}\,\biggl[(\partial_- \mu)(\partial_+\overbar{\mu})-\frac{1}{2}\,\overbar{\mu}(\partial_-\mu)^{2}-\frac{1}{2}\,\mu(\partial_+\overbar{\mu})^{2}\biggr].
\end{gather}
The term $ \widetilde{K} $ has previously appeared in \cite{Verlinde:1989hv,verlinde_conformal_1990,nguyen_holographic_2021} and is known as the QBK term \cite{quillen_determinants_1985,belavin_algebraic_1986}.

\subsubsection{Including a background Weyl factor}

The background Weyl factor can also be included in the Beltrami parametrization which was done in \cite{lazzarini_integrating_1998}. For completeness, we will explain how to get the result of \cite{lazzarini_integrating_1998} in the special case when the background Weyl factor is a sum of right- and left-moving terms (we have not been able to reproduce the general result).

We write the metric \eqref{general2D} as
\begin{equation}
ds^2 = g_{AB}\,d\phi^{A}d\phi^{B} = e^{\hat{\tilde{\omega}}}\,\hat{g}_{AB}\,d\phi^{A}d\phi^{B}
\end{equation}
where $\hat{\tilde{\omega}} = \tilde{\omega}-\chi_+-\chi_-$ so that instead of \eqref{tildexmu} we now have
\begin{equation}
\hat{g}_{AB}\,d\phi^{A}d\phi^{B} = e^{\hat{\varphi}}e^{\chi_-+\chi_+}\,d x^{-}d x^{+}.
\label{newghatmu}
\end{equation}
In equation \eqref{startingdecompositionmu}, this amounts to the replacement
\begin{equation}
	\hat{\varphi}_{\pm}\rightarrow \hat{\varphi}_{\pm} + \chi_{\pm}, \quad \chi_{\pm} = \log{(\partial_\pm D_{\pm})},
\end{equation}
where $\hat{\varphi}$ is given by \eqref{hatvarphimu}. The resulting formulae \eqref{muidentitiescomputed} are the same up to the replacement
\begin{align}
	\partial_- \mu &\rightarrow \mathcal{D}_- \mu - \partial_- \chi_+\nonumber\\
	\partial_+ \overbar{\mu} &\rightarrow \mathcal{D}_+ \overbar{\mu} - \partial_+ \chi_-\nonumber\\
	\partial_{\pm}\log{(\partial_{\pm}F^{\pm})} &\rightarrow \partial_{\pm}\log{(\partial_{\pm}F^{\pm})}- \partial_{\pm}\chi_{\pm},
\end{align}
where the covariant derivatives are defined as
\begin{equation}
	\mathcal{D}_-\mu = \partial_-\mu + (\partial_- \chi_-)\,\mu, \quad \mathcal{D}_+\overbar{\mu} = \partial_+\overbar{\mu}  + (\partial_+\chi_+)\,\overbar{\mu}.
\end{equation}
Now considering the special case $\partial_{\mp}\chi_{\pm} = 0 $, we get that the Polyakov action of the metric \eqref{general2D} is given by
\begin{equation}
W[g] = \widetilde{\Gamma}[\mu,\chi_-] + \widetilde{\overbar{\Gamma}}[\overbar{\mu},\chi_+] + \widetilde{K}[\mu,\overbar{\mu},\chi_-,\chi_+] + I_{\text{L}}[\hat{\tilde{\omega}};\mu,\overbar{\mu},\chi_-,\chi_+]
\end{equation}
where $ I_{\text{L}}[\hat{\tilde{\omega}};\mu,\overbar{\mu},\chi_-,\chi_+] $ is the Liouville action of $\hat{\tilde{\omega}} = \tilde{\omega} - \chi_--\chi_+$ in the background metric \eqref{newghatmu} and
\begin{align}
\widetilde{\Gamma}[\mu,\chi_-] &= -\frac{c}{24\pi}\int d\phi^{-}d\phi^{+}\,\mu\,\biggl(\frac{1}{2}\,\partial_-^{2}\log{(\partial_-F^{-})} - \{D_-(x^{-}),x^{-}\}\biggr)\\
 \widetilde{\overbar{\Gamma}}[\overbar{\mu},\chi_+] &= -\frac{c}{24\pi}\int d\phi^{-}d\phi^{+}\,\overbar{\mu}\,\biggl(\frac{1}{2}\,\partial_+^{2}\log{(\partial_+F^{+})} - \{D_+(x^{+}),x^{+}\}\biggr)\\
\widetilde{K}[\mu,\overbar{\mu},\chi_-,\chi_+] &= -\frac{c}{24\pi}\int d\phi^{-}d\phi^{+}\,\frac{1}{1-\mu\overbar{\mu}}\,\biggl[(\mathcal{D}_- \mu)(\mathcal{D}_+ \overbar{\mu}) -\frac{1}{2}\,\overbar{\mu}\,(\mathcal{D}_- \mu)^{2}-\frac{1}{2}\,\mu\,(\mathcal{D}_+\overbar{\mu})^{2}\biggr]
\end{align}
which matches with \cite{lazzarini_integrating_1998}.

\section{Non-negativity of relative entropy between Virasoro states}\label{app:nonneg}

Relative entropy between two Virasoro states was computed in Section \ref{subsec:BKM} with the result
\begin{equation}
S(\sigma_{f_2}\lVert \sigma_{f_1}) = \frac{c\beta}{24\pi}\int_{0}^{2\pi}d\phi\,\biggl(\frac{\gamma}{2}\,[\mathcal{F}'(\phi)^{2}-1] - \{\mathcal{F}(\phi),\phi\}\biggr)
\label{1strelcircleapp}
\end{equation}
where we have defined
\begin{equation}
\gamma \equiv \frac{48\pi\langle T\rangle_{\beta}}{c}.
\end{equation}	
To prove non-negativity, we employ the \textit{average lemma} \cite{schwartz_projectively_1992,balog_coadjoint_1998} which is the inequality
\begin{equation}
\int_{0}^{2\pi}d\phi\,\{\mathcal{F}(\phi),\phi\} \leq \frac{1}{2}\int_{0}^{2\pi}d\phi\,[1-\mathcal{F}'(\phi)^{2}], \quad \mathcal{F} \in \widetilde{\text{Diff}}_+S^{1},
\label{averagelemma}
\end{equation}
with an equality if and only if $ \mathcal{F} \in \r{SL}(2,\mathbb{R}) $. Using \eqref{averagelemma} in \eqref{1strelcircleapp}, we get the lower bound
\begin{equation}
S(\sigma_{f_2}\lVert \sigma_{f_1}) \geq  \frac{c\beta}{24\pi}\frac{\gamma+1}{2}\int_{0}^{2\pi}d\phi\,[\mathcal{F}'(\phi)^{2}-1].
\label{rellowerbound}
\end{equation}
By using
\begin{equation}
\int_{0}^{2\pi}d\phi\,[\mathcal{F}'(\phi)^{2}-1] = \int_{0}^{2\pi}d\phi\,[\mathcal{F}'(\phi)-1]^{2},
\end{equation}
we get
\begin{equation}
S(\sigma_{f_2}\lVert \sigma_{f_1}) \geq  \frac{c\beta}{24\pi}\frac{\gamma+1}{2}\int_{0}^{2\pi}d\phi\,[\mathcal{F}'(\phi)-1]^{2} \geq 0,
\end{equation}
where the last inequality follows from the unitarity bound\footnote{By equation \eqref{Tbeta}, we have $\langle T \rangle_\beta = \frac{1}{2\pi}\tr{(\sigma_\beta\,L_0^{\text{rad}})} - \frac{c}{24\pi} $ from which \eqref{gammabound} follows assuming the theory admits only positive energy states $L_0^{\text{rad}}\geq 0$.}
\begin{equation}
\gamma \geq -1.
\label{gammabound}
\end{equation}
Hence relative entropy on the circle is non-negative. For $ \mathcal{F} \in \r{SL}(2,\mathbb{R}) $, we have an equality in \eqref{rellowerbound} which implies that $ S(\sigma_{f_2}\lVert \sigma_{f_1}) = 0 $ if and only if $ \mathcal{F} = \text{id} $ as required.

\addcontentsline{toc}{section}{References}
\bibliography{refs.bib}
\bibliographystyle{JHEP}

\end{document}